\documentclass[pra,twocolumn,showpacs,amsmath,amssymb]{revtex4}

\usepackage{graphicx}
\usepackage{dcolumn}
\usepackage{bm}

\begin{document}


\title{Some properties of the resonant state in quantum mechanics and its computation}

\author{Naomichi Hatano}
\affiliation{Institute of Industrial Science, University of Tokyo, Komaba, Meguro, Tokyo 153-8505, Japan}
\email{hatano@iis.u-tokyo.ac.jp}
\author{Keita Sasada}
\affiliation{Department of Physics, University of Tokyo, Komaba, Meguro, Tokyo 153-8505, Japan}
\email{sasada@iis.u-tokyo.ac.jp}
\author{Hiroaki Nakamura}%
\affiliation{Theory and Computer Simulation Center,
National Institute for Fusion Science,
Oroshi, Toki, Gifu 509-5292, Japan}
\email{nakamura.hiroaki@nifs.ac.jp}
\author{Tomio Petrosky}
\affiliation{Center for Complex Quantum Systems, University of Texas at Austin, 1 University Station, C1609, Austin, TX 78712, USA}
\email{petrosky@physics.utexas.edu}


\date{\today}

\begin{abstract}
The resonant state of the open quantum system is studied from the viewpoint of the outgoing momentum flux.
We show that the number of particles is conserved for a resonant state, if we use an expanding volume of integration in order to take account of the outgoing momentum flux;
the number of particles would decay exponentially in a fixed volume of integration.
Moreover, we introduce new numerical methods of treating the resonant state with the use of the effective potential.
We first give a numerical method of finding a resonance pole in the complex energy plane.
The method seeks an energy eigenvalue iteratively.
We found that our method leads to a super-convergence, the convergence exponential with respect to the iteration step.
The present method is completely independent of commonly used complex scaling.
We also give a numerical trick for computing the time evolution of the resonant state in a limited spatial area.
Since the wave function of the resonant state is diverging away from the scattering potential, it has been previously difficult to follow its time evolution numerically in a finite area.
\end{abstract}

\maketitle

\section{Introduction}
\label{sec1}

The resonance has been studied extensively for quite a long time~\cite{Gamow28,Siegert39,Peierls59,leCouteur60,Zeldovich60,Humblet61,Rosenfeld61,Humblet62,Humblet64-1,Jeukenne64,Humblet64-2,Mahaux65,Rosenfeld65,Landau77,Brandas89,Kukulin89}.
It appears in almost every field of physics from classical mechanics to quantum mechanics.
In spite of this fact, however, many fundamental aspects remain to be investigated.
Particularly in condensed-matter and statistical physics, many textbooks indeed introduce the complex eigenvalue of the resonant state only phenomenologically.

Meanwhile, the resonant phenomenon is of increasing importance especially in the quantum mechanics of mesoscopic devices.
When we use nano-devices, we inevitably attach leads to them.
Hence the devices are always open systems and have resonant states;
an electron comes into the device through a lead, is trapped in the confining potential of the device for a while with a finite lifetime, and may come out through another lead.
This resonant conduction has been intensely studied experimentally;
for example, the Fano resonance~\cite{Kobayashi02,Kobayashi03,Kobayashi04,Sato05} has attracted much attention.

In the present paper, we study the resonant state of the open quantum system in two parts.
In the first part, from Section~\ref{sec2} to Section~\ref{sec4}, we discuss the physical significance of the imaginary part of the resonant eigenvalue in a somewhat pedagogical manner, having in view condensed-matter and statistical physicists as a target reader.
In the course of presenting our results in various places, we have often received many questions regarding the well-established knowledge on the resonant state from condensed-matter and statistical physicists, and hence we find it necessary to review here fundamental properties of the resonant state for self-containedness.

We nevertheless include two new points in the discussions.
First, we show that 
\begin{quote}
the imaginary part of the energy expectation is proportional to a momentum flux going out of the system for \textit{arbitrary} wave functions.
\end{quote}
The relation has appeared in the literature but for \textit{each} resonant state~\cite{Berggren87,Moiseyev90,Masui99}, in which case the imaginary part of the energy expectation is reduced to the lifetime, or the half-width of the resonance and the momentum flux is reduced to the real part of the wave number of the eigenstate.

Second, we show that
\begin{quote}
the number of particles in a resonant state is \textit{conserved} when we count the number in an expanding volume.
\end{quote}
The resonant state represents a decaying state and its particle number decreases exponentially, \textit{if we count the number in a fixed volume}.
Since the decay rate is related to the outgoing momentum flux as stated above, we can keep track of leaking particles by expanding the integration volume constantly.
As far as we know, the above fact has never been stated in the literature albeit its simpleness.

In the second part of the present paper, from Section~\ref{sec5} to Section~\ref{sec6}, we introduce new numerical methods of treating the resonant state with the use of the effective potential~\cite{Livsic57,Feshbach58,Feshbach62,Okolowicz03,Datta95,Albeverio96,Fyodorov97,Dittes00,Pichugin01,Sadreev03,Sasada05,Sasada07}.
A logical consequence of the arguments in Sections~\ref{sec2} and~\ref{sec3} is that the eigenfunction of the resonant state is diverging away from the scattering potential.
For this reason, treating resonant states numerically can involve difficulties.
Naive treatment by chopping off the infinite space would fail inevitably;
the resonant state never appears in a closed system and the diverging eigenfunction is never represented in a finite space.
A conventional way is suppressing the divergence by the complex scaling, or the complex rotation~\cite{Masui99,Aguilar71,Baslev71,Simon72,Moiseyev78,Moiseyev88,Csoto90,Moiseyev98,Ho83,Homma97,Myo97,Myo98,Suzuki05,Aoyama06}.
We here use the effective potential, an energy-dependent boundary condition, in order to cut off the infinite space.
We can thereby find resonance poles in the complex energy plane, working in a finite space.
The present method seeks the energy eigenvalue iteratively.
We numerically demonstrate for a simple model that we can indeed find a complex eigenvalue of the resonance pole.
We found that our method is a rapidly converging method as in the KAM theory~\cite{Kolmogorov54,Barrar70} in nonlinear dynamics or in the Newton method of finding the solution of the nonlinear equation; \textit{i.e.}\ the convergence is exponential with respect to the iteration step.
We can also compute the time evolution of the resonant state in a finite space.
We numerically demonstrate for the Friedrichs model that we calculate the dynamics of the central part of the diverging resonant eigenfunction.
Numerical calculations for a ladder lattice finding a quasi-stable resonant states with very long lifetime will be reported elsewhere~\cite{Nakamura07}.


\section{The imaginary part of the energy expectation}
\label{sec2}

In the present section, we show that the imaginary part of the energy expectation is proportional to the flux out of a volume as well as the lifetime of the resonant state.
Thus the energy is generally complex in an open quantum system.
The relation has been obtained only for each resonant state previously.

Hereafter, we discuss the one-dimensional case just for explanatory simplicity.
See Appendix~\ref{appA-1} for the corresponding expressions in the three-dimensional case.
Consider the one-dimensional one-body Hamiltonian
\begin{equation}\label{eq2-10}
\hat{\mathcal{H}}=\hat{\mathcal{K}}+\hat{\mathcal{V}},
\end{equation}
where
\begin{equation}\label{eq2-20}
\hat{\mathcal{K}}\equiv
\frac{\hat{p}^2}{2m}=-\frac{\hbar^2}{2m}\frac{d^2}{dx^2}
\quad\mbox{and}\quad
\hat{\mathcal{V}}\equiv V(x).
\end{equation}
We assume that the potential operator $\hat{\mathcal{V}}$ has a finite support
\begin{equation}\label{eq2-25}
\Omega_\mathrm{pot}\equiv\{x\mid  -l\leq x\leq l\};
\end{equation}
that is, we assume that the potential function is reasonably localized around the origin.
We also assume that the potential operator is Hermitian; in other words, the potential function is real:
\begin{equation}\label{eq2-30}
V(x)^\ast\equiv V(x).
\end{equation}
We will see below that the kinetic energy is not necessarily a Hermitian operator.

\subsection{The imaginary part and the momentum flux}
\label{sec2-1}

Let us check whether the Hamiltonian~(\ref{eq2-10}) is a Hermitian operator or not.
We can do this by seeing whether the expectation value of the Hamiltonian operator is real for an arbitrary wave function $\psi(x)$.
We define the Hamiltonian expectation as follows:
\begin{equation}\label{eq2-40}
\langle\psi|\hat{\mathcal{H}}|\psi\rangle_\Omega
\equiv
\int_{-L}^{L}
\psi(x)^\ast \hat{\mathcal{H}} \psi(x) dx.
\end{equation}
Note here that the volume of the integration is limited to a segment
\begin{equation}\label{eq2-45}
\Omega\equiv\{x\mid -L\leq x\leq L\}
\end{equation}
that includes the support of the potential operator $\hat{\mathcal{V}}$; that is, we assume $\Omega\supset\Omega_\mathrm{pot}$, or $L>l$.
The reason for the trick of limiting the integration volume will be self-evident below.
The usual definition of the expectation is recovered in the infinite-segment limit after normalization:
\begin{equation}\label{eq2-50}
\langle\psi|\hat{\mathcal{H}}|\psi\rangle
\equiv
\lim_{|\Omega|\to\infty}
\frac{\langle\psi|\hat{\mathcal{H}}|\psi\rangle_\Omega}{\langle\psi|\psi\rangle_\Omega}.
\end{equation}

We now compute the imaginary part of the expectation,
\begin{equation}\label{eq2-60}
2i\mathop{\mathrm{Im}}
\langle\psi|\hat{\mathcal{H}}|\psi\rangle_\Omega
=
\langle\psi|\hat{\mathcal{H}}|\psi\rangle_\Omega
-\left(\langle\psi|\hat{\mathcal{H}}|\psi\rangle_\Omega\right)^\ast,
\end{equation}
and see in what case it remains.
Owing to the assumption that the potential is a real and localized function, the expectation of the potential term is real.
Only the kinetic term therefore contributes to Eq.~(\ref{eq2-60}).
The expectation of the kinetic term is transformed by the partial integration as
\begin{eqnarray}\label{eq2-70}
\langle\psi|\hat{\mathcal{K}}|\psi\rangle_\Omega
&=&
-\frac{\hbar^2}{2m}\int_{-L}^{L}
\psi(x)^\ast\psi''(x)dx
\nonumber\\
&=&\frac{\hbar^2}{2m}\int_{-L}^{L}
\psi'(x)^\ast\psi'(x)dx
\nonumber\\
&&-\frac{\hbar^2}{2m}
\left[
\psi(x)^\ast\psi'(x)
\right]_{x=-L}^{L}.
\end{eqnarray}
The first term of the right-hand side of Eq.~(\ref{eq2-70}) is real.
The second term can be complex, however, which yields the imaginary part of the Hamiltonian expectation.

Subtracting from Eq.~(\ref{eq2-70}) its complex conjugate, we have
\begin{eqnarray}\label{eq2-90}
\lefteqn{
2i\mathop{\mathrm{Im}}
\langle\psi|\hat{\mathcal{H}}|\psi\rangle_\Omega
}
\nonumber\\
&=&
-\frac{\hbar^2}{2m}
\left[
\psi(x)^\ast\psi'(x)-\psi(x)\psi'(x)^\ast
\right]_{x=-L}^{L}
\nonumber\\
&=&
-\frac{i\hbar}{m}
\mathop{\mathrm{Re}}
\left(
\psi(x)^\ast \hat{p}\psi(x)
\bigr|_{x=L}
-
\psi(x)^\ast \hat{p}\psi(x)
\bigr|_{x=-L}
\right),
\quad
\end{eqnarray}
where $\hat{p}$ is the momentum operator.
In light of generalizing the expression to higher-dimensional cases, we denote the last line as
\begin{equation}\label{eq2-100}
\langle\psi|\hat{p}_\mathrm{n}|\psi\rangle_{\partial\Omega}
\equiv
\psi(x)^\ast \hat{p}\psi(x)
\bigr|_{x=L}
-
\psi(x)^\ast \hat{p}\psi(x)
\bigr|_{x=-L},
\end{equation}
where $\hat{p}_\mathrm{n}$ is the normal component of the momentum on the surface $\partial\Omega$.
This represents the momentum flux going out of the segment $\Omega$.
Thus we arrive at
\begin{equation}\label{eq2-110}
\mathop{\mathrm{Im}}
\langle\psi|\hat{\mathcal{H}}|\psi\rangle_\Omega
=-\frac{\hbar}{2m}\mathop{\mathrm{Re}}
\langle\psi|\hat{p}_\mathrm{n}|\psi\rangle_{\partial\Omega}.
\end{equation}
This is a very insightful formula in analyzing resonant states.

We have shown this general result for arbitrary wave functions.
This relation, however, has been already derived for individual resonant states;
see \textit{e.g.}\ Eq.~(2.13) of Ref.~\cite{Berggren87}, Eq.~(24) of Ref.~\cite{Moiseyev90} and Eq.~(3.19) of Ref.~\cite{Masui99}.
For each resonant state $\psi_{\mathrm{res},n}$, the energy expectation is reduced to the complex eigenvalue
\begin{equation}\label{eq2-112}
\lim_{|\Omega|\to\infty}
\frac{\langle\psi|\hat{\mathcal{H}}|\psi\rangle_\Omega}{\langle\psi|\psi\rangle_\Omega}
=\varepsilon_n-i\frac{\Gamma_n}{2}
\end{equation}
as will be reviewed below in Section~\ref{sec4-1}.
The momentum flux, on the other hand, is reduced to
\begin{equation}\label{eq2-114}
\mathop{\mathrm{Re}}
\langle\psi|\hat{p}_\mathrm{n}|\psi\rangle_{\partial\Omega}.
=\hbar k_n,
\end{equation}
again as will be reviewed in Section~\ref{sec4-1}.
Taking account of the normalization factor $1/\sqrt{2\kappa_n}$ of the resonant wave function, we see that the relation~(\ref{eq2-110}) is reduced to
\begin{equation}\label{eq2-116}
\frac{\Gamma}{4\kappa_n}=\frac{\hbar^2}{2m}k_n,
\end{equation}
which has been obtained previously~\cite{Berggren87,Moiseyev90,Masui99} and will be derived in a different context in Section~\ref{sec3-2}.

The formula~(\ref{eq2-110}) claims that the imaginary part of the energy expectation indicates a momentum flux going out of the system.
This leads to the following statement:
\begin{quote}
The Hamiltonian (the kinetic term, in particular) is a Hermitian operator if the system is closed,
but is generally a non-Hermitian operator if it is open.
\end{quote}
The Hamiltonian of an open system is Hermitian only when the flux into the system is balanced with the flux out of the system.
Conversely, when we \textit{assume} that the Hamiltonian of an open system is Hermitian from the beginning, the flux must be automatically conserved.
We see indeed in the textbook example of the tunneling phenomenon that the sum of the reflection flux and the transmission flux is equal to the incident flux.
In fact, this physically reasonable phenomenon of the flux conservation is, from an algebraic point of view, a consequence of assuming that the energy variable is real.

\subsection{The imaginary part and the lifetime}
\label{sec2-2}

Now we mention the familiar relation between the imaginary part of an eigenvalue and the decay rate of the resonant state (see \textit{e.g.}~Ref.~\cite{Bohm89}).
We now consider the time-dependent Schr\"{o}dinger equation
\begin{equation}\label{eq2-120}
i\hbar\frac{\partial}{\partial t}\Psi(x,t)=\hat{\mathcal{H}}\Psi(x,t).
\end{equation}
We compute the time dependence of the particle number in the segment $\Omega$, given by
\begin{equation}\label{eq2-130}
N_\Omega(t)\equiv\langle\Psi|\Psi\rangle_\Omega.
\end{equation}
Its time derivative is transformed as
\begin{eqnarray}\label{eq2-140}
\lefteqn{
\frac{d}{dt}N_\Omega(t)
}
\nonumber\\
&=&
\int_{-L}^{L}
\left(
\Psi(x,t)^\ast
\frac{\partial\Psi(x,t)}{\partial t}
+
\frac{\partial\Psi(x,t)^\ast}{\partial t}\Psi(x,t)
\right)dx
\nonumber\\
&=&
-\frac{i}{\hbar}
\int_{-L}^{L}
\left[
\Psi(x,t)^\ast\hat{\mathcal{H}}\Psi(x,t)
-\Psi(x,t)\left(\hat{\mathcal{H}}\Psi(x,t)\right)^\ast
\right]dx
\nonumber\\
&=&\frac{2}{\hbar}\mathop{\mathrm{Im}}
\langle\Psi|\hat{\mathcal{H}}|\Psi\rangle_\Omega.
\end{eqnarray}

\begin{figure}
\begin{center}
\includegraphics[width=0.8\columnwidth]{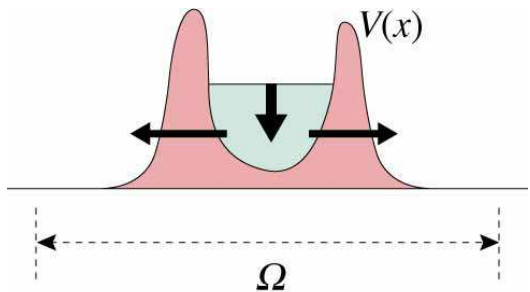}
\end{center}
\caption{The decrease of the particle number in the system is equal to the leak from the system.}
\label{fig2-10}
\end{figure}

We notice that the last line of Eq.~(\ref{eq2-140}) is related to the momentum flux in Eq.~(\ref{eq2-110}) as
\begin{equation}\label{eq2-150}
\frac{d}{dt}N_\Omega(t)=
\frac{2}{\hbar}\mathop{\mathrm{Im}}
\langle\Psi|\hat{\mathcal{H}}|\Psi\rangle_\Omega
=
-\frac{1}{m}\mathop{\mathrm{Re}}
\langle\Psi|\hat{p}_\mathrm{n}|\Psi\rangle_{\partial\Omega}.
\end{equation}
We thus conclude that the imaginary part of the Hamiltonian expectation is related to two quantities,
the momentum flux going out of the system and the decay rate of the particle number in the system.
This is a very natural conclusion;
the decrease of the particle number in the system is due to the leak from the system (Fig.~\ref{fig2-10}).

Equation~(\ref{eq2-150}) also indicates that dealing with the complex energy eigenvalue is closely related to dealing with the complex momentum of the eigenstate.
In fact, we have found that in some cases it is more convenient to solve the problem in the complex momentum plane than in the complex energy plane, which we will demonstrate below.

\section{Resonant state as a solution of the stationary Schr\"{o}dinger equation}
\label{sec3}

The facts established in the previous section indicate that the open quantum system can possess an eigenstate with a complex eigenvalue.
We generally refer to such a state as a resonant state.
In the present section, we review that the resonant state is an eigenfunction of the Hamiltonian of the open quantum system under the boundary condition
that we have only outgoing waves.
The statements in this section are known except for a comment on a resonance peak of the transmission probability presented in Sections~\ref{sec3-2} and~\ref{sec3-3}.
Readers familiar with the resonant state may skip to Section~\ref{sec3-2}.

The perhaps new point in the present section is the observation that a resonance peak of the transmission probability does not necessarily correspond to a resonant eigenstate.
We show in a tutorial example, specifically in Fig.~4(d) and~(e), that the lowest peak of the transmission probability, or the conductance, cannot be called a resonance peak in the sense that the corresponding resonant eigenstate is missing.

\subsection{Defining the resonant state as an eigenfunction}
\label{sec3-1}

The quantum resonant state is, in most textbooks, defined as a singularity of the $S$ matrix.
We show that this definition is actually equivalent to solving the stationary Schr\"{o}dinger equation under a certain boundary condition.
Specifically, the boundary condition is to have outgoing waves only, no incoming waves.
The momentum is obviously not conserved and hence the eigenvalue is complex as is indicated by Eq.~(\ref{eq2-110}).

We again discuss the one-dimensional case.
As in the previous section, we have a potential $V(x)$ in a finite region $\Omega_\mathrm{pot}$ defined in Eq.~(\ref{eq2-25}).
Outside the region, we have only the kinetic term in the Hamiltonian,
\begin{equation}\label{eq3-3}
V(x)=0\quad\mbox{for}\quad x\notin \Omega_\textrm{pot},
\end{equation}
and hence any eigenfunction is composed of $\exp(\pm iKx)$ with
\begin{equation}\label{eq3-5}
K=\frac{\sqrt{2mE}}{\hbar}.
\end{equation}

The argument in the previous section shows that a momentum flux going out of the central area ($\Omega$ defined in Eq.~(\ref{eq2-45})) may yield a state with a complex eigenvalue.
This motivates us to seek an eigenstate with boundary conditions with outgoing waves only:
\begin{equation}\label{eq3-70}
\psi_\textrm{res}(x)=\left\{
\begin{array}{ll}
Be^{-iKx} & \quad\mbox{for $\displaystyle x<-L$,} \\
Ce^{iKx}  & \quad\mbox{for $\displaystyle x>L$.}
\end{array}
\right.
\end{equation}
This set of boundary conditions is often called the Siegert condition~\cite{Siegert39,Landau77} and has been used occasionally in the literature.
An eigenfunction satisfying Eq.~(\ref{eq3-70}), if it exists, obviously has a finite momentum flux going out of the central area, and hence its energy eigenvalue $E$ must be complex because of the formula~(\ref{eq2-110}).

In the present paper, we
define the resonant state in the following way:
\begin{quote}
A resonant state is a solution of the Schr\"{o}dinger equation (an eigenfunction of the Hamiltonian) under the boundary conditions that only the outgoing waves exist outside the segment $\Omega$.
\end{quote}
We show below that this definition is equivalent to seeking poles of the $S$ matrix of the potential.

In order to define the $S$ matrix, we consider an eigenfunction satisfying the following boundary conditions (Fig.~\ref{fig3-10}) instead of Eq.~(\ref{eq3-70}):
\begin{figure}
\begin{center}
\includegraphics[width=0.8\columnwidth]{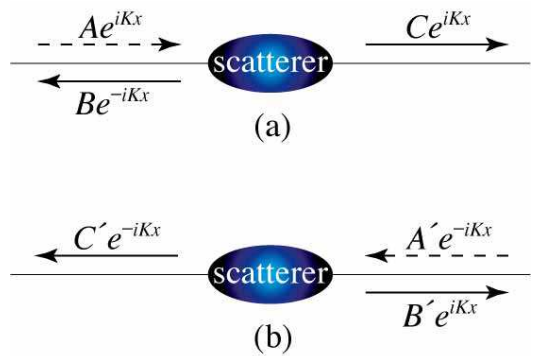}
\end{center}
\caption{The $S$ matrix of a scattering potential is defined from two scattering solutions as in (a) and (b).
The singularities of the $S$ matrix occur when the incident waves are missing.}
\label{fig3-10}
\end{figure}
\begin{equation}\label{eq3-10}
\psi(x)=\left\{
\begin{array}{ll}
Ae^{iKx}+Be^{-iKx} & \quad\mbox{for $\displaystyle x<-L$,} \\
Ce^{iKx}                 & \quad\mbox{for $\displaystyle x>L$.}
\end{array}
\right.
\end{equation}
In order for Eq.~(\ref{eq3-10}) to be a solution of the stationary Schr\"{o}dinger equation $\hat{\mathcal{H}}\psi(x)=E\psi(x)$, the coefficients $A$, $B$ and $C$ must depend on $E$.
The momentum flux out of the central area $\Omega$ can vanish for this set of boundary conditions whenever the outgoing flux is balanced by the incoming flux as in $|A(E)|^2=|B(E)|^2+|C(E)|^2$.
Then, an eigenstate satisfying Eq.~(\ref{eq3-10}) has a real energy eigenvalue.

The $S$ matrix in one dimension is defined as follows.
The reflection amplitude from the left back to the left is given by
\begin{equation}\label{eq3-20}
r_\mathrm{LL}(E)\equiv \frac{B(E)}{A(E)},
\end{equation}
while transmission amplitude from the left through to the right is given by
\begin{equation}\label{eq3-30}
t_\mathrm{RL}(E)\equiv \frac{C(E)}{A(E)}.
\end{equation}
Likewise, we can consider an eigenfunction satisfying
\begin{equation}\label{eq3-35}
\psi(x)=\left\{
\begin{array}{ll}
C'e^{-iKx}                 & \quad\mbox{for $\displaystyle x<-L$,} \\
A'e^{-iKx}+B'e^{iKx} & \quad\mbox{for $\displaystyle x>L$,}
\end{array}
\right.
\end{equation}
and have the reflection amplitude from the right back to the right and the transmission amplitude from the right through to the left as
\begin{equation}\label{eq3-40}
r_\mathrm{RR}(E)\equiv \frac{B'(E)}{A'(E)}
\quad\mbox{and}\quad
t_\mathrm{LR}(E)\equiv \frac{C'(E)}{A'(E)}.
\end{equation}
The $S$ matrix in this case is then defined in the form
\begin{equation}\label{eq3-50}
S\equiv\left(
\begin{array}{cc}
r_\mathrm{LL} & t_\mathrm{LR} \\
t_\mathrm{RL} & r_\mathrm{RR}
\end{array}
\right),
\end{equation}
which relates the incoming waves from the left and the right, to the outgoing waves to the left and the right.

Many textbooks define resonant states phenomenologically as singularities of the $S$ matrix analytically continued onto the complex $E$ plane;
the singularities are not associated to the true complex eigenstates of the Schr\"{o}dinger equation.
The elements of the $S$ matrix in fact diverges whenever
\begin{equation}\label{eq3-60}
A(E)=0\quad\mbox{or}\quad A'(E)=0.
\end{equation}
We notice that the set of boundary conditions~(\ref{eq3-10}) is reduced to Eq.~(\ref{eq3-70}) when $A(E)=0$.
Instead of solving the Schr\"{o}dinger equation under the boundary conditions~(\ref{eq3-10}) \textit{and then} seeking the zeros of $A(E)$,
Eq.~(\ref{eq3-70}) means that we solve the Schr\"{o}dinger equation with $A(E)=0$ from the very beginning.
Thus we see that the phenomenological definition of the resonant state given in many textbooks is equivalent to our definition of the resonant state as an eigenfunction.

%

Some might think that the outgoing waves in Eq.~(\ref{eq3-70}) spring out of nowhere; it is of course not true.
The particle as the source of the outgoing waves is there in the trapping potential from the very beginning.
The particle probability leaks from the potential at a rate proportional to the particle probability itself;
notice that the right-hand side of Eq.~(\ref{eq2-150}) has the factor $\exp(-t\mathop{\mathrm{Im}}E/\hbar)$.
Hence the particle probability in the trapping potential decreases, so does the leaking probability.
The particle probability keeps decreasing exponentially forever, as the leak keeps decreasing forever;
see Sec.~\ref{sec4} below for more details.
Since we here define the resonant state as a stationary state, we can only describe the resonance as above.
In this picture, we do not employ the dynamic description of the resonance that a particle comes into the scattering potential from outside, is trapped by the potential for a while before it escapes.

\subsection{Tutorial example}
\label{sec3-2}

We here confirm the above arguments, solving a simple example in one dimension.
Let us consider the potential with two delta functions:
\begin{equation}\label{eq3-105}
V(x)=V_0\left(\delta\left(x+l\right)+\delta\left(x-l\right)\right).
\end{equation}
In order to obtain the resonant state for the potential, we assume the form
\begin{equation}\label{eq3-110}
\psi_\textrm{res}(x)=\left\{
\begin{array}{ll}
Be^{-iKx} & \quad\mbox{for $\displaystyle x<-l$,} \\
Fe^{iKx}+Ge^{-iKx} & \quad\mbox{for $\displaystyle -l<x<l$,}\\
Ce^{iKx}  & \quad\mbox{for $\displaystyle x>l$.}
\end{array}
\right.
\end{equation}
Note that only outgoing waves exist outside the potential.
The connection conditions at $x=\pm l$ are given by
\begin{eqnarray}\label{eq3-120}
\psi_\textrm{res}\left(\pm l+0\right)
-\psi_\textrm{res}\left(\pm l-0\right)&=&0, \\
\label{eq3-130}
\psi_\textrm{res}'\left(\pm l+0\right)
-\psi_\textrm{res}'\left(\pm l-0\right)
&=&a^{-1}\psi_\textrm{res}\left(\pm l\right),
\end{eqnarray}
where the length scale $a$ is defined by
\begin{equation}
a\equiv\frac{\hbar^2}{2mV_0}.
\end{equation}
The solution~(\ref{eq3-110}) has four undetermined constants (the energy eigenvalue $E$ and the three ratios of the coefficients $B$, $C$, $F$, $G$), while we have four connection conditions~(\ref{eq3-120})--(\ref{eq3-130}).
Hence we can fix the energy eigenvalues to discrete values.

The solution is simplified further.
Since the potential~(\ref{eq3-105}) is an even function, the solutions can be classified according to their parities.
The even and odd solutions are given by
\begin{equation}\label{eq3-140}
B=\pm C\quad\mbox{and}\quad G=\pm F,
\end{equation}
respectively.
The solutions of a definite parity have two undetermined constants (the energy eigenvalue $E$ and the ratio between $C$ and $F$), while we have now two independent connection conditions at $x=+l$.
Eliminating the coefficients $C$ and $F$, we arrive at
\begin{equation}\label{eq3-210}
1-2iKa=\mp e^{2iKl},
\end{equation}
where the upper sign gives the even solutions and the lower sign gives the odd ones.
This equation determines discrete wave numbers $K_n$ and hence does discrete energy eigenvalues $E_n=\hbar^2 K_n^2/2m$ in the complex energy plane.

\begin{figure*}
\begin{center}
\begin{minipage}[t]{0.75\columnwidth}
\vspace{0mm}
\centering
\includegraphics[width=\textwidth]{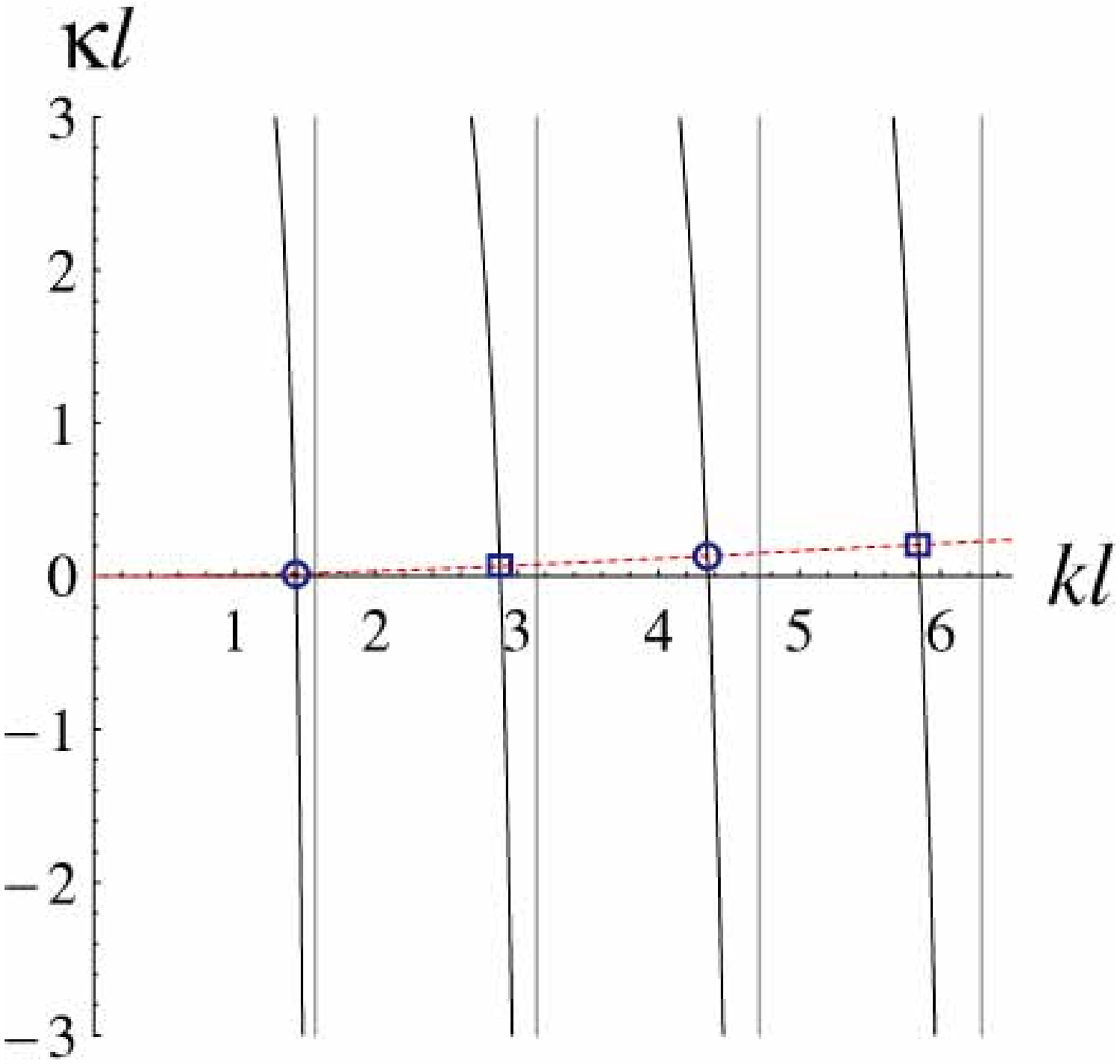}
(a)
\end{minipage}
\hspace{0.2\columnwidth}
\begin{minipage}[t]{0.75\columnwidth}
\vspace{0mm}
\centering
\includegraphics[width=\textwidth]{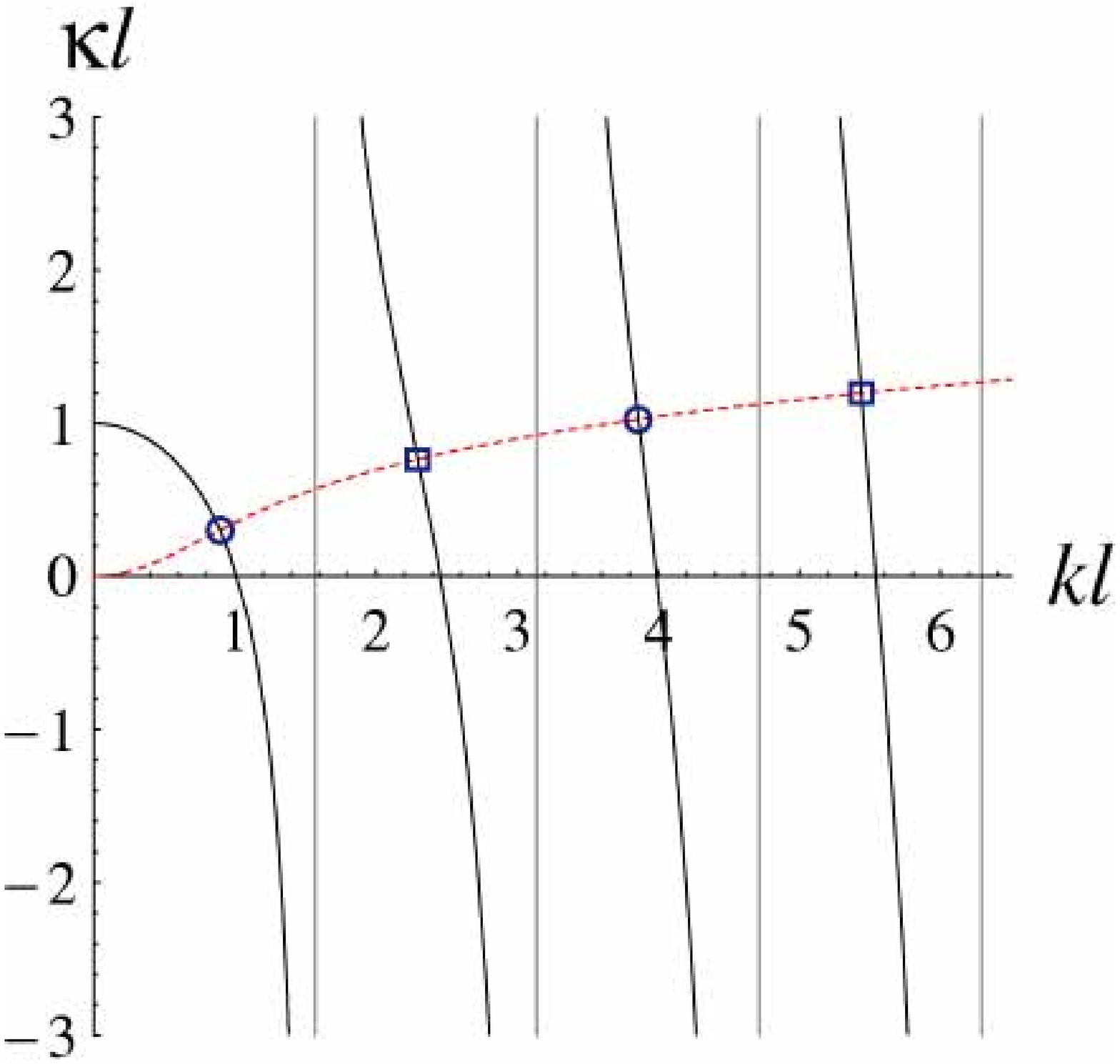}
(b)
\end{minipage}
\\
\begin{minipage}[t]{0.75\columnwidth}
\vspace{0mm}
\centering
\includegraphics[width=\textwidth]{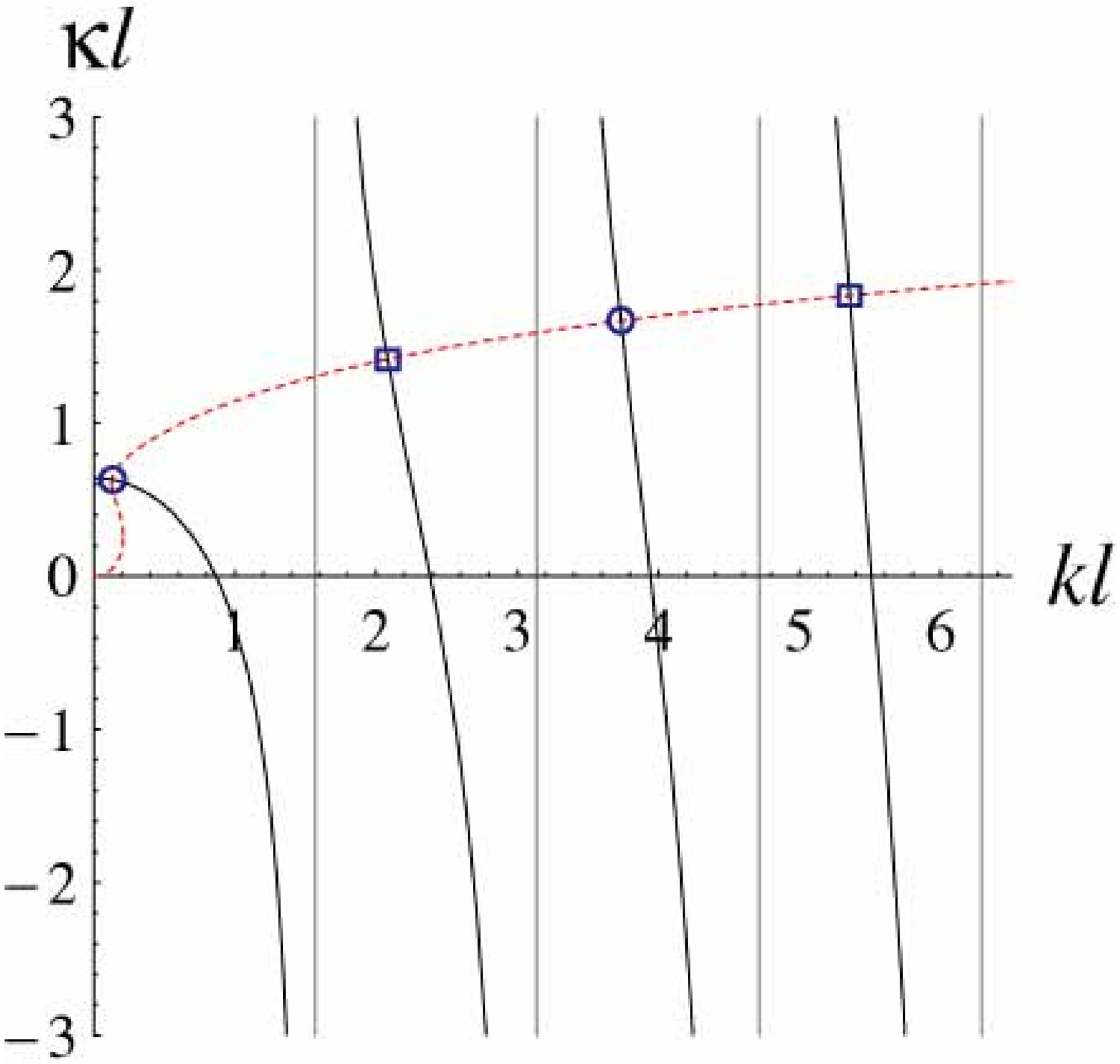}
(c)
\end{minipage}
\hspace{0.2\columnwidth}
\begin{minipage}[t]{0.75\columnwidth}
\vspace{0mm}
\centering
\includegraphics[width=\textwidth]{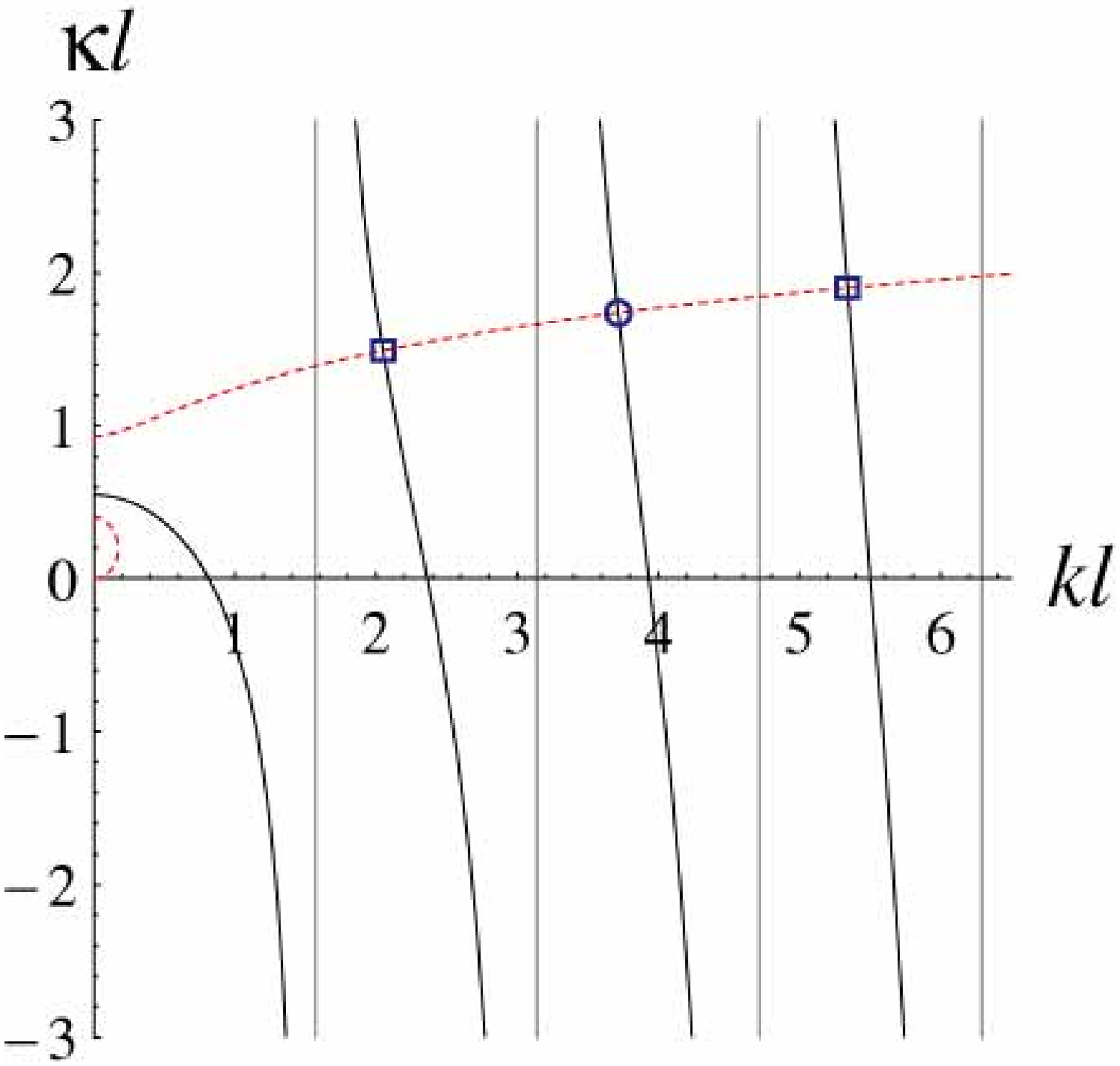}
(d)
\end{minipage}
\\
\begin{minipage}[t]{0.75\columnwidth}
\vspace{0mm}
\centering
\includegraphics[width=\textwidth]{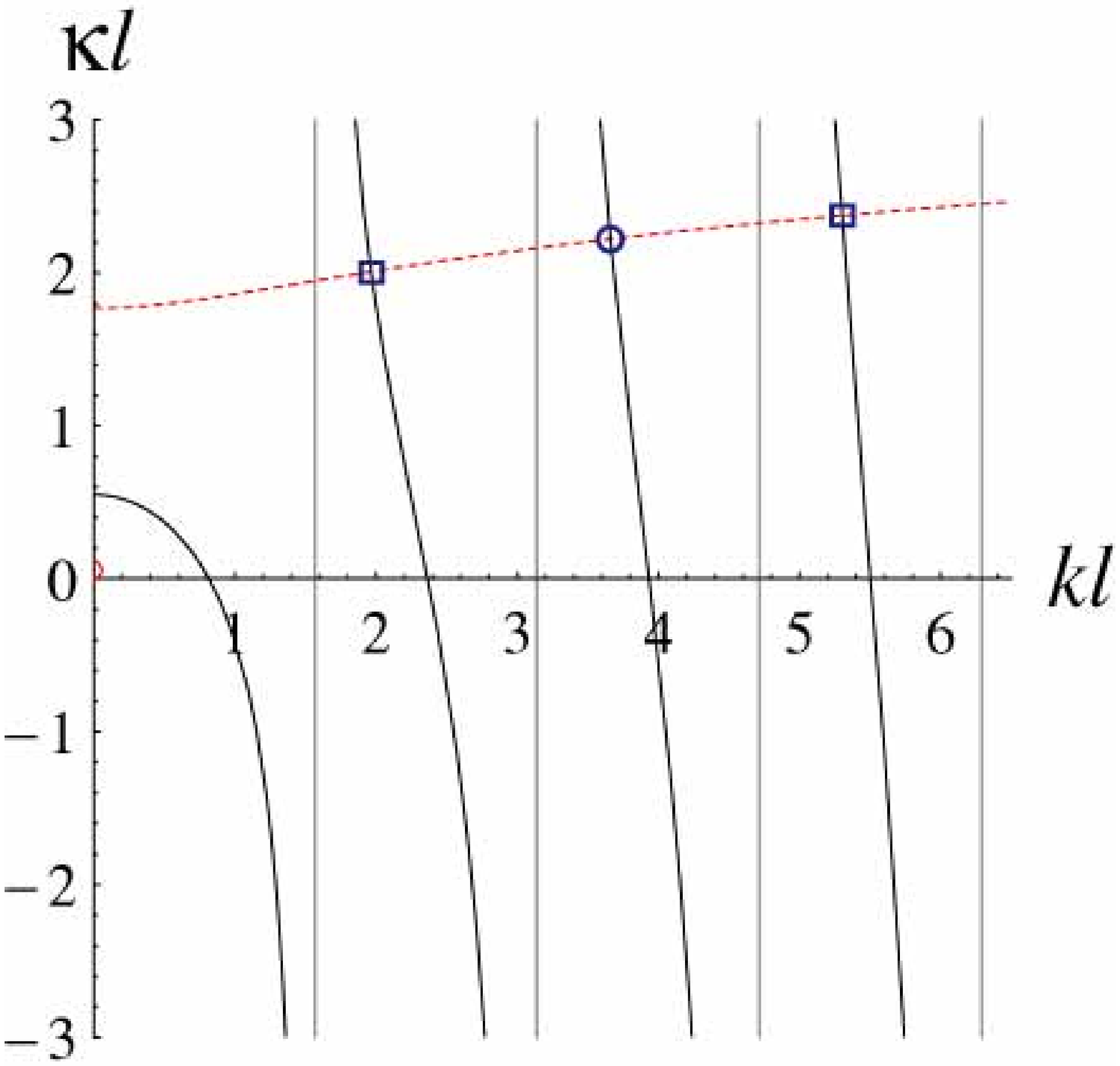}
(e)
\end{minipage}
\end{center}
\caption{Plots of Eq.~(\ref{eq3-240}) (the solid lines) and Eq.~(\ref{eq3-250}) (the broken lines) in the cases (a) $a/l=0.1$, (b) $a/l=1$, (c) $a/l=3.5$, (d) $a/l=4$ and (e) $a/l=10$.
The vertical thin gray lines indicate $kl=n\pi$, which are the bound states at $V_0\to\infty$ or $a=0$.
The circles indicate even solutions and the squares indicate odd solutions.
Note that the first even solution disappears in~(d) and~(e).}
\label{fig3-20}
\end{figure*}

For numerical calculations, we set
\begin{equation}\label{eq3-215}
K_n\equiv k_n-i\kappa_n,
\end{equation}
where $k_n>0$.
Equation~(\ref{eq3-210}) is then rewritten as
\begin{eqnarray}\label{eq3-240}
\eta_n&=&\frac{l}{2a}+\frac{\xi_n}{\tan 2\xi_n},\\
\label{eq3-250}
\xi_n&=&\pm\frac{l}{2a}\sqrt{e^{4\eta_n}-\left(1-\frac{2a}{l}\eta_n\right)^2},
\end{eqnarray}
where $\xi_n\equiv k_n l$ and $\eta_n\equiv \kappa_n l$.
We plot in Fig.~\ref{fig3-20} the functions~(\ref{eq3-240}) and~(\ref{eq3-250}) in the region $\xi=kl>0$ in the cases $a/l=0.1$, $1$, $4$ and $10$.
The crossing points are the solutions; the circles denote the solutions with even parity and the squares denote the solutions with odd parity.
There are also solutions in the region $\xi=kl<0$ with the negative real part $-\xi_n$ and the same imaginary part $\eta_n$;
Eqs.~(\ref{eq3-240}) and~(\ref{eq3-250}) are symmetric with respect to the $\eta$ axis if we take into account the sign of the right-hand side of Eq.~(\ref{eq3-250}).

The lowest even solution disappears for $a/l > 3.59112\cdots$.
The curve of Eq.~(\ref{eq3-250}) splits into two branches for $a/l > 3.59112\cdots$, one circling near the origin, the other continuing to infinity.
In the cases Fig.~\ref{fig3-20}~(d) and (e), there is a small branch near the origin, which does not intersect with Eq.~(\ref{eq3-240}) at all.
The remnant of the resonance peak corresponding to this missing pole nevertheless exists in the transmission probability; see the last paragraph of Section~\ref{sec3-3} below.
We have never seen this phenomenon reported previously.

Notice here that all solutions (the crossing points in Fig.~\ref{fig3-20}) satisfy
\begin{equation}\label{eq3-255}
\kappa_n=-\mathop{\mathrm{Im}}K_n>0.
\end{equation}
We list in Table~\ref{tab3-10} numerical estimates (by Mathematica) of the solutions $K_n$ and the resulting energy eigenvalues
\begin{equation}\label{eq3-260}
E_n=\frac{\hbar^2 K_n^2}{2m}\equiv \varepsilon_n-i\frac{\Gamma_n}{2}.
\end{equation}
\begin{table*}
\caption{Numerical estimates of the complex wave numbers and the complex eigenvalues of the first few resonant states of the double delta potential~(\ref{eq3-105}).
The solutions with $\mathop{\mathrm{Re}}K_n>0$ and $\mathrm{\mathrm{Im}}E_n<0$ are shown in Fig.~\ref{fig3-20}.
Note that the first even solution disappears in the cases $a/l=4$ and $a/l=10$.}
\label{tab3-10}
\begin{center}
\begin{tabular}{rclcl}
\hline
$a/l$ & state (parity) & \multicolumn{1}{c}{$K_n=k_n-i\kappa_n$ [$1/l$]} 
                       && \multicolumn{1}{c}{$E_n=\varepsilon_n-i\Gamma_n/2$ [$\hbar^2/ml^2$]} \\
\hline\hline
0.1 & 1st (even) & $\pm 1.4309486581029545770-i0.0180132370706695616$    
               && $1.023644792708441123\mp i0.025776017414365005$    \\
\cline{2-5}
    & 2nd (odd)  & $\pm 2.8775774584575874315-i0.0665106724899688980$    
               && $4.138014179934080207\mp i0.191389611903989679$    \\
\cline{2-5}
    & 3rd (even) & $\pm 4.3478216485135269076-i0.1331827632067771627$    
               && $9.442907719433745119\mp i0.579054901079276581$    \\
\cline{2-5}
    & 4th (odd)  & $\pm 5.8413795860760520688-i0.2064800963021565454$    
               && $17.03954071922854077\mp i1.206128619470434570$     \\
\hline
  1 & 1st (even) & $\pm 0.8940940206918146011-i0.3025104586463533055$    
               && $0.353945770123213981\mp i0.270472792272442937$    \\
\cline{2-5}
    & 2nd (odd)  & $\pm 2.2985790066512866386-i0.7660460609931899527$    
               && $2.348319441127416761\mp i1.760817393926857498$     \\
\cline{2-5}
    & 3rd (even) & $\pm 3.8592068943854588960-i1.0264132410357959781$
               && $6.919976856149325741\mp i3.961141056293867654$    \\
\cline{2-5}
    & 4th (odd)  & $\pm 5.4340030287668464008-i1.1969911205792216173$
               && $14.04780058695087972\mp i6.504453374634511715$     \\
\hline
 3.5 & 1st (even) & $\pm 0.1281226970608689462-i0.6318653191999576592$    
               && $-0.191419178052756592\mp i0.080956288875125433$    \\
\cline{2-5}
    & 2nd (odd)  & $\pm 2.0811197436940274902-i1.4192306059503943142$    
               && $1.158421937363385626\mp i2.953588834898203941$     \\
\cline{2-5}
    & 3rd (even) & $\pm 3.7327941962053519687-i1.6702340594138141162$
               && $5.572035348999205673\mp i6.234640003284390347$    \\
\cline{2-5}
    & 4th (odd)  & $\pm 5.3444838317948077028-i1.8348671938103419869$
               && $12.59838490469733564\mp i9.806418050810082608$     \\
\hline
 4 & 1st (odd)  & $\pm 2.0634804406374274361-i1.4929932341806764916$
               && $1.014461365791977689\mp i3.080762336735840215$    \\
\cline{2-5}
    & 2nd (even) & $\pm 3.7223094302121584017-i1.7400125856792561725$
               && $5.413971847962076268\mp i6.476865256361736495$    \\
\cline{2-5}
    & 3rd (odd)  & $\pm 5.3369637981977019936-i1.9033720492900427597$
               && $12.43017871262713230\mp i10.15822772156233025$     \\
\hline
 10 & 1st (odd)  & $\pm 1.9643116049421679709-i2.0079089942082502035$
               && $-0.086589223855955588\mp i3.944158938991022210$    \\
\cline{2-5}
    & 2nd (even) & $\pm 3.6591545130696776716-i2.2219369473965324296$
               && $4.226203976156184490\mp i8.130410608822284633$    \\
\cline{2-5}
    & 3rd (odd)  & $\pm 5.2907672454808363886-i2.3749885684814657674$
               && $11.17582367271761742\mp i12.56551172651315939$     \\
\hline
\end{tabular}
\end{center}
\end{table*}

The solutions with $k_n\equiv\mathop{\mathrm{Re}}K_n>0$ have outgoing waves only, whereas the solutions with $k_n<0$ (not shown in Fig.~\ref{fig3-20}) have incoming waves only.
We note here that Eq.~(\ref{eq3-260}) gives
\begin{eqnarray}\label{eq3-265}
\varepsilon_n&=&\frac{\hbar^2}{2m}(k_n^2-\kappa_n^2)
\quad\mbox{and}
\\
\label{eq3-266}
\Gamma_n&=&\frac{2\hbar^2}{m}k_n\kappa_n.
\end{eqnarray}
The latter equation is equivalent to Eq.~(\ref{eq2-116}).
Since we always have
\begin{equation}\label{eq3-270}
-\kappa_n\equiv\mathop{\mathrm{Im}}K_n<0,
\end{equation}
we have $\mathop{\mathrm{Im}}E_n<0$, or
\begin{equation}\label{eq3-280}
\Gamma_n>0
\quad\mbox{for the solutions with $k_n>0$}.
\end{equation}
This indicates that the states with outgoing waves only are indeed decaying.
For the solutions with $k_n<0$, on the other hand, we have $\mathop{\mathrm{Im}}E_n>0$, or
\begin{equation}\label{eq3-290}
\Gamma_n<0
\quad\mbox{for the solutions with $k_n<0$}.
\end{equation}
This indicates that the states are growing because of the incoming waves;
such states are often called anti-resonant states.
We argue these facts more in Sec.~\ref{sec4} below.

A decaying resonant state and a growing anti-resonant state always appear in a pair.
Both of the states break the time-reversal symmetry.
The latter states are the time-reversed states of the former.

\subsection{Relation to the $S$ matrix}
\label{sec3-3}

Let us now demonstrate for the double delta potential~(\ref{eq3-105}) that the pole of the $S$ matrix gives the same answer as those in Table~\ref{tab3-10}.
Computing the $S$ matrix, or computing the transmission and reflection amplitudes, differs from the computation in Sec.~\ref{sec3-2} in two points:
First, the wave function assumes the form
\begin{equation}\label{eq3-500}
\psi(x)=\left\{
\begin{array}{ll}
Ae^{iKx}+Be^{-iKx} & \quad\mbox{for $x<-l$,} \\
Fe^{iKx}+Ge^{-iKx} & \quad\mbox{for $-l<x<l$,}\\
Ce^{iKx}  & \quad\mbox{for $x>l$.}
\end{array}
\right.
\end{equation}
rather than Eq.~(\ref{eq3-110}).
Second, the wave number $K$ (and hence the energy eigenvalue $E$) is a given real number.
(We later carry out analytic continuation onto the complex $K$ plane.)
The undetermined constants are four ratios of the coefficients $A$, $B$, $C$, $F$, and $G$, 
while we have four connection conditions~(\ref{eq3-120})--(\ref{eq3-130}).
Hence the four ratios can be obtained as functions of $K$.
We can rephrase this point as follows:
we have five undetermined constants, the wave number $K$ and the four ratios, whereas we have only four conditions, and hence the solutions exist continuously on the real $K$ axis.

The connection conditions~(\ref{eq3-120})--(\ref{eq3-130}) relates the five coefficients in the forms
\begin{eqnarray}\label{eq3-510}
\lefteqn{
\left(
\begin{array}{cc}
e^{-iKl} & e^{iKl} \\
(iK+a^{-1})e^{-iKl} & -(iK-a^{-1})e^{iKl}
\end{array}
\right)
\left(
\begin{array}{cc}
A \\
B
\end{array}
\right)
}
\nonumber\\
&&\qquad\quad
=
\left(
\begin{array}{cc}
e^{-iKl} & e^{iKl} \\
iKe^{-iKl} & -iKe^{iKl}
\end{array}
\right)
\left(
\begin{array}{cc}
F \\
G
\end{array}
\right),
\end{eqnarray}
\begin{eqnarray}\label{eq3-520}
\lefteqn{
\left(
\begin{array}{cc}
e^{iKl} & e^{-iKl} \\
iKe^{iKl} & -iKe^{-iKl}
\end{array}
\right)
\left(
\begin{array}{cc}
F \\
G
\end{array}
\right)
}
\nonumber\\
&&\qquad\qquad\qquad
=
\left(
\begin{array}{c}
e^{iKl} \\
(iK-a^{-1})e^{iKl}
\end{array}
\right)C.
\end{eqnarray}
By eliminating the vector $\left(\begin{array}{c}F\\ G\end{array}\right)$, we have
\begin{eqnarray}\label{eq3-530}
\lefteqn{
\left(
\begin{array}{cc}
A \\
B
\end{array}
\right)
=\frac{-1}{4K^2a^2}\times
}
\nonumber\\
&&\quad
\left(
\begin{array}{c}
(2iKa-1)^2-e^{4iKl} \\
(2iKa-1)e^{-2iKl}+(2iKa+1)e^{2iKl}
\end{array}
\right)C.
\quad
\end{eqnarray}
The $S$ matrix is then given by
\begin{equation}\label{eq3-540}
S(K)=\left(
\begin{array}{cc}
r(K) & t(K) \\
t(K) & r(K)
\end{array}
\right)
\end{equation}
with
\begin{eqnarray}\label{eq3-550}
r(K)&\equiv&\frac{B}{A}
=\frac{4iKa \cos 2Kl + 2i\sin 2Kl}{(2iKa-1)^2-e^{4iKl}},
\\ \label{eq3-560}
t(K)&\equiv&\frac{C}{A}
=\frac{-4K^2a^2}{(2iKa-1)^2-e^{4iKl}}.
\end{eqnarray}

Now we carry out the analytic continuation onto the complex $K$ plane.
The poles of the elements of the $S$ matrix are determined by the zeros of the denominator,
\begin{equation}\label{eq3-570}
(2iKa-1)^2-e^{4iKl}=0,
\end{equation}
which is equivalent to Eq.~(\ref{eq3-210}) in Sec.~\ref{sec3-2}.
(It is notable, however, that the parity of the eigenstate is not obtained from the $S$ matrix.)

Note in Eq.~(\ref{eq3-530}) that the coefficient $A$ vanishes at the complex zeros of Eq.~(\ref{eq3-570}) and the wave function~(\ref{eq3-500}) is reduced to the resonant wave function~(\ref{eq3-110}).
(The flux is conserved only when $K$ is real.)
Thus we show that the solutions in Table~\ref{tab3-10} are identical to the resonant states given by the conventional definition.

\begin{figure*}
\begin{center}
\begin{minipage}[t]{0.85\columnwidth}
\vspace{0mm}
\centering
\includegraphics[width=\textwidth]{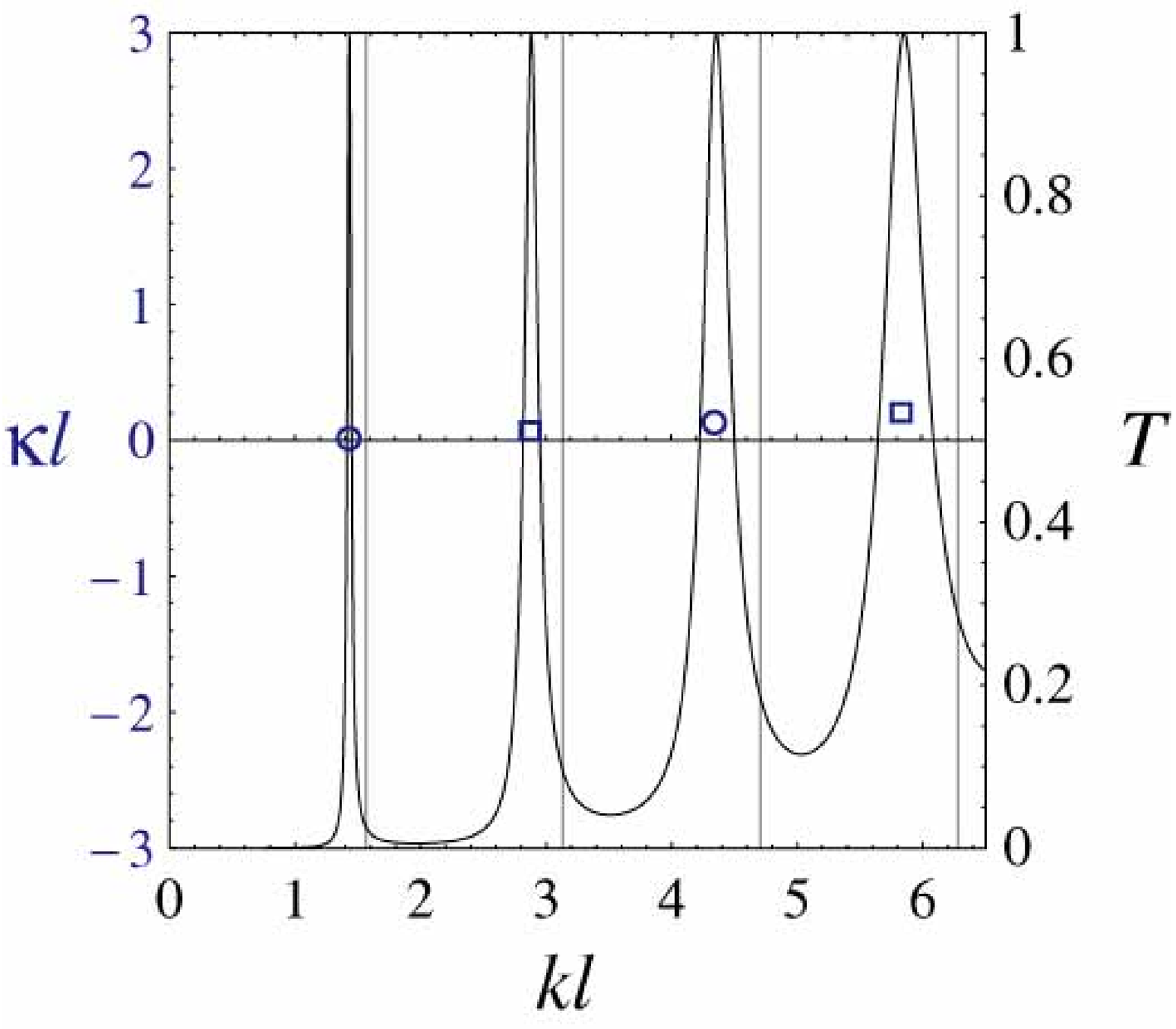}
(a)
\end{minipage}
\hspace{0.2\columnwidth}
\begin{minipage}[t]{0.85\columnwidth}
\vspace{0mm}
\centering
\includegraphics[width=\textwidth]{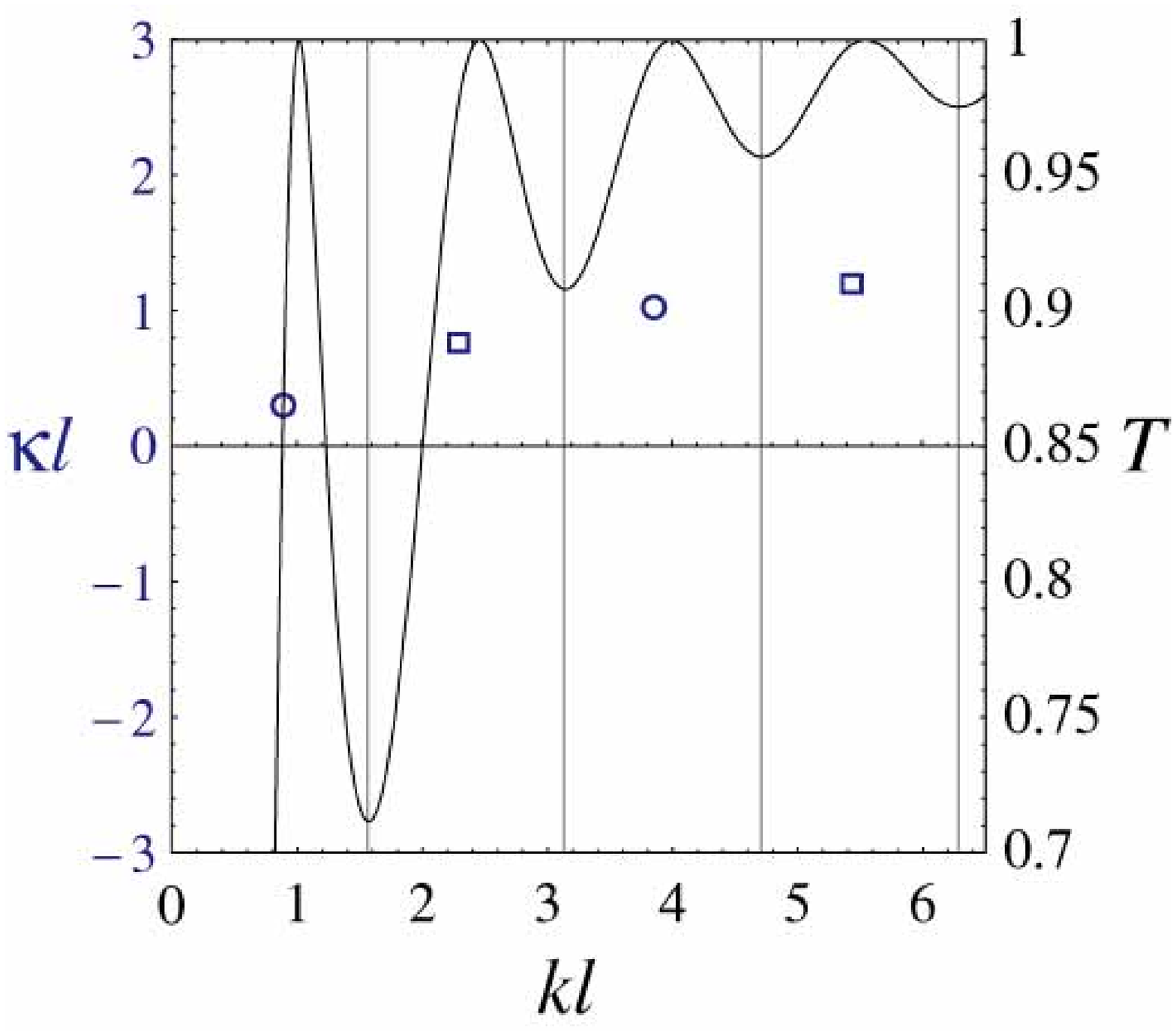}
(b)
\end{minipage}
\\
\begin{minipage}[t]{0.85\columnwidth}
\vspace{0mm}
\centering
\includegraphics[width=\textwidth]{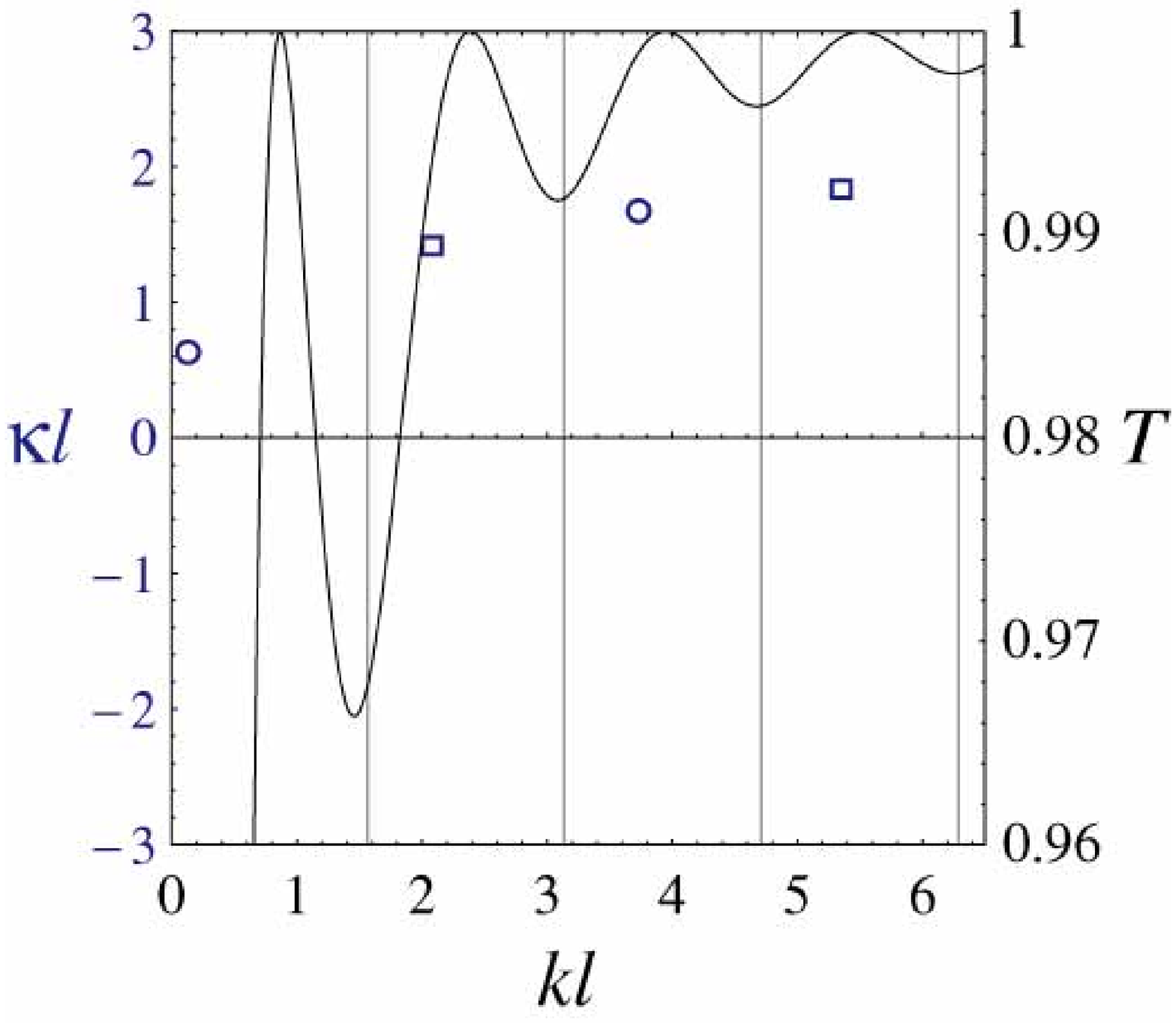}
(c)
\end{minipage}
\hspace{0.2\columnwidth}
\begin{minipage}[t]{0.85\columnwidth}
\vspace{0mm}
\centering
\includegraphics[width=\textwidth]{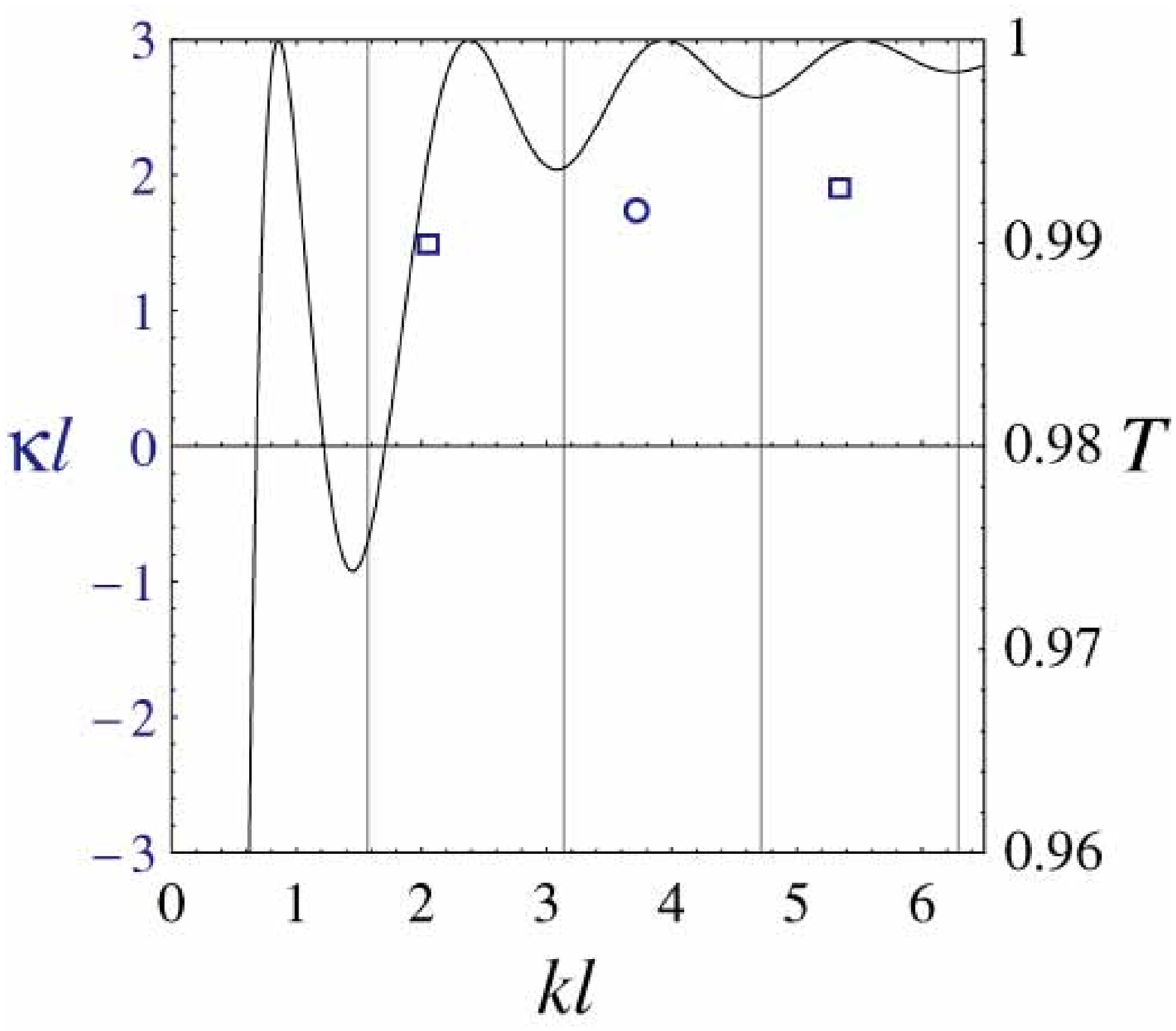}
(d)
\end{minipage}
\\
\begin{minipage}[t]{0.85\columnwidth}
\vspace{0mm}
\centering
\includegraphics[width=\textwidth]{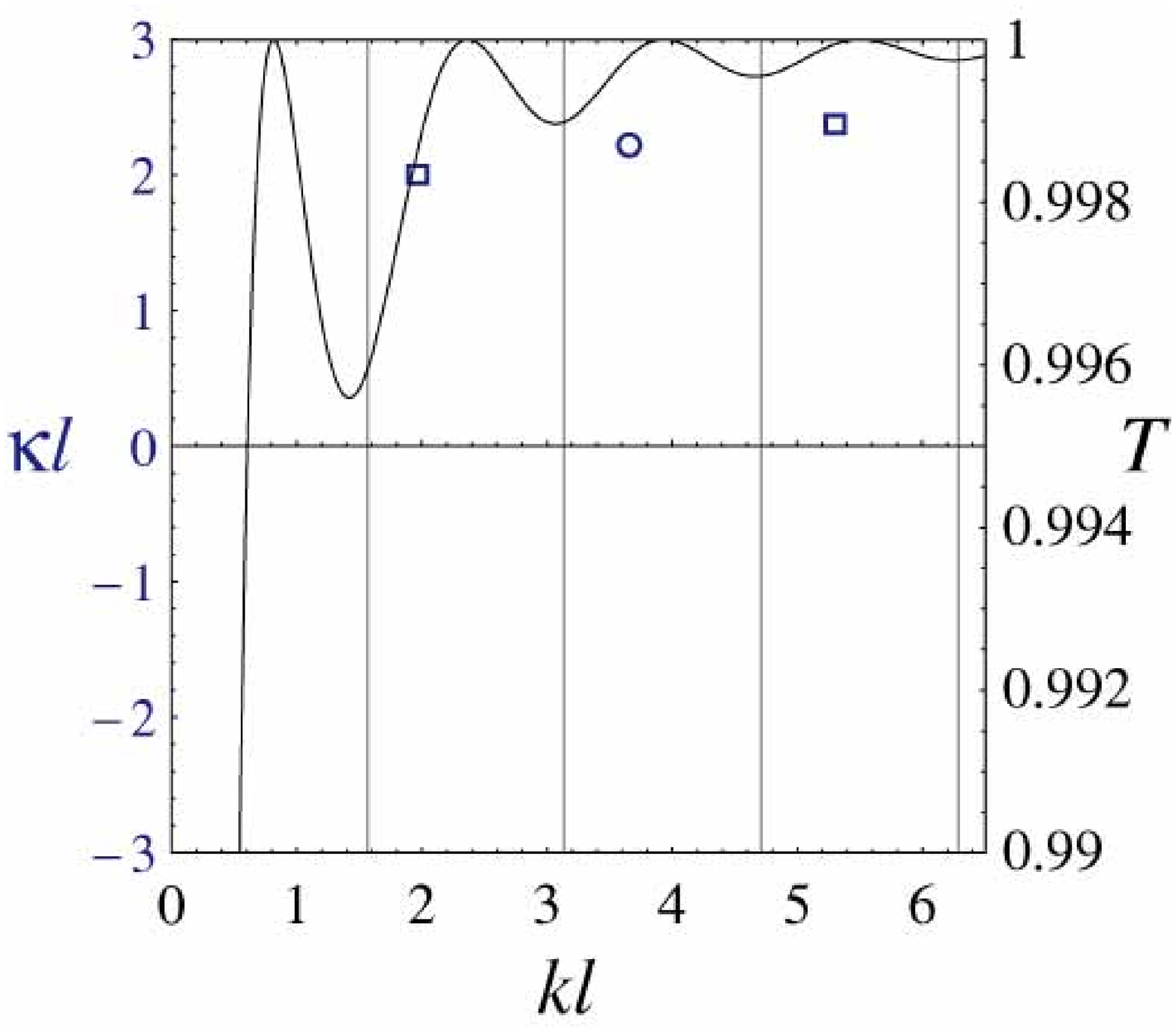}
(e)
\end{minipage}
\end{center}
\caption{The $k$ dependence of the transmission probability $T$ (the right axis of each panel) in the cases (a) $a/l=0.1$, (b) $a/l=1$, (c) $a/l=3.5$, (d) $a/l=4$, and (e) $a/l=10$.
The vertical thin gray lines indicate $kl=n\pi$.
The locations of resonant states given in Fig.~\ref{fig3-20} are superimposed (the left axis of each panel).
(The circles indicate the even solutions and the squares indicate the odd solutions.)
Note in~(d) and~(e) that the first ``resonance'' peak does not have a corresponding resonance pole.}
\label{fig3-30}
\end{figure*}

The transmission probability $T\equiv |t(K)|^2$ has ``resonance peaks" on the real $k$ axis (and on the real $\varepsilon$ axis) as the skirts of the poles of the function $t(K)$.
In Fig.~\ref{fig3-30}, we superimposed the locations of the resonant states on the $k$ dependence of the transmission probability $T$.
A resonant state with a small imaginary part yields a resonance peak near its real part, but a resonant state with a large imaginary part and the corresponding resonance peak are relatively apart from each other.

Note in Fig.~\ref{fig3-30}(d) and~(e) that the first ``resonance peak" does not have a corresponding resonant state as mentioned in Section~\ref{sec3-2};
perhaps we should not call this peak a resonance peak.

\section{Diverging eigenfunction and particle-number conservation}
\label{sec4}

Following the above tutorial example, we now discuss the resonant wave function in general cases.
We first show that the resonant eigenfunction is diverging away from the scattering potential.
In spite of, or rather, thanks to the divergence, the resonant wave function indeed conserves the particle number.

\subsection{Divergence of the resonant eigenfunction}
\label{sec4-1}

The general form of the one-dimensional resonant wave function~(\ref{eq3-70}) now reads
\begin{equation}\label{eq3-990}
\psi_{\mathrm{res},n}(x)=\left\{
\begin{array}{ll}
Be^{-ik_nx-\kappa_nx} & \quad\mbox{for $\displaystyle x<-L$,} \\
Ce^{ik_nx+\kappa_nx}   & \quad\mbox{for $\displaystyle x>L$.}
\end{array}
\right.
\end{equation}
We argue in the following that the above eigenfunction is diverging away from the potential.

The solution~(\ref{eq3-990}) is an eigenfunction of the stationary Schr\"{o}dinger equation under the boundary conditions of the outgoing waves only:
\begin{equation}\label{eq3-1000}
\hat{\mathcal{H}}\psi_{\mathrm{res},n}=\left(\varepsilon_n-i\frac{\Gamma_n}{2}\right)\psi_{\mathrm{res},n}.
\end{equation}
The solution of the time-dependent Schr\"{o}dinger equation~(\ref{eq2-120}) is then given by
\begin{equation}\label{eq3-1010}
\Psi_{\mathrm{res},n}(x,t)
=\psi_{\mathrm{res},n}(x)
\exp\left[-\frac{i}{\hbar}\left(\varepsilon_n-i\frac{\Gamma_n}{2}\right)t\right].
\end{equation}
Then the particle number of this particular state in the segment $\Omega$,
\begin{equation}\label{eq3-1020}
N_{n,\Omega}(t)\equiv\langle\Psi_{\mathrm{res},n}|\Psi_{\mathrm{res},n}\rangle_\Omega,
\end{equation}
decays as in Eq.~(\ref{eq2-140}), or
\begin{eqnarray}\label{eq3-1030}
\frac{d}{dt}N_{n,\Omega}(t)
&=&-e^{-\Gamma_n t/\hbar}\times\frac{\Gamma_n}{\hbar}
\langle\psi_{\mathrm{res},n}|\psi_{\mathrm{res},n}\rangle_\Omega
\nonumber\\
&=&-e^{-\Gamma_n t/\hbar}\times\frac{\Gamma_n}{\hbar}
N_{n,\Omega}(0).
\end{eqnarray}

If the eigenfunction has outgoing waves only, or $k_n\equiv\mathop{\mathrm{Re}}K_n>0$, then the state should decay in time, or $\Gamma_n>0$, because the particles leak from the central segment $\Omega$.
If the eigenfunction satisfies $k_n<0$, on the other hand, the state should grow in time, or $\Gamma_n<0$, because the particles gather into the central segment $\Omega$.
As illustrated in Sec.~\ref{sec3-2} and particularly in Eqs.~(\ref{eq3-265})--(\ref{eq3-280}), 
the above physical conditions
\begin{eqnarray}\label{eq3-1040}
k_n>0 & \Leftrightarrow & \Gamma_n>0,
\\
\label{eq3-2001}
k_n<0 & \Leftrightarrow & \Gamma_n<0,
\end{eqnarray}
are necessary and sufficient conditions of
\begin{equation}\label{eq3-1050}
\kappa_n\equiv -\mathop{\mathrm{Im}}K_n>0
\end{equation}
owing to the exact relation~(\ref{eq2-116}), or~(\ref{eq3-266}),
\begin{equation}\label{eq3-1060}
\Gamma_n=\frac{2\hbar^2}{m}k_n\kappa_n.
\end{equation}
Upon applying the inequality~(\ref{eq3-1050}) to the wave function~(\ref{eq3-990}), we realize that the resonant eigenfunction must be diverging as $|x|\to\infty$.

We can also show the necessity of the divergence of the resonant eigenfunction~(\ref{eq3-990}) by the following dimensional analysis.
The formula~(\ref{eq2-110}) reads for the eigenfunction~(\ref{eq3-990}),
\begin{equation}\label{eq3-400}
\frac{\Gamma_n}{2}
\langle\psi_{\mathrm{res},n}|\psi_{\mathrm{res},n}\rangle_\Omega
=\frac{\hbar^2k_n}{2m}
\langle\psi_{\mathrm{res},n}|\psi_{\mathrm{res},n}\rangle_{\partial\Omega}.
\end{equation}
As we expand the volume of the integration, $\Omega$, the left-hand side of Eq.~(\ref{eq3-400}) would increase in the order of the volume of the region $|\Omega|$, whereas the right-hand side would increase in the order of $|\Omega|^{(d-1)/d}$, where $d$ is the dimensionality of the space.
This apparent inconsistency comes from the implicit assumption that the wave function is almost constant over the volume of the integration.
A way out of the inconsistency is that the wave function in Eq.~(\ref{eq3-400}) grows or decays exponentially as we expand the integration volume $\Omega$.

Indeed, the solution~(\ref{eq3-990}) gives the left-hand side of the formula~(\ref{eq3-400}) as
\begin{equation}\label{eq3-2000}
\frac{\Gamma_n}{2}\langle\psi_{\mathrm{res},n}|\psi_{\mathrm{res},n}\rangle_\Omega
\simeq
\frac{\Gamma_n}{2}\int_{-L}^{L}
e^{2\kappa_n |x|}dx
=\frac{\Gamma_n}{2\kappa_n}e^{2\kappa_n L},
\end{equation}
where we have neglected the details of the wave function around the origin.
Meanwhile, the solution~(\ref{eq3-990}) gives the right-hand side of the formula~(\ref{eq3-400}) as
\begin{equation}\label{eq3-2010}
\frac{\hbar^2k_n}{2m}
\langle\psi_{\mathrm{res},n}|\psi_{\mathrm{res},n}\rangle_{\partial\Omega}
=
\frac{\hbar^2 k_n}{m}e^{2\kappa_n L},
\end{equation}
which are equal to each other thanks to the relation~(\ref{eq3-1060}).
(In fact, it is customary to normalize the diverging wave function by first introducing a Gaussian convergence factor $\exp(-\alpha x^2)$ in the volume integration in Eq.~(\ref{eq3-2000}) and then, after integration, putting $\alpha$ to 0.
This gives the normalization factor $1/\sqrt{2\kappa_n}$ for the eigenfunction.)

Thus we here characterize the resonant state as \textit{an eigenfunction in the form of an exponentially divergent outgoing wave and an eigenvalue with a negative imaginary part}.
We show in the next subsection that the divergence is physically plausible; the resonant eigenfunction must diverge in order to conserve the particle number.
How to deal with the divergence is also the central issue of Secs.~\ref{sec5} and~\ref{sec6};
we propose methods of getting rid of the divergence computationally in Sections~\ref{sec5} and~\ref{sec6}.

\subsection{Particle-number conservation}
\label{sec4-2}

We here show that we can intuitively understand the divergence of the resonant eigenfunction from the point of view of the particle-number conservation.
Because the energy eigenvalue is complex, the particle number is not conserved in the conventional sense.
We show that the particle number in a volume that expands in time is indeed conserved.
We then see that the resonant eigenfunction must be diverging in order to conserve the particle number.

As we show in Fig.~\ref{fig3-10}, the number of the particles in the trapping potential decreases because some of the particles escape the potential.
The particles that escaped run at the velocity $\hbar k_n/m$.
The idea here is to count the particle number in a volume that expands at the same velocity;
in the simple one-dimensional case, for example, let the boundary $L$ expand as
\begin{equation}\label{eq3-360}
L(t)=\frac{\hbar k_n}{m} t.
\end{equation}
Then the particle number included in the expanding volume $\Omega(t)$ should be conserved.

\begin{figure}
\begin{center}
\includegraphics[width=0.8\columnwidth]{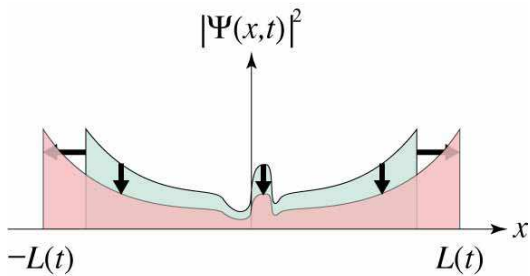}
\end{center}
\caption{The wave amplitude decreases exponentially at every point, whereas it increases exponentially as $|x|\to\infty$.
The integral is constant.}
\label{fig3-40}
\end{figure}

We can estimate the particle number~(\ref{eq3-1020}) as
\begin{eqnarray}\label{eq3-370}
N_{n,\Omega(t)}(t)&=&
e^{-\Gamma_nt/\hbar}
\int_{-L(t)}^{L(t)}
\left|\psi_{\mathrm{res},n}\right|^2 dx
\nonumber \\
&\simeq&
e^{-\Gamma_nt/\hbar}
\times\frac{1}{\kappa_n}
e^{2\kappa_nL(t)}
\nonumber\\
&=&\frac{1}{\kappa_n}
\exp\left[
\left(-\frac{\Gamma_n}{\hbar}+\frac{2\hbar k_n\kappa_n}{m}\right)t
\right],
\end{eqnarray}
where we have neglected the details of the wave function around the origin.
This is indeed constant because of the relation~(\ref{eq3-1060}).
The exponential temporal decrease of the wave amplitude is supplemented by its exponential spatial increase (Fig.~\ref{fig3-40}).

In short, the divergence of the resonant eigenfunction indicates that the particle eventually runs away to infinity.
The divergence is not just plausible but necessary for the particle number to be conserved.

We note that the edge of the expanding volume is well defined and never spread in spite of the fact that we have a nonlinear dispersion relation $E=\hbar^2 K^2/2m$.
This is because we are considering here properties of a single resonant state, and hence are not considering spreading of an arbitrary wave packet.

\section{Numerical method of finding resonant eigenvalues}
\label{sec5}

As we argued in Section~\ref{sec4}, the eigenfunction of the resonant state necessarily diverges away from the scattering potential.
It can be difficult to treat resonant states numerically because of this divergence.
The resonant state would disappear if we naively cut off the infinite space for numerical calculations.
A conventional method is to suppress the divergence by the complex scaling, or the complex rotation~\cite{Masui99,Aguilar71,Baslev71,Simon72,Moiseyev78,Moiseyev88,Csoto90,Moiseyev98,Ho83,Homma97,Myo97,Myo98,Suzuki05,Aoyama06}.
The complex rotation modifies the Hamiltonian by
\begin{eqnarray}\label{eq4-0022}
\hat{p}&\to&\hat{p}(\theta)\equiv\hat{p}e^{-i\theta},
\\
\label{eq4-0024}
x&\to&x(\theta)\equiv xe^{i\theta}.
\end{eqnarray}
The modification of the momentum operator, Eq.~(\ref{eq4-0022}), makes the resonant eigenfunctions in the $L^2$ functional space.
The modification of the coordinate operator, Eq.~(\ref{eq4-0024}), which must be introduced to conserve the commutation relation, can be problematic, however.
The potential $V(x)$ must be an analytic function in order for the modification~(\ref{eq4-0024}) to be applied~\cite{Aguilar71,Baslev71,Simon72};
we cannot apply the complex scaling to a box potential nor to any lattice models.
When the potential $V(x)$ is a Gaussian scattering potential, the complex scaling is applicable only for $\theta<\pi/4$;
otherwise, the potential becomes divergent.

In the present section, we propose a new numerical method of finding resonant eigenvalues iteratively in the complex energy plane.
Our method is completely independent of the complex scaling and free from the above restrictions;
in fact, we demonstrate the method for a lattice model.
The method utilizes the effective potential, which is called the self-energy of the leads in the context of the condensed-matter physics.
The effective potential is nothing but an energy-dependent boundary condition and is useful in cutting off the infinite space of the open system.

The resonant eigenvalue is exactly obtained once the energy-dependent effective potential is fixed exactly.
To fix the effective potential exactly, however, we need to know the resonant eigenvalue exactly.
In other words, the resonant eigenvalue and the effective potential must be self-consistent.
We make a postulated value of the eigenvalue converge to the exact value by seeking the self-consistency between the eigenvalue and the effective potential.

\subsection{Effective potential}
\label{sec5-1}

Before presenting our numerical method, let us introduce the effective potential by following Ref.~\cite{Sasada07}.
The effective potential was first introduced in nuclear physics half a century ago~\cite{Livsic57,Feshbach58,Feshbach62,Okolowicz03}.
It is used in physics of mesoscopic systems from time to time in recent years~\cite{Datta95,Albeverio96,Fyodorov97,Dittes00,Pichugin01,Sadreev03,Sasada05,Sasada07}.

We demonstrated in Section~\ref{sec3} that we can obtain the resonant eigenvalue using the boundary condition of purely outgoing waves.
When the scattering potential is complicated, we may need to solve the Schr\"{o}dinger equation numerically.
Numerical solution of the Schr\"{o}dinger equation often begins with discretization of the space, if the space is continuous.
Regardless of whether the system is the tight-binding model or the continuous model after discretization, the Schr\"{o}dinger equation takes the form
\begin{eqnarray}\label{eq5-20}
\lefteqn{
\hat{\mathcal{H}}\psi_{\mathrm{res},n}(x)
}
\nonumber\\
&=&-\frac{t_\mathrm{h}}{2}\left(
\psi_{\mathrm{res},n}(x-\Delta x)
+\psi_{\mathrm{res},n}(x+\Delta x)
\right)
\nonumber\\
&&+V(x)\psi_{\mathrm{res},n}(x)
\nonumber\\
&=&E_n \psi_{\mathrm{res},n}(x),
\end{eqnarray}
where $\Delta x$ denotes either the lattice constant or the grid size of the discretization mesh.
We dropped in Eq.~(\ref{eq5-20}) the constant potential term $t_\mathrm{h}\psi_{\mathrm{res},n}(x)$ that arises from the discretization.

Suppose that we try to solve the Schr\"{o}dinger equation in the finite region $-L\leq x \leq L$, which contains the support of the potential; in other words, the potential is zero at the boundaries and outside the region:
\begin{equation}\label{eq5-25}
V(x)\equiv 0
\quad\mbox{for}\quad
|x|\geq L.
\end{equation}
Then, the resonant eigenfunction at the boundaries must contain the outgoing wave only:
\begin{equation}\label{eq5-30}
\psi_{\mathrm{res},n}(x)\propto e^{iK_n |x|}
\quad\mbox{for}\quad
|x|\geq L.
\end{equation}
By using the form~(\ref{eq5-30}) in the Schr\"{o}dinger equation in the region without the potential, we find that the wave number $K_n$ is related to the eigenvalue $E_n$ as 
\begin{equation}\label{eq5-35}
E_n=-t_\mathrm{h}\cos (K_n \Delta x),
\end{equation}
where $E_n$ and $K_n$ are generally complex numbers;
see Eqs.~(\ref{eq5-1000})--(\ref{eq5-1030}) below for the derivation.

We here cast Eq.~(\ref{eq5-30}) into the form
\begin{equation}\label{eq5-40}
\psi_{\mathrm{res},n}\left(\pm \left(L+\Delta x\right)\right)=
e^{iK_n\Delta x}
\psi_{\mathrm{res},n}\left(\pm L\right)
\end{equation}
and substitute it for the term in Eq.~(\ref{eq5-20}) outside the region in question.
We thus have 
\begin{eqnarray}\label{eq5-50}
\lefteqn{
\hat{\mathcal{H}}\psi_{\mathrm{res},n}\left(\pm L\right)
}
\nonumber\\
&=&
-\frac{t_\mathrm{h}}{2}
\psi_{\mathrm{res},n}\left(\pm L-\Delta x\right)
-\frac{t_\mathrm{h}}{2}
\psi_{\mathrm{res},n}\left(\pm L+\Delta x\right)
\nonumber\\
&=&-\frac{t_\mathrm{h}}{2}
\psi_{\mathrm{res},n}\left(\pm \left(L-\Delta x\right)\right)
\nonumber\\
&&+V_{\mathrm{eff},n}\left(\pm L\right)
\psi_{\mathrm{res},n}\left(\pm L\right)
\nonumber\\
&=&E_n \psi_{\mathrm{res},n}(x),
\end{eqnarray}
where the effective potential is given by
\begin{equation}\label{eq5-60}
V_{\mathrm{eff},n}\left(\pm L\right)\equiv -\frac{t_\mathrm{h}}{2}e^{iK_n\Delta x}
\end{equation}
Using Eq.~(\ref{eq5-35}), this is rewritten in terms of the energy eigenvalue as
\begin{equation}\label{eq5-65}
V_{\mathrm{eff},n}\left(\pm L\right)=\frac{1}{2}\left(E_n-i\sqrt{{t_\mathrm{h}}^2-{E_n}^2}\right),
\end{equation}
where we assumed $\mathop{\mathrm{Re}}K_n>0$ in order to fix the sign of the imaginary part.

Equation~(\ref{eq5-65}) is the form often found in the literature (\textit{e.g.}~Refs.~\cite{Datta95,Sadreev03}), where the effective potential is obtained as the self-energy that comes into the energy denominator of the Green's function.
The easier derivation above is noted in Ref.~\cite{Sasada07}.

The effective potential~(\ref{eq5-65}) is an energy-dependent complex potential;
the potential is different for different resonant states.
As a result, the Hamiltonian matrix (such as Eq.~(\ref{eq5-1110}) below) with the effective potential is a non-Hermitian operator and may have a complex eigenvalue.
The point is that Eq.~(\ref{eq5-50}) contains only the eigenfunctions inside the region $-L\leq x\leq L$.
Hence we can solve the Schr\"{o}dinger equation by restricting ourselves to the region in question.

\subsection{Self-consistent solution}
\label{sec5-2}

We now propose a numerical method of finding a resonant eigenvalue by utilizing the self-consistency between the resonant eigenvalue and the effective potential.
The method starts with a postulate of a complex resonant eigenvalue $E^{(0)}$.
This gives the corresponding complex effective potential
\begin{equation}\label{eq5-100}
V_{\mathrm{eff}}^{(0)}\left(\pm L\right)=\frac{1}{2}\left(E^{(0)}-i\sqrt{{t_\mathrm{h}}^2-{E^{(0)}}^2}\right).
\end{equation}
We can then diagonalize the Hamiltonian in the region $-L\leq x \leq L$ with the effective potential $V_{\mathrm{eff}}^{(0)}$.
From the resulting eigenvalues we choose the one closest to the postulate $E^{(0)}$ and refer to it as an updated postulate $E^{(1)}$.
This again gives the corresponding effective potential $V_{\mathrm{eff}}^{(1)}$.
We repeat the procedure of
\begin{enumerate}
\item obtaining an updated postulate $E^{(q)}$ from the diagonalization of the Hamiltonian with the effective potential $V_{\mathrm{eff}}^{(q-1)}$
\item and setting the effective potential $V_{\mathrm{eff}}^{(q)}$ from the postulate $E^{(q)}$.
\end{enumerate}

For a few simple examples that we have tested, the convergence is surprisingly good even with the above straightforward iteration.
A more sophisticated iteration may be necessary in more complicated cases.
In the following, we demonstrate that the straightforward iteration for a simple model Hamiltonian yields an exponentially rapid convergence.

We solve the Schr\"{o}dinger equation~(\ref{eq5-20}) with the potential
\begin{equation}\label{eq5-1000}
V(x)=
\left\{
\begin{array}{ll}
V_0 > 0 & \quad\mbox{for $x=\pm \Delta x$,} \\
0 & \quad\mbox{otherwise.}
\end{array}
\right.
\end{equation}
Let us first solve the problem exactly.
As we showed in Section~\ref{sec3}, we assume that the eigenfunction has outgoing waves only.
We have solutions with even parity and solutions with odd parity.
Equations~(\ref{eq3-70}) and~(\ref{eq3-140}) show that we can assume the form
\begin{equation}\label{eq5-1010}
\psi_{\mathrm{res},n}(x)=
\left\{\begin{array}{ll}
\pm B_n e^{-iK_nx}& \quad\mbox{for $x\leq -\Delta x$,} \\
F_n & \quad\mbox{for $x=0$},\\
B_n e^{iK_nx} & \quad\mbox{for $x\geq \Delta x$,}
\end{array}\right.
\end{equation}
where $B_n$ and $F_n$ are constant.
The Schr\"{o}dinger equation for $|x|>\Delta x$,
\begin{equation}\label{eq5-1020}
-\frac{t_\mathrm{h}}{2}\left(\psi_{\mathrm{res},n}(x-\Delta x)+\psi_{\mathrm{res},n}(x+\Delta x)\right)
=E_n\psi_{\mathrm{res},n}(x),
\end{equation}
gives the dispersion relation
\begin{equation}\label{eq5-1030}
E_n=-t_\mathrm{h}\cos (K_n \Delta x).
\end{equation}
The Schr\"{o}dinger equation at $x=\Delta x$ and at $x=0$ reads
\begin{eqnarray}\label{eq5-1040}
&&-\frac{t_\mathrm{h}}{2}\left(\psi_{\mathrm{res},n}(0)+\psi_{\mathrm{res},n}(2\Delta x)\right)
+V(1)\psi_{\mathrm{res},n}(\Delta x)
\nonumber\\
&&\qquad\qquad\qquad\qquad\qquad\qquad
=E_n\psi_{\mathrm{res},n}(\Delta x),
\\
\label{eq5-1041}
&&-\frac{t_\mathrm{h}}{2}\left(\psi_{\mathrm{res},n}(-\Delta x)+\psi_{\mathrm{res},n}(\Delta x)\right)
\nonumber\\
&&\qquad\qquad\qquad\qquad\qquad\qquad
=E_n\psi_{\mathrm{res},n}(0).
\end{eqnarray}

The two equations are reduced to
\begin{eqnarray}\label{eq5-1050}
&&
-\left(F_n+B_ne^{2iK_n\Delta x}\right)+2\tilde{V}_0B_ne^{iK_n\Delta x}
\nonumber\\
&&\qquad\qquad
=-B_ne^{iK_n\Delta x}\left(e^{iK_n\Delta x}+e^{-iK_n\Delta x}\right),
\\
\label{eq5-1051}
&&
-2B_ne^{iK_n\Delta x}=-F_n\left(e^{iK_n\Delta x}+e^{-iK_n\Delta x}\right)
\nonumber\\
&&\qquad\qquad\qquad\qquad\qquad\qquad\qquad
\mbox{for even parity},
\qquad
\\
\label{eq5-1052}
&&
0=-F_n\left(e^{iK_n\Delta x}+e^{-iK_n\Delta x}\right)
\nonumber\\
&&\qquad\qquad\qquad\qquad\qquad\qquad\qquad
\mbox{for odd parity},
\qquad
\end{eqnarray}
where $\tilde{V}_0\equiv V_0/t_\mathrm{h}$, or
\begin{eqnarray}\label{eq5-1060}
(2\tilde{V}_0z_n+1)B_n-F_n &=& 0,
\\
\label{eq5-1061}
2z_nB_n-(z_n+{z_n}^{-1})F_n&=&0
\quad\mbox{for even parity},
\\
\label{eq5-1062}
F_n&=&0
\quad\mbox{for odd parity},
\end{eqnarray}
where $z_n\equiv e^{iK_n\Delta x}$.
For the odd parity, we have the solution
\begin{eqnarray}\label{eq5-1070}
&&z_n=-\frac{1}{2\tilde{V}_0},\quad\mbox{or}
\\
&&
K_n\Delta x=\pi+i\log 2\tilde{V}_0
\quad\mbox{and}\quad
E_n=\tilde{V}_0+\frac{1}{4\tilde{V}_0}
\qquad
\end{eqnarray}
which is a bound state with a real energy.
For the even parity, we eliminate the degree of freedom of $B_n/F_n$, arriving at the third-order equation
\begin{equation}
\label{eq5-1080}
2\tilde{V}_0{z_n}^3-{z_n}^2 +2\tilde{V}_0 z_n+1=0.
\end{equation}
The three solutions for $\tilde{V}_0=1$, for example, are
\begin{eqnarray}\label{eq5-1090}
E_n/t_\mathrm{h}&=&1.517526485679543\ldots,
\\
\label{eq5-1091}
E_n/t_\mathrm{h}&=&-0.383763242839771\ldots 
\nonumber\\
&&- i0.132164836187054\ldots
\\
\label{eq5-1092}
E_n/t_\mathrm{h}&=&-0.383763242839771\ldots 
\nonumber\\
&&+ i0.132164836187054\ldots
\end{eqnarray}
The first solution is another bound state.
The second solution is a resonant state and the third solution an anti-resonant state.

These are the exact solutions of the real and complex energy eigenvalues.
We now demonstrate our numerical iterative method with the rapid convergence to obtain the resonant state with the complex eigenvalue.

\begin{figure}
\includegraphics[width=0.85\columnwidth]{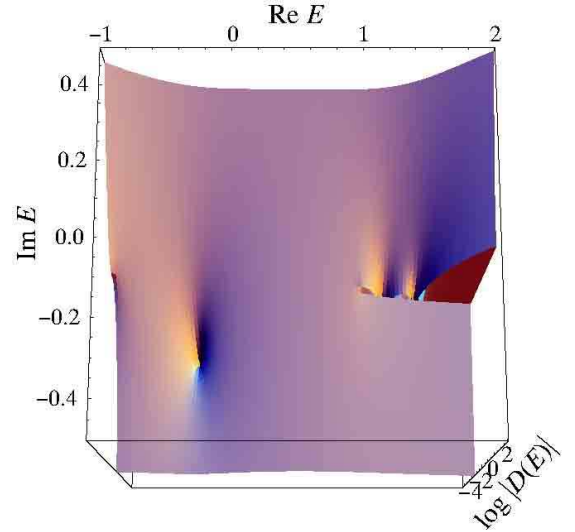}
\caption{A three-dimensional plot of $\log |D(E)|$ in the complex energy plane.}
\label{fig-Dfunc}
\end{figure}

We set $L=2$ for the numerical procedure proposed here.
In order to set the first postulate, it may be useful to scan the quantity
\begin{equation}\label{eq5-1100}
D(E)=\left|\det \left(
\mathcal{H}_\mathrm{eff}(E)-EI
\right)
\right|
\end{equation}
in the complex $E$ plane, where the effective Hamiltonian $\mathcal{H}_\mathrm{eff}(E)$ is defined on the region $-L\leq x\leq L$ with the effective potential;
for $L=2$, for example, we have
\begin{equation}\label{eq5-1110}
\mathcal{H}_\mathrm{eff}(E)
=\left(
\begin{array}{ccccc}
V_\mathrm{eff}(E) & -\frac{t_\mathrm{h}}{2} & 0 & 0 & 0 \\
-\frac{t_\mathrm{h}}{2} & V_0 & -\frac{t_\mathrm{h}}{2} & 0 & 0 \\
0 & -\frac{t_\mathrm{h}}{2} & 0 & -\frac{t_\mathrm{h}}{2} & 0 \\
0 & 0 & -\frac{t_\mathrm{h}}{2} & V_0 & -\frac{t_\mathrm{h}}{2} \\
0 & 0 & 0 & -\frac{t_\mathrm{h}}{2} & V_\mathrm{eff}(E)
\end{array}\right),
\end{equation}
where 
\begin{equation}\label{eq5-1120}
V_\mathrm{eff}(E)\equiv \frac{1}{2}\left(
E-i\sqrt{{t_\mathrm{h}}^2-E^2}
\right).
\end{equation}
The zeros of $D(E)$ indicate the eigenvalues, although it is not practical to use the zeros as numerical estimators of the eigenvalues.
Figure~\ref{fig-Dfunc} shows a three-dimensional plot of $\log | D(E)|$ in the complex energy plane.
The dimple in the lower left corner is the resonance pole~(\ref{eq5-1091}).

\begin{figure}
\includegraphics[width=0.85\columnwidth]{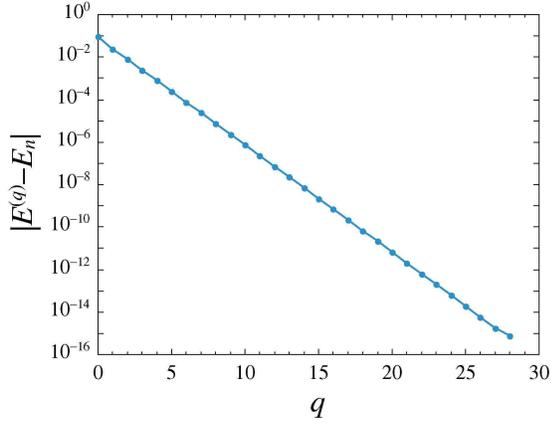}
\caption{The convergence of the numerical procedure proposed here for the model~(\ref{eq5-1000}) with $V_0/t_\mathrm{h}=1$.
The postulate  $E^{(q)}$ approaches the exact value exponentially with respect to the step number $q$.}
\label{fig-conv}
\end{figure}

On the basis of rough estimation in Fig.~\ref{fig-Dfunc}, we set the first postulate as $E^{(0)}=-0.3-i0.1$.
The postulate $E^{(q)}$ converged to the eigenvalue~(\ref{eq5-1091}) exponentially with respect to the number of iterations, $q$; see Fig.~\ref{fig-conv}.
The numerical calculation involves diagonalization of $5\times5$ non-Hermitian matrices.

\section{Numerical computation of dynamics of resonant eigenfunctions}
\label{sec6}

We finally introduce a trick for numerical computation of the dynamics of diverging resonant eigenfunctions with the use of the effective potential.
The trick introduced below enables us to calculate the time evolution of the divergent eigenfunction precisely in a finite space, when we discretize the space for the numerical calculation or when we consider a tight-binding model.

Now, we are interested in integrating the time-dependent Schr\"{o}dinger equation
\begin{equation}\label{eq6-10}
i\hbar\frac{\partial}{\partial t}\Psi_{\mathrm{res},n}(x,t)
=\hat{\mathcal{H}}\Psi_{\mathrm{res},n}(x,t).
\end{equation}
In the right-hand side, we again cutoff the space $|x|>L$ with the use of the effective potential:
\begin{eqnarray}\label{eq6-50}
\hat{\mathcal{H}}\Psi_{\mathrm{res},n}\left(\pm L,t\right)
&=&-\frac{t_\mathrm{h}}{2}
\Psi_{\mathrm{res},n}\left(\pm \left(L-\Delta x\right),t\right)
\nonumber\\
&&+V_{\mathrm{eff},n}\left(\pm L\right)
\Psi_{\mathrm{res},n}\left(\pm L,t\right).
\nonumber\\
\end{eqnarray}
Since we already know energy eigenvalues by the method in Section~\ref{sec5}, we can fix the effective potential $V_{\mathrm{eff},n}\left(\pm L\right)$ for each resonant state.
The point is that Eq.~(\ref{eq6-50}) contains only the eigenfunctions inside the region $-L\leq x\leq L$.
Hence we can integrate Eq.~(\ref{eq6-10}) by restricting ourselves to the region in question.


Let us demonstrate the above trick in the computation of a version of the Friedrichs-Fano (or Newns-Anderson) model (Fig.~\ref{fig-Friedrichs})~\cite{Friedrichs48,Petrosky91,Ordonez01,Tanaka06}:
\begin{figure}
\centering
\includegraphics[width=0.8\columnwidth]{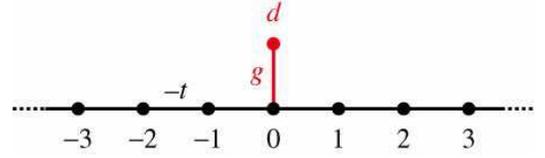}
\caption{The tight-binding model on a chain with an adatom.}
\label{fig-Friedrichs}
\end{figure}
\begin{eqnarray}\label{eq6-100}
\hat{\mathcal{H}}&=&
-\frac{t_\mathrm{h}}{2}\sum_{x}\left(
\left|x+\Delta x\right\rangle \left\langle x \right|
+\left| x\right\rangle \left\langle x+\Delta x \right|
\right)
\nonumber\\
&&+g \left(
\left| \mathrm{d} \right\rangle \left\langle 0 \right|
+\left| 0 \right\rangle \left\langle \mathrm{d} \right|
\right)
+E_\mathrm{d} \left| \mathrm{d} \right\rangle \left\langle \mathrm{d} \right|,
\end{eqnarray}
where $\mathrm{d}$ denotes the site of an adatom.
The first term denotes the tight-binding hopping on a chain, the second term denotes the hopping between the origin of the chain $x=0$ and the adatom, and the third term denotes the impurity level of the adatom.

\begin{figure}
\centering
\includegraphics[width=0.74\columnwidth]{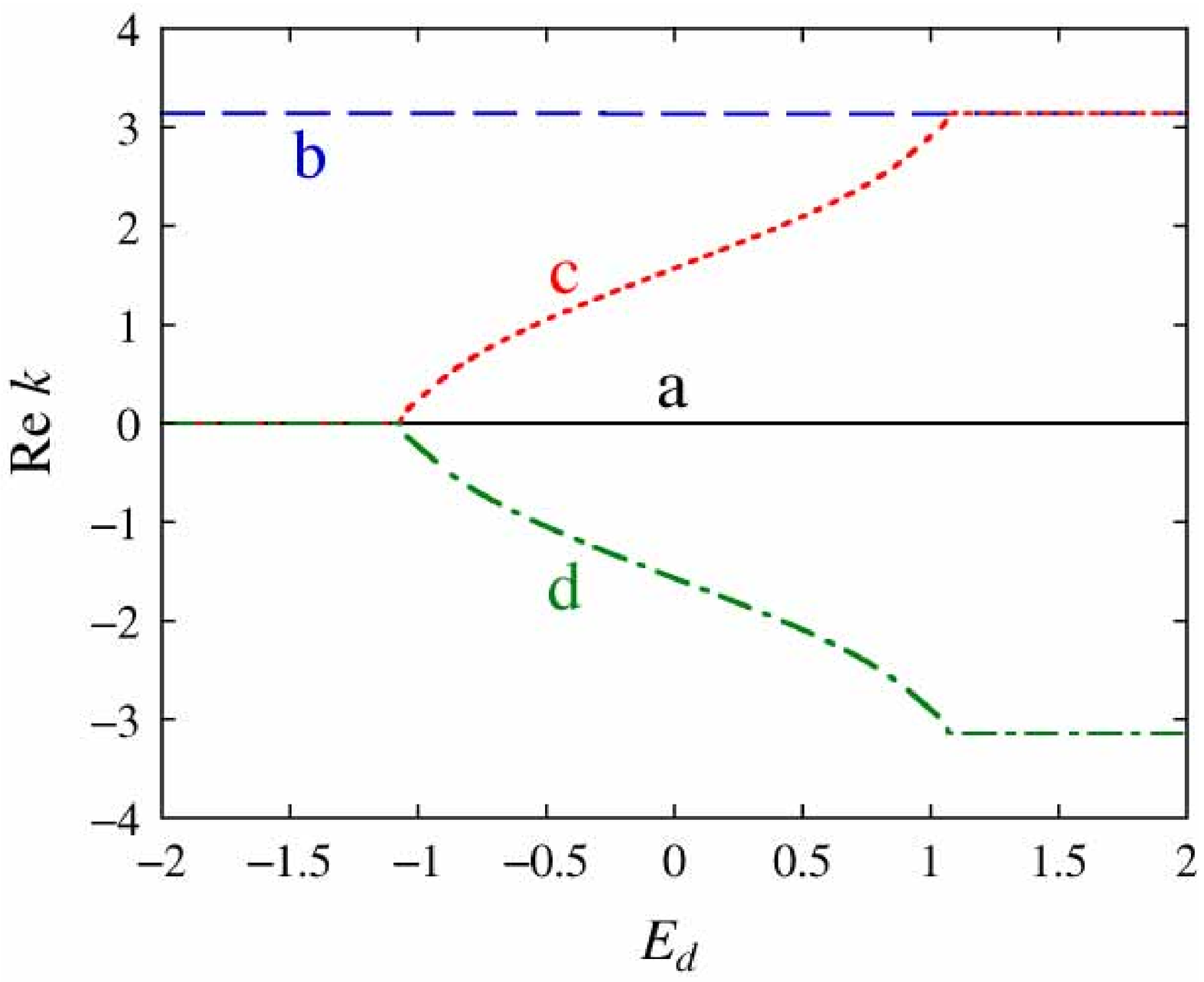}
\includegraphics[width=0.74\columnwidth]{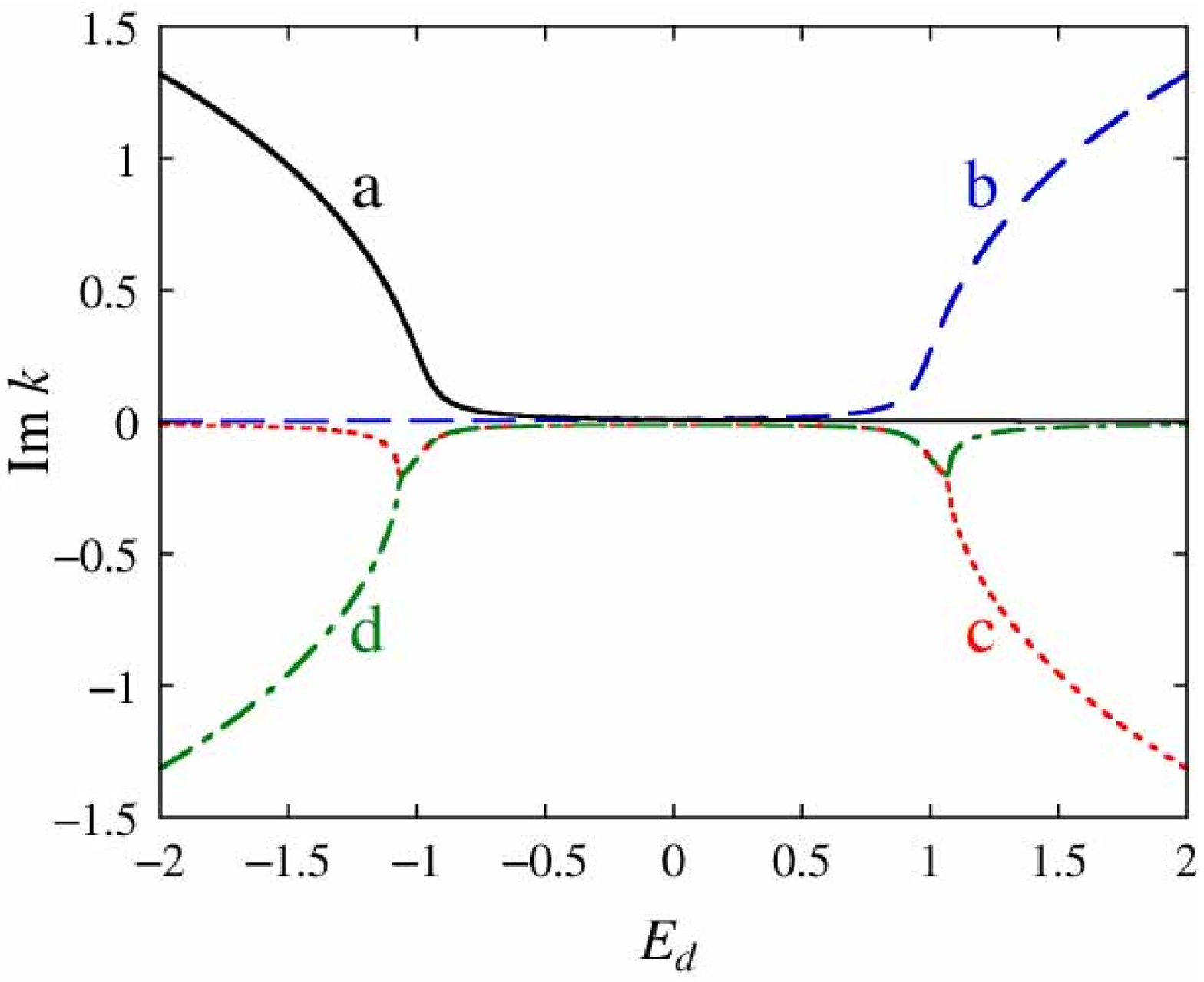}
\includegraphics[width=0.74\columnwidth]{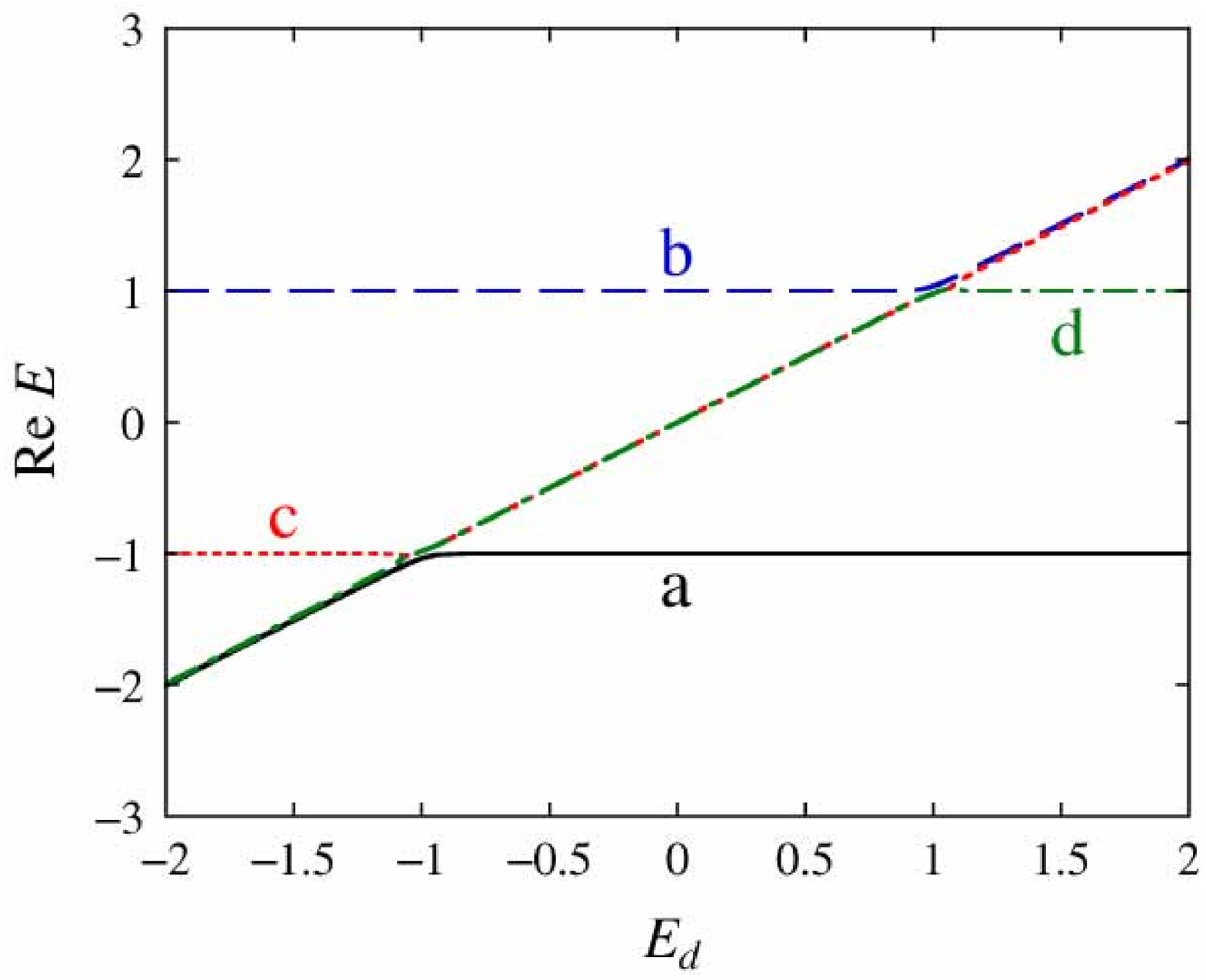}
\includegraphics[width=0.74\columnwidth]{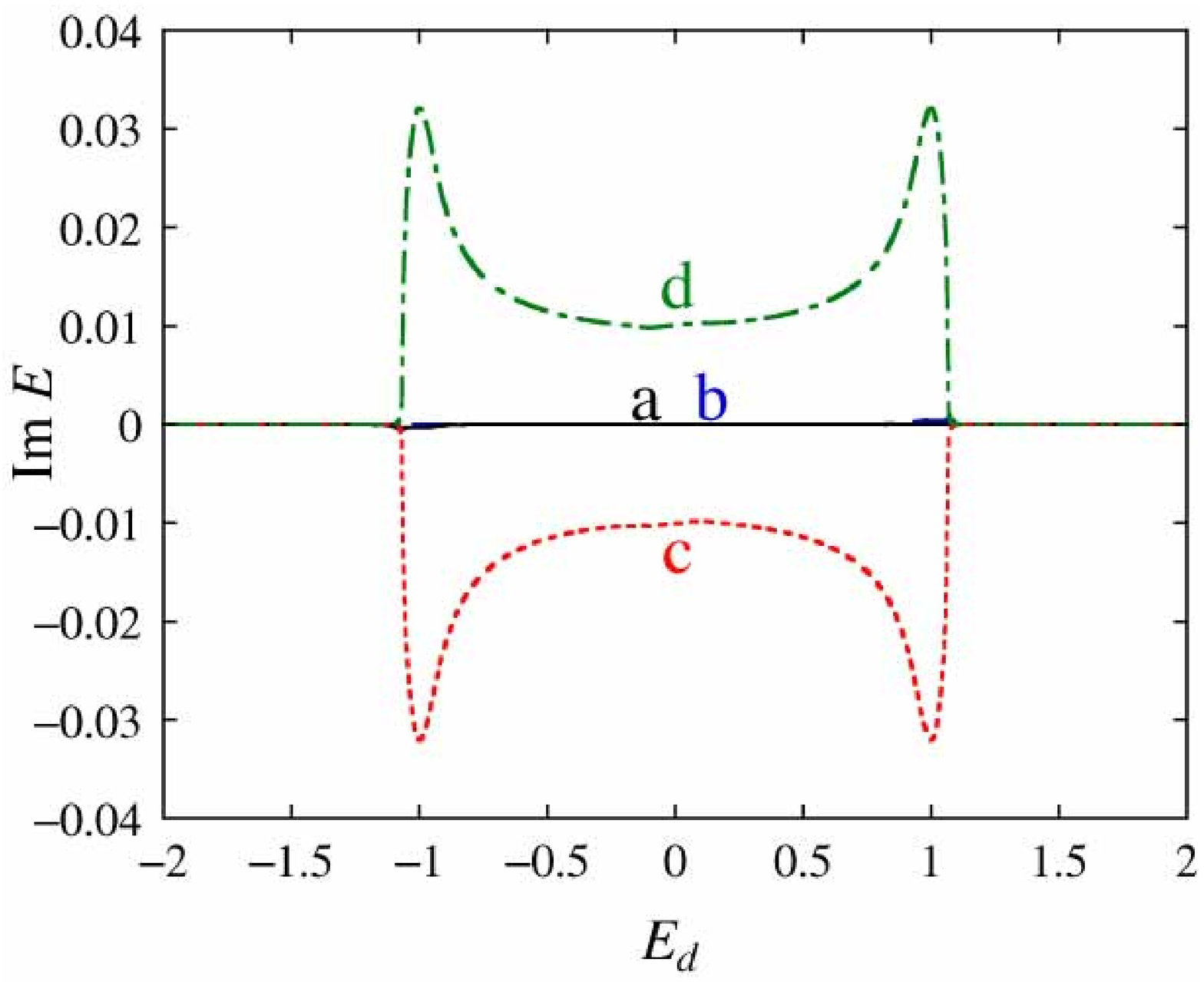}
\caption{The four solutions of Eq.~(\ref{eq6-170}) for $\tilde{g}=0.1$.
The dependence on the impurity level of, from above to the bottom, the real part and the imaginary part of $\{K_n\}$ and the real part and the imaginary part of $\{\tilde{E}_n\}$ for $\tilde{g}=0.1$.
The solid line (a) and the broken line (b) indicate bound states, while the dotted line (c) and the chained line (d) indicate resonant states.}
\label{fig-sols}
\end{figure}

Let us first find the resonance poles exactly.
As we showed in Section~\ref{sec3}, we assume that the eigenfunction has outgoing waves only.
We have solutions with even parity and solutions with odd parity.
The odd solutions vanish at $x=0$, and hence do not couple to the adatom.
We consider only even solutions hereafter.
We assume the form
\begin{equation}\label{eq6-110}
\psi_{\mathrm{res},n}(x)=B_n e^{iK_n|x|}
\end{equation}
for the chain and
\begin{equation}\label{eq6-120}
\psi_{\mathrm{res},n}(\mathrm{d})=F_n
\end{equation}
for the adatom.
The same procedure as in Section~\ref{sec5-2} yields the fourth-order equation
\begin{equation}
\label{eq6-170}
{z_n}^4+2\tilde{E}_d{z_n}^3 +4\tilde{g}^2{z_n}^2-2\tilde{E}_d{z_n}-1=0,
\end{equation}
where $z_n=e^{iK_n\Delta x}$, $\tilde{E}_d\equiv E_d/t_\mathrm{h}$ and $\tilde{g}\equiv g/t_\mathrm{h}$.
Hence there are four solutions, as shown in Fig.~\ref{fig-sols} for $\tilde{g}=0.1$.

\begin{figure}
\centering
\includegraphics[width=0.8\columnwidth]{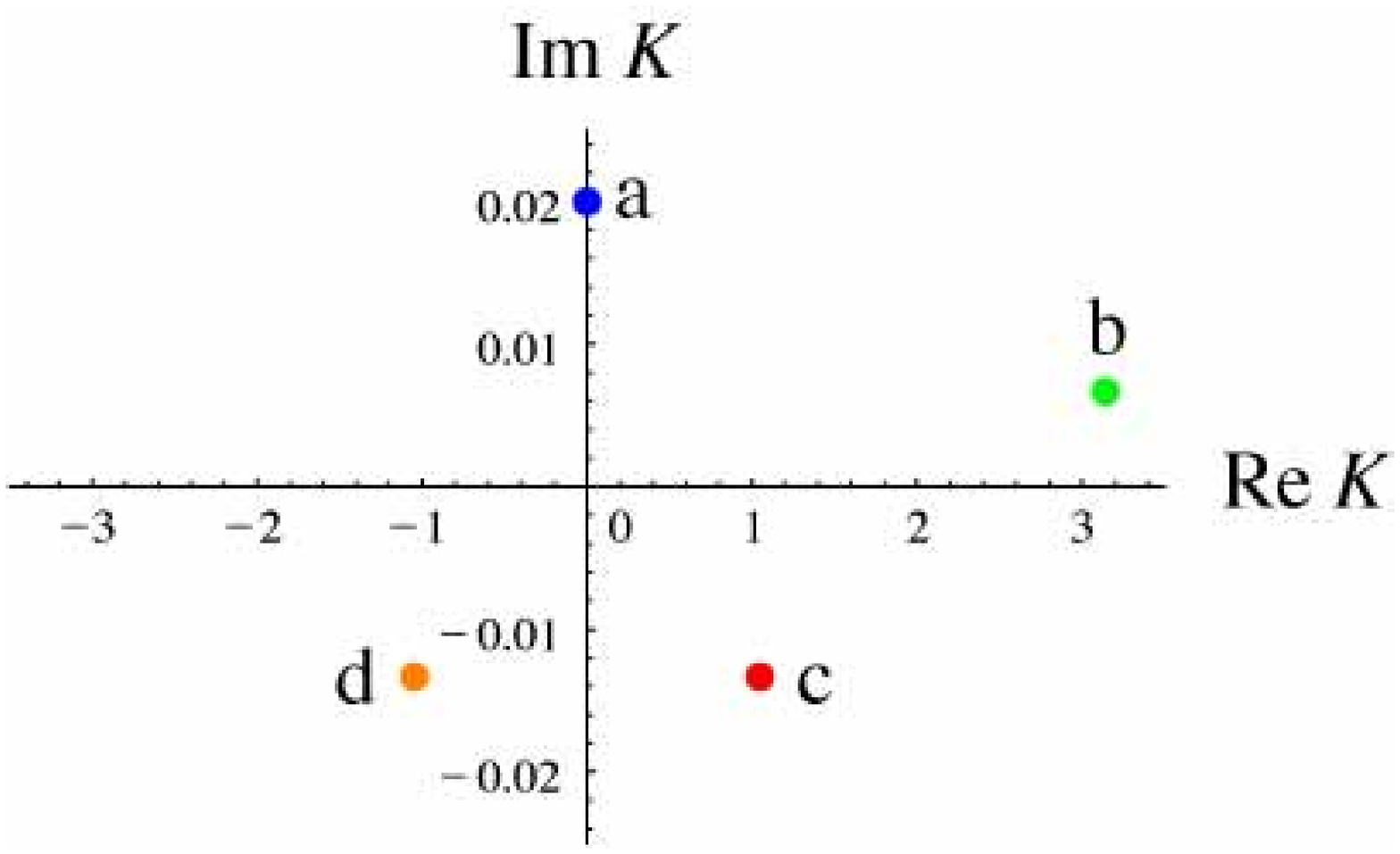}
\vskip \baselineskip
\includegraphics[width=0.8\columnwidth]{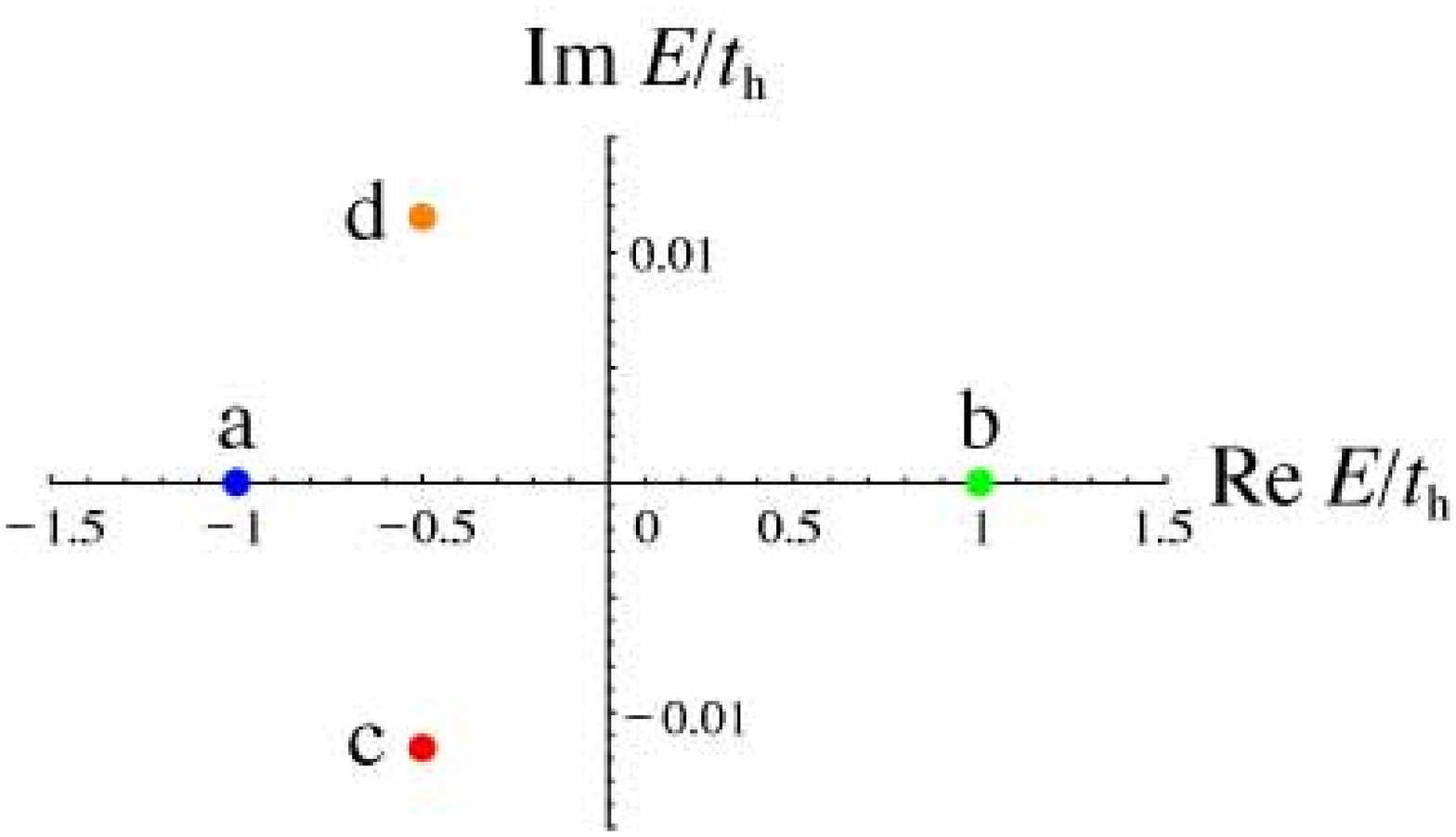}
\caption{The four solutions for $\tilde{g}=0.1$ and $\tilde{E}_d=-1/2$ on the complex momentum plane and the complex energy plane.
In each panel, the symbols a, b, c and d indicate the solutions of the solid line, the broken line, the dotted line and the chained line in Fig.~\ref{fig-sols}, respectively.
The solutions a and b are bound states, while the solutions c and d are resonant states.}
\label{fig-sol4}
\end{figure}

Through the dispersion relation $E_n=-t_\mathrm{h}\cos (K_n\Delta x)$, or the relations
\begin{eqnarray}
\label{eq6-175}
t_\mathrm{h}\left(z_n+\frac{1}{z_n}\right)&=&-2E_n, \\
\label{eq6-176}
t_\mathrm{h}\left(z_n-\frac{1}{z_n}\right)&=&\pm 2 \sqrt{{E_n}^2-{t_\mathrm{h}}^2}, \\
\label{eq6-177}
{t_\mathrm{h}}^2\left({z_n}^2-\frac{1}{{z_n}^2}\right) &=& \mp 4E_n\sqrt{{E_n}^2-{t_\mathrm{h}}^2},
\end{eqnarray}
the fourth-order equation~(\ref{eq6-170}) is transformed to an equation with respect to the energy:
\begin{equation}
\label{eq6-178}
\pm \left(E_n-E_d\right) \sqrt{{E_n}^2-{t_\mathrm{h}}^2}=g^2.
\end{equation}
This is equivalent to the dispersion equation obtained in Ref.~\cite{Tanaka06} for the Friedrichs model;
see the pole of Eq.~(7) of the article.
(The correspondence of the notations is as follows: $E_n \Leftrightarrow z$, $E_d \Leftrightarrow E_0$, $t_\mathrm{h} \Leftrightarrow B$, $g \Leftrightarrow gB$.)
We stress that the equation is easier to obtain in the momentum space than in the energy space, because we need not perform a complicated integration of the self-energy part as in the case of the energy space.

Two of the solutions (the solid and broken lines, a and b, in Fig.~\ref{fig-sols}) are bound states (positive $\mathop{\mathrm{Im}} K_n$ and real $E_n$) all through the range and are referred to as superstable states in Ref.~\cite{Tanaka06}.
The other two solutions (the dotted and chained lines, c and d, in Fig.~\ref{fig-sols}) have negative $\mathop{\mathrm{Im}}K_n$, which indicates that they are resonant states.
In some regions, the solutions are pure imaginary and hence the energy eigenvalues are real.
The states with pure imaginary wave number are  often called anti-bound states~\cite{Ohanian74}, but we here include them in resonant states.

Figure~\ref{fig-sol4} shows the four solutions for $\tilde{g}=0.1$ and $\tilde{E}_d=-1/2$ on the complex momentum plane and the complex energy plane.
The two solutions a and b are bound states.
The solution a has $\mathop{\mathrm{Re}} K_n=\pi$ and the solution b has $\mathop{\mathrm{Re}} K_n=0$.
They are both on the first (so-called \textit{physical}) Riemann sheet of the complex energy plane.
The other two solutions are resonant states.
The solution c has $\mathop{\mathrm{Re}}K_n>0$ and $\mathop{\mathrm{Im}}E_n<0$;
this is a decaying state with an outgoing wave only.
The solution d has $\mathop{\mathrm{Re}}K_n<0$ and $\mathop{\mathrm{Im}}E_n>0$;
this is a growing state with an incoming wave only.
These two are conjugate to each other and always appear in a pair, as was also demonstrated in Sec.~\ref{sec3-2}.
They are both on the second (so-called \textit{unphysical}) Riemann sheet of the complex energy plane.

We stress here again that it is generally more convenient to solve the problem in the complex momentum plane than in the complex energy plane.
This is because of the added complexity in identifying the Riemann sheet for a given solution of the complex energy eigenvalue in the case where there are several Riemann sheets.
It is generally not easy to tell whether a state is on the first or the second Riemann sheet before we find the location of the state in the complex momentum plane.

\begin{figure}[ht]
\centering
\includegraphics[width=0.77\columnwidth]{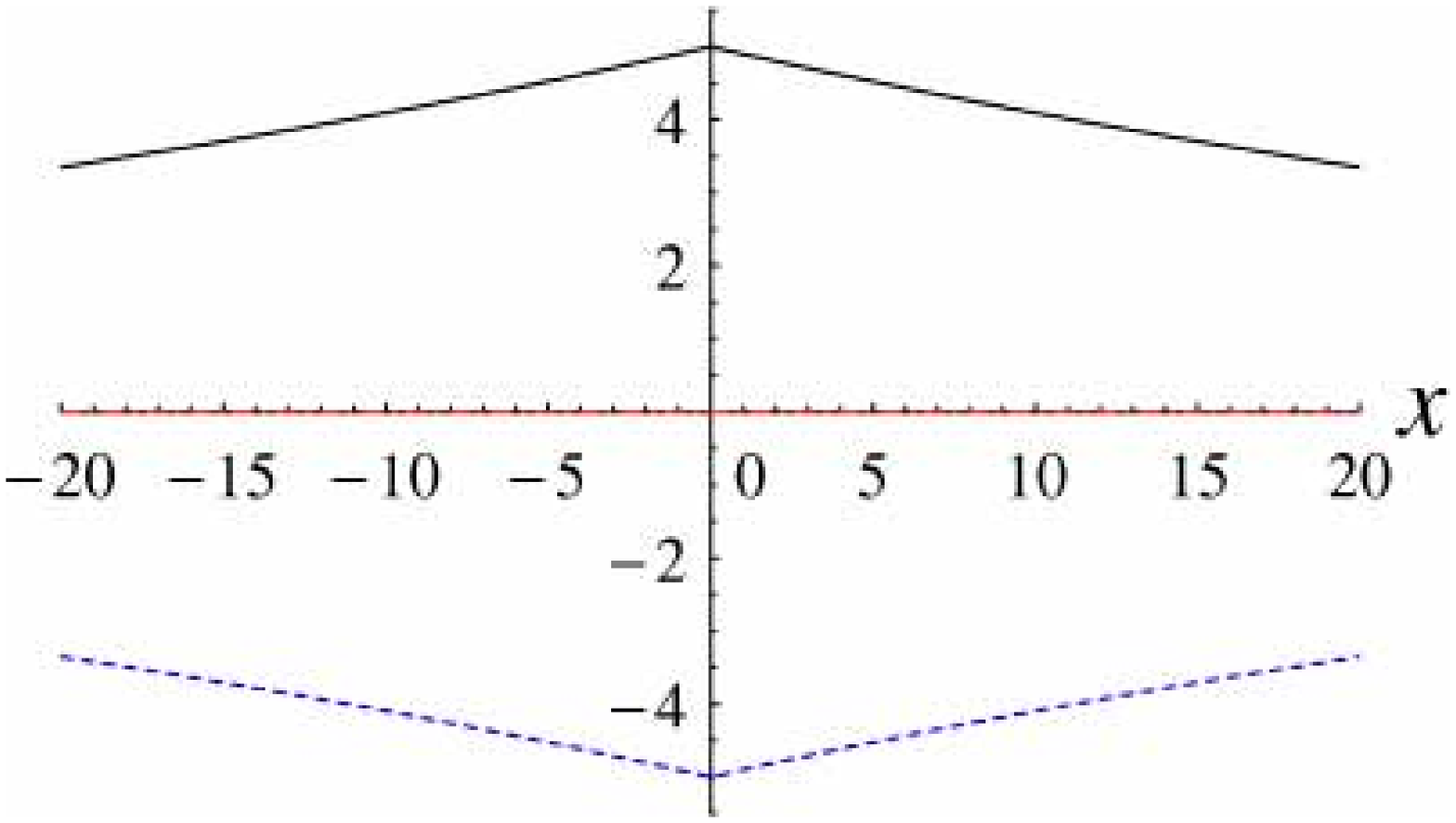}
\vskip \baselineskip
\includegraphics[width=0.77\columnwidth]{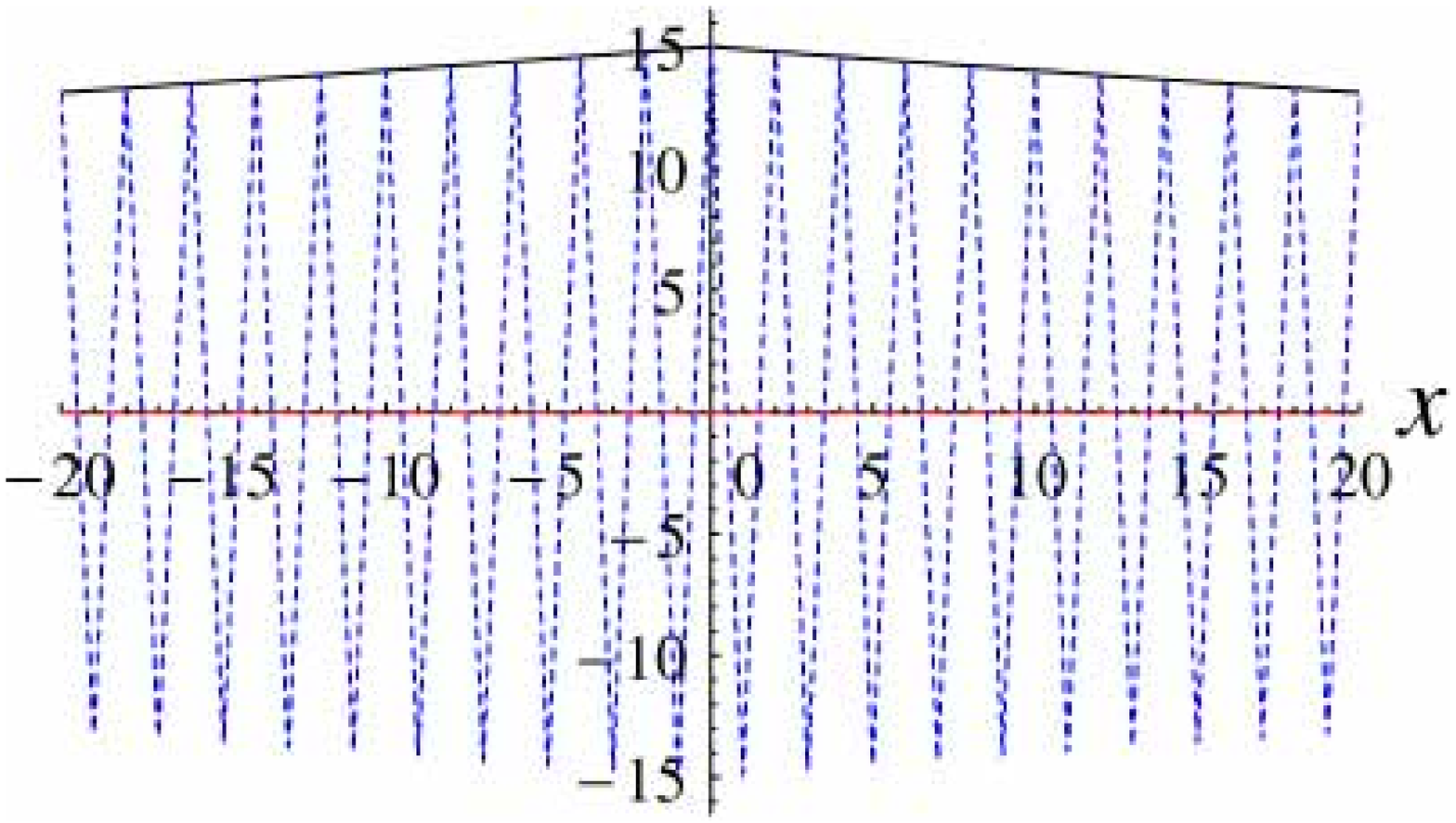}
\vskip \baselineskip
\includegraphics[width=0.77\columnwidth]{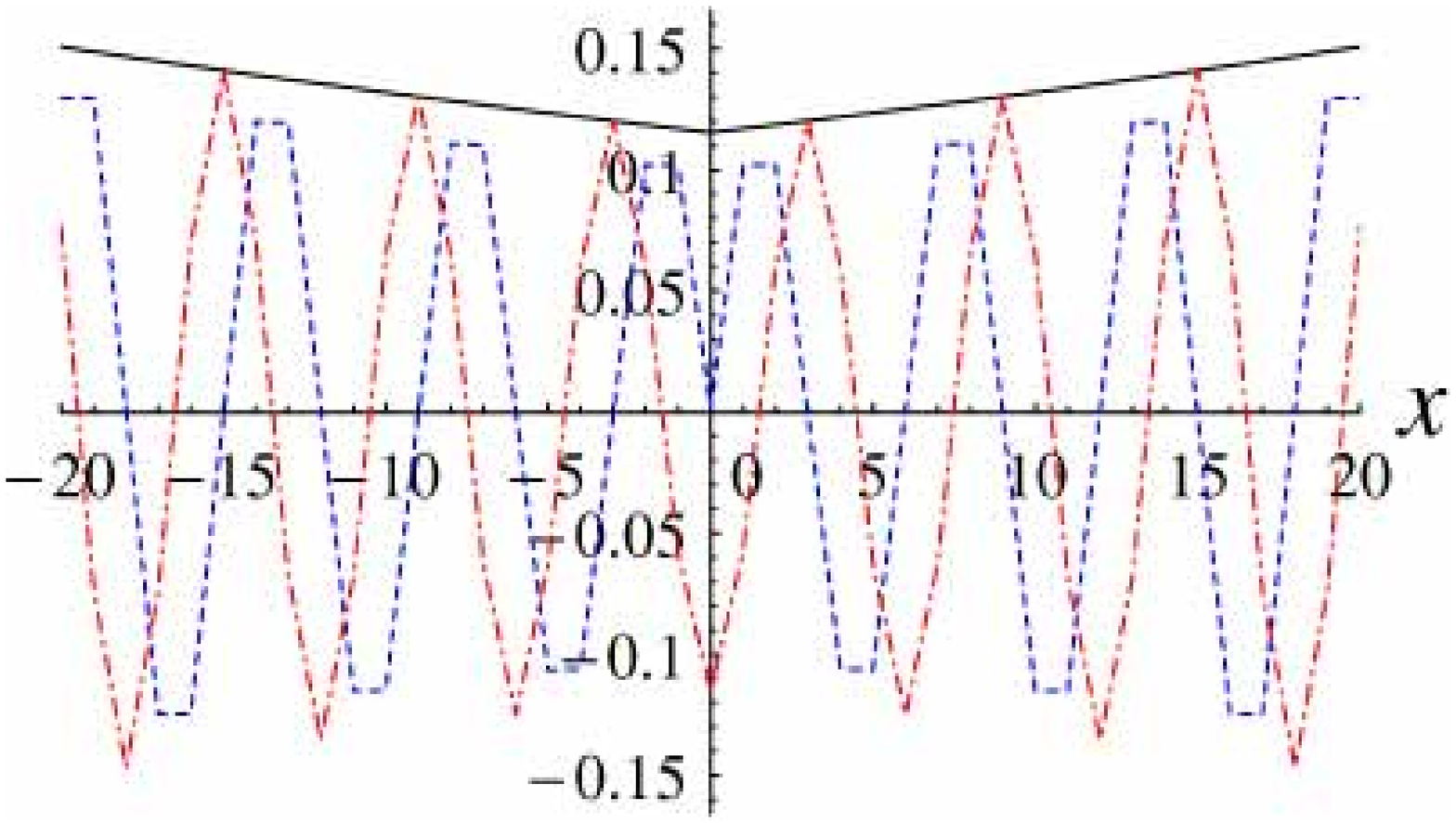}
\vskip \baselineskip
\includegraphics[width=0.77\columnwidth]{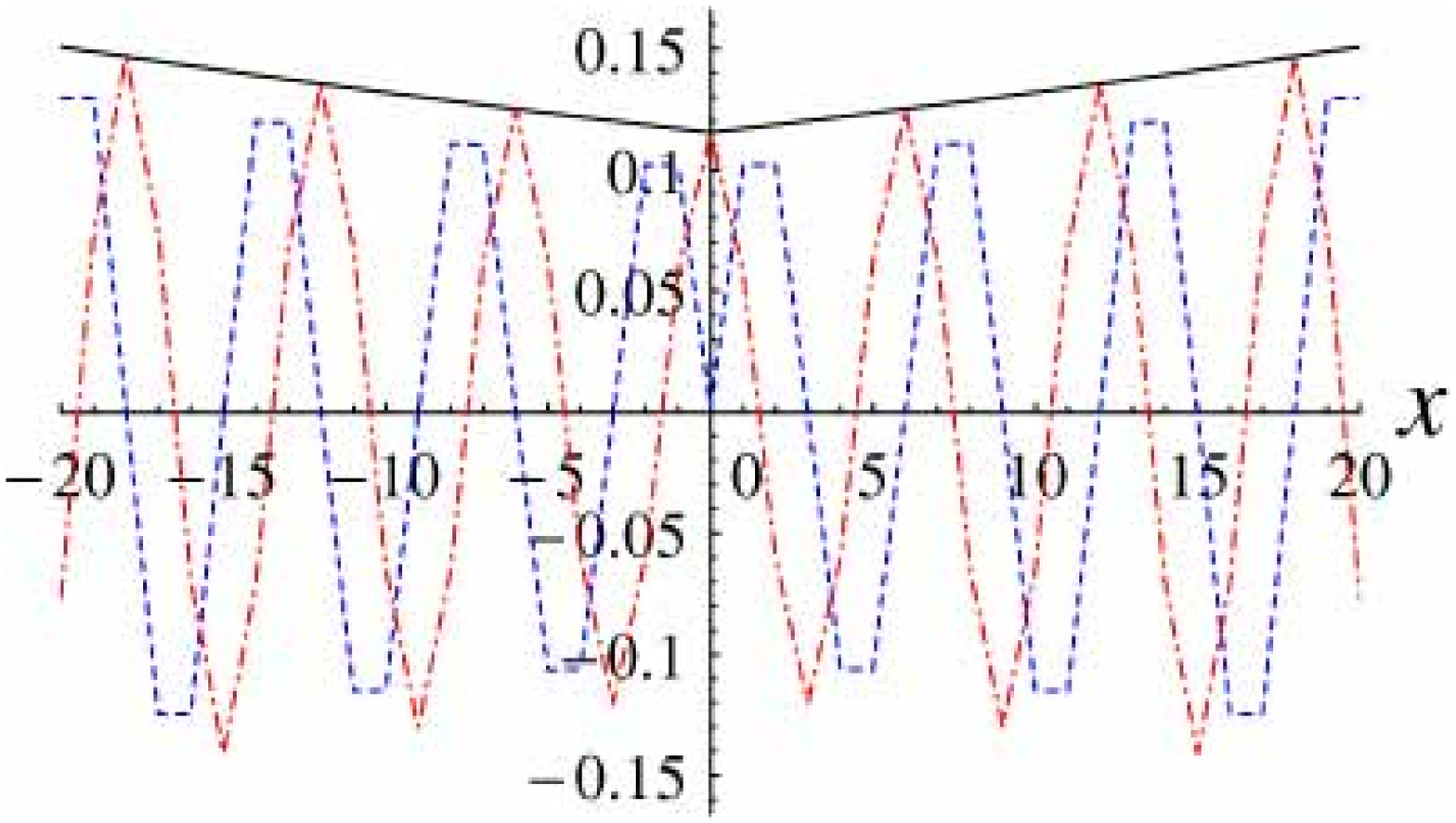}
\caption{The eigenfunctions at $t=0$ for $\tilde{g}=0.1$ and $E_d=-1/2$: each panel shows the solution of, from top to bottom, the solid line, the broken line, the dotted line and the chained line in Fig.~\ref{fig-sols}, or the solutions a, b, c and d in Fig.~\ref{fig-sol4}.
The solutions a and b are bound states, while the solutions c and d are resonant states.
In each panel, the absolute value (the solid line), the real part (the broken line) and the imaginary part (the chained line) are displayed in the region $-20\leq x\leq 20$.
(In the first and second panels, the imaginary part is zero.)}
\label{fig-init}
\end{figure}

The eigenfunctions $\{\psi_{\mathrm{res},n}(x)\}$ of the solutions a, b, c and d for $\tilde{g}=0.1$ and $\tilde{E}_d=-1/2$ are shown in Fig.~\ref{fig-init}, where the normalization is always fixed by $\psi_{\mathrm{res},n}(d)=1$.
The eigenfunctions from the top to the bottom of Fig.~\ref{fig-init} correspond to the solutions a, b, c and d in Fig.~\ref{fig-sol4}, respectively.
It is evident that the solutions a and b are bound states decaying in space whereas the solutions c and d are resonant states diverging in space.

\begin{figure*}[t]
\includegraphics[width=0.24\textwidth]{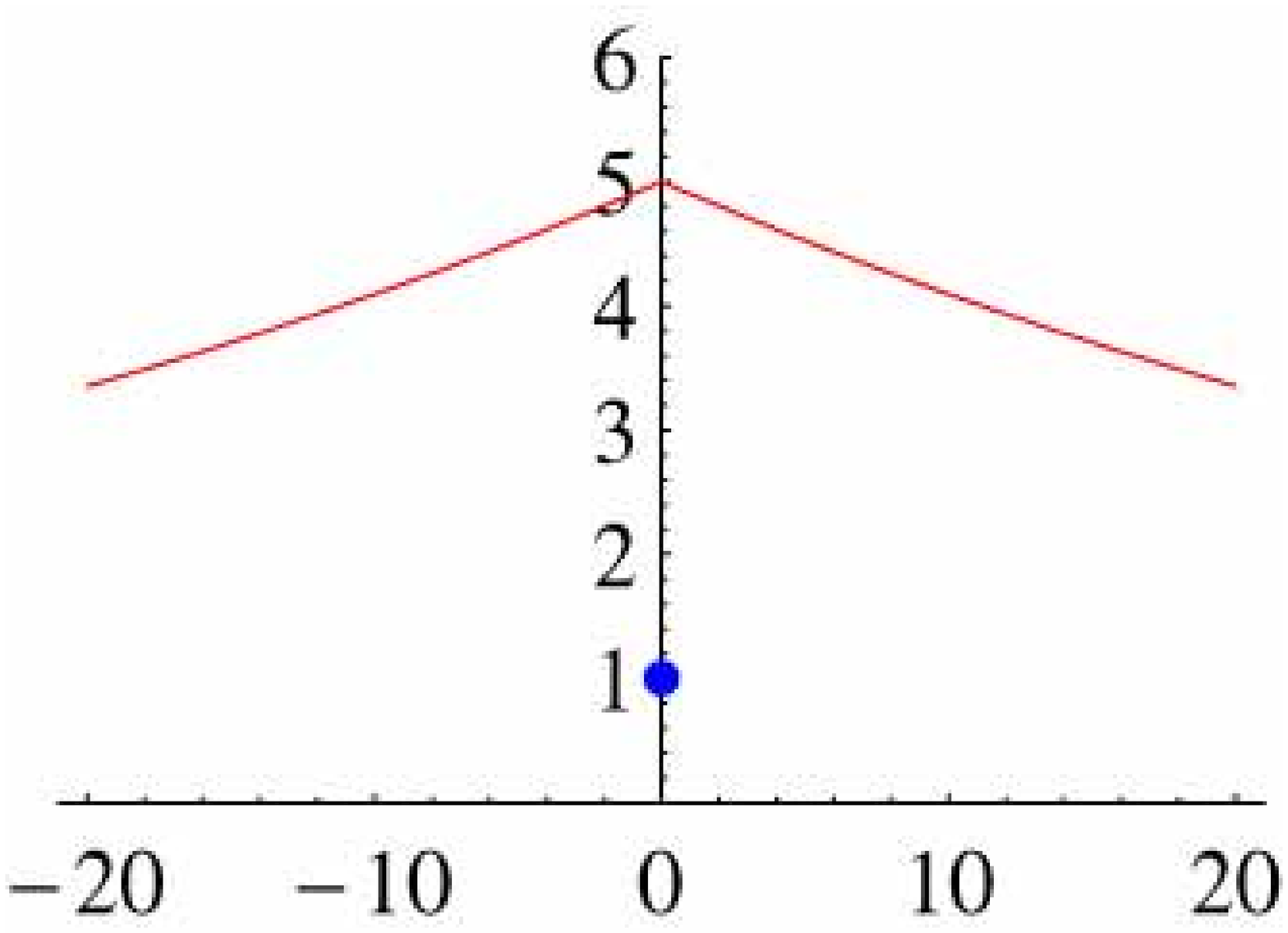}
\hfill
\includegraphics[width=0.24\textwidth]{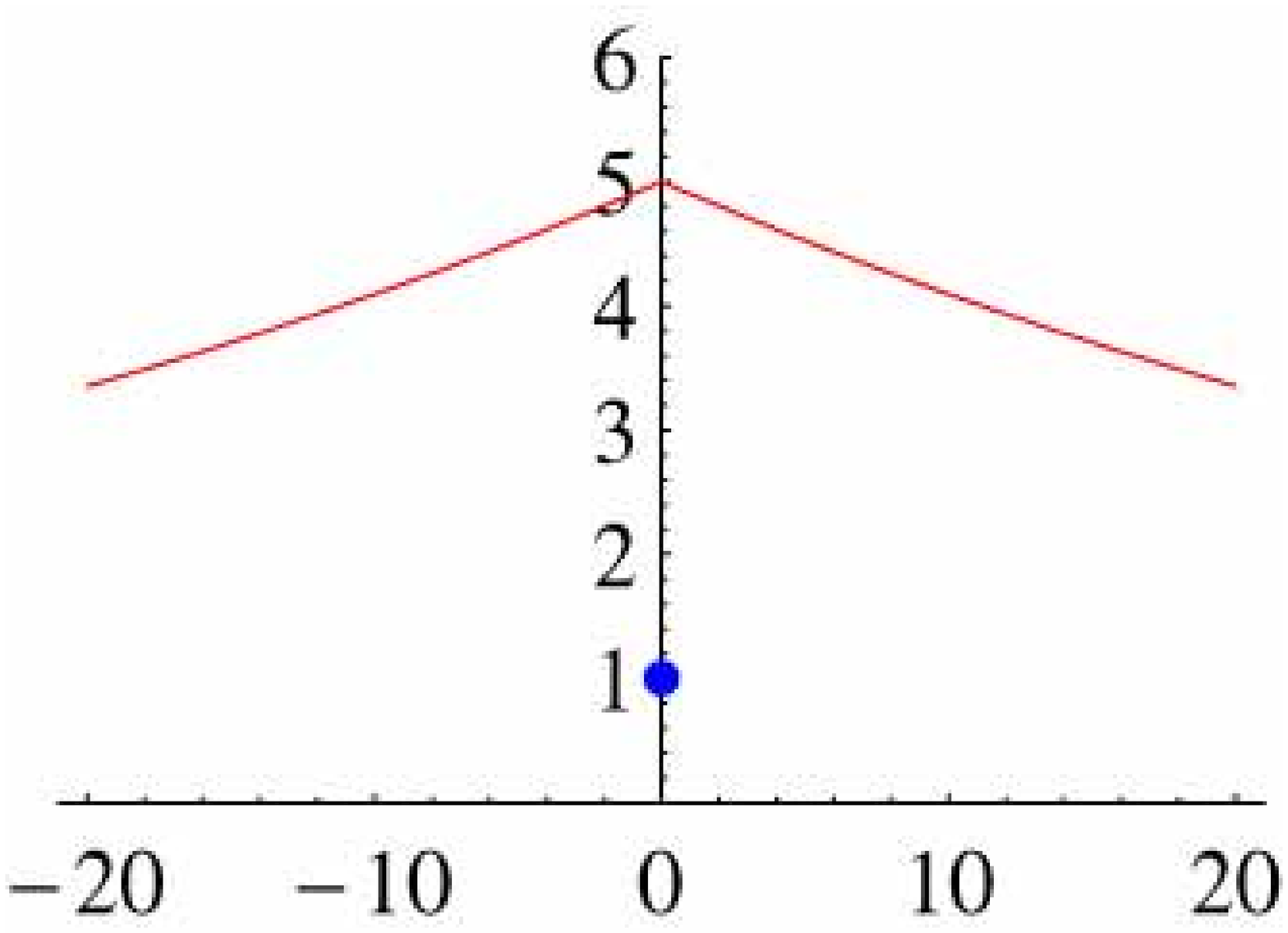}
\hfill
\includegraphics[width=0.24\textwidth]{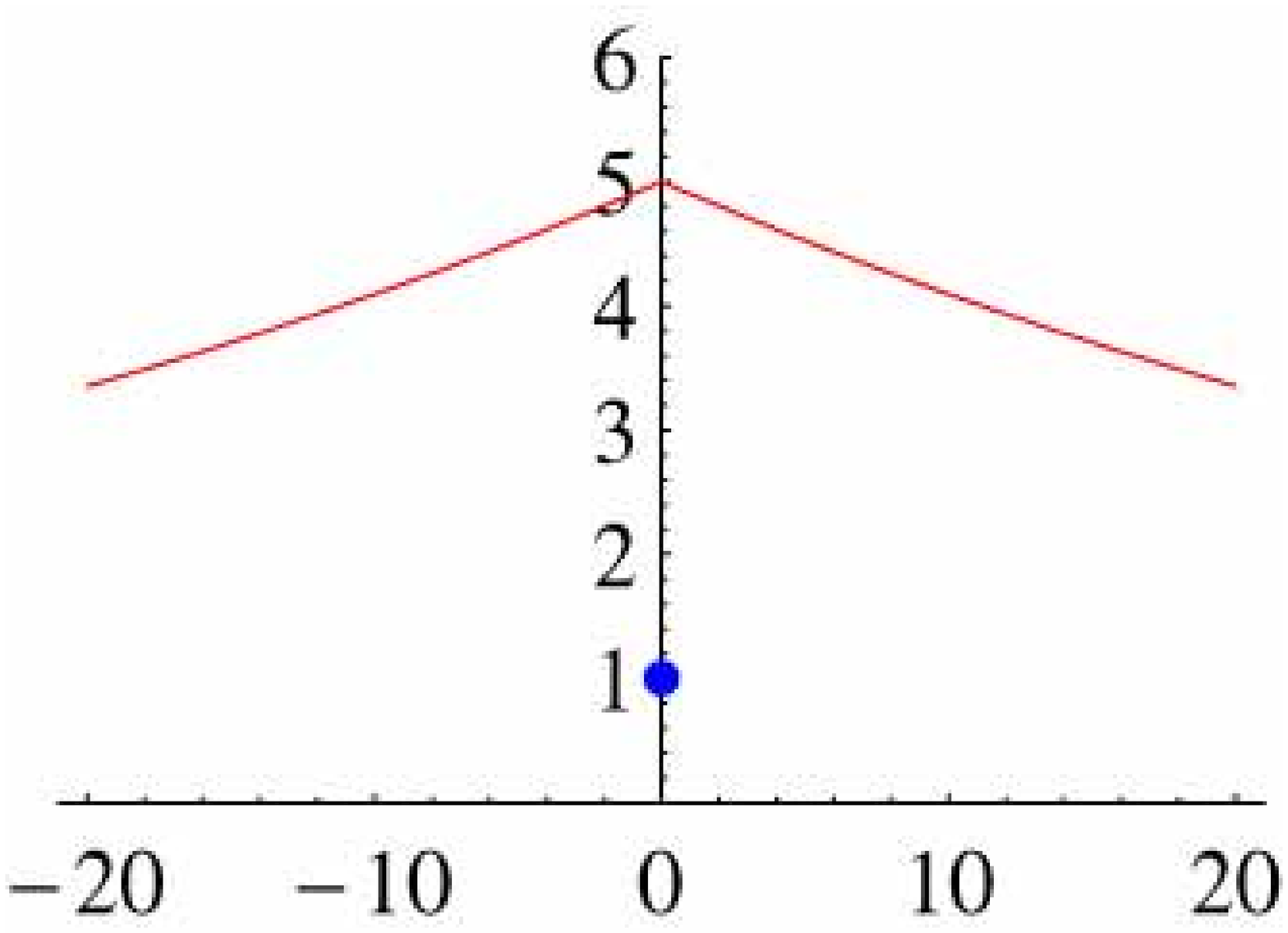}
\hfill
\includegraphics[width=0.24\textwidth]{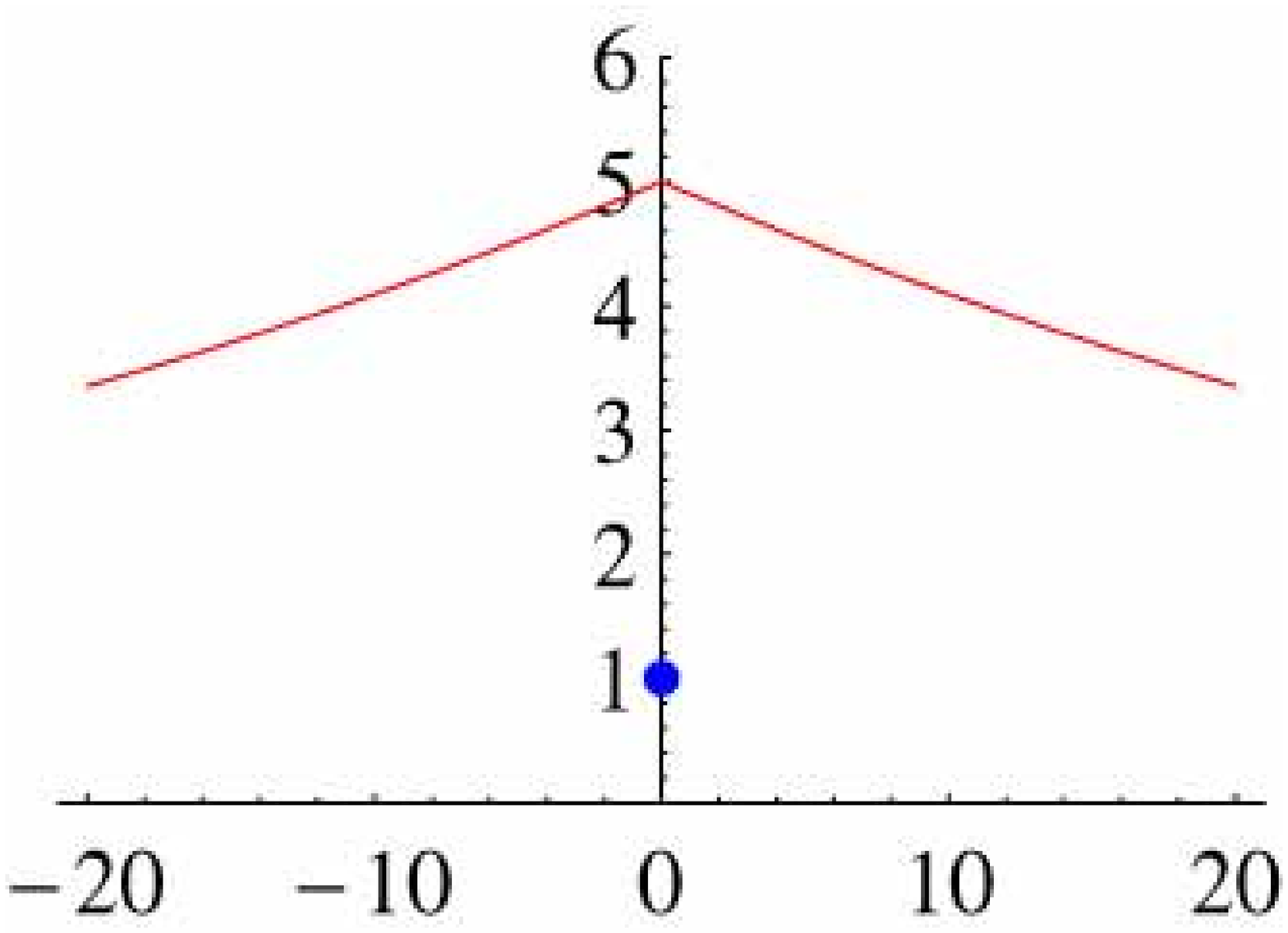}
\\
\vskip \baselineskip
\includegraphics[width=0.24\textwidth]{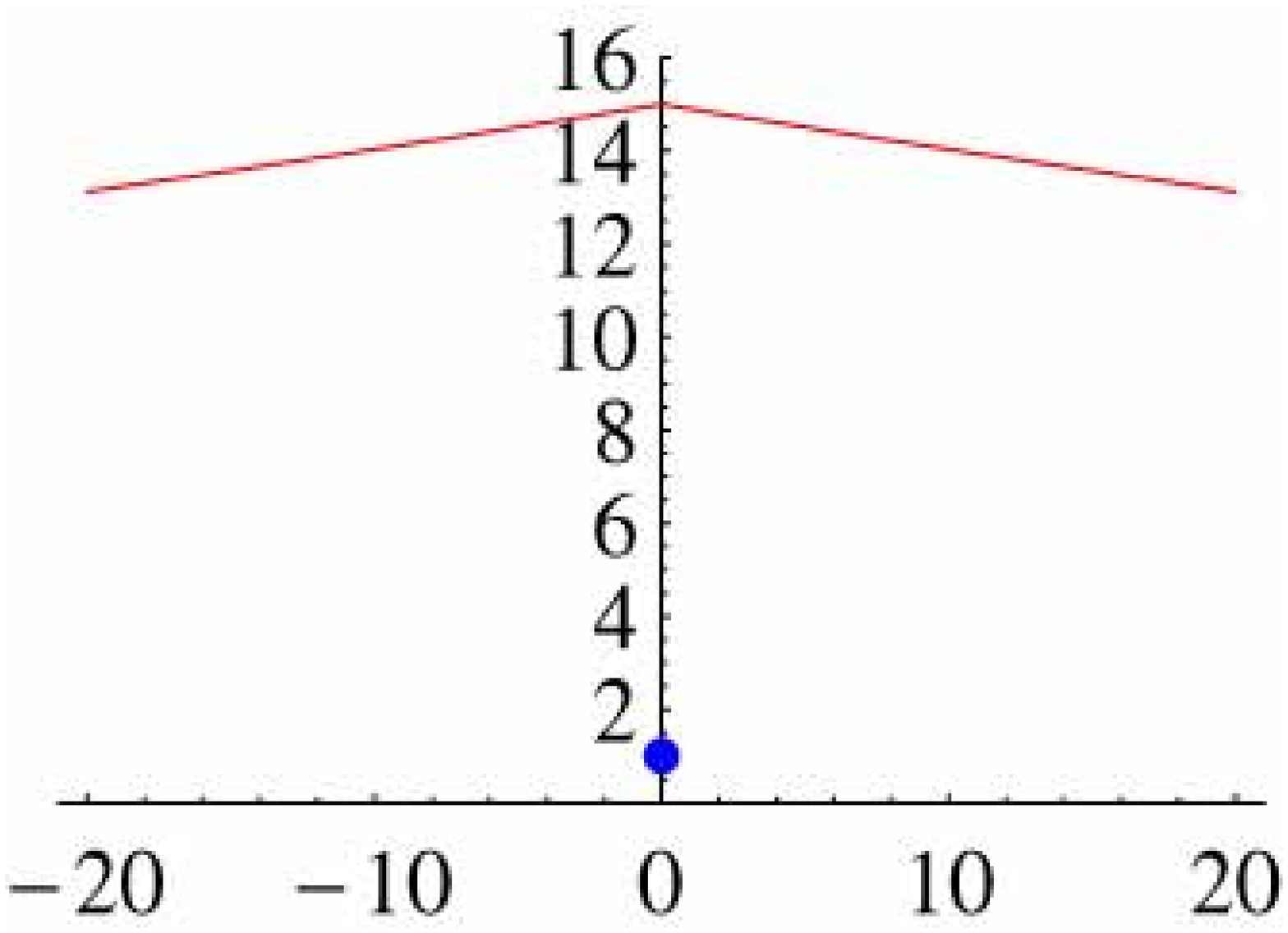}
\hfill
\includegraphics[width=0.24\textwidth]{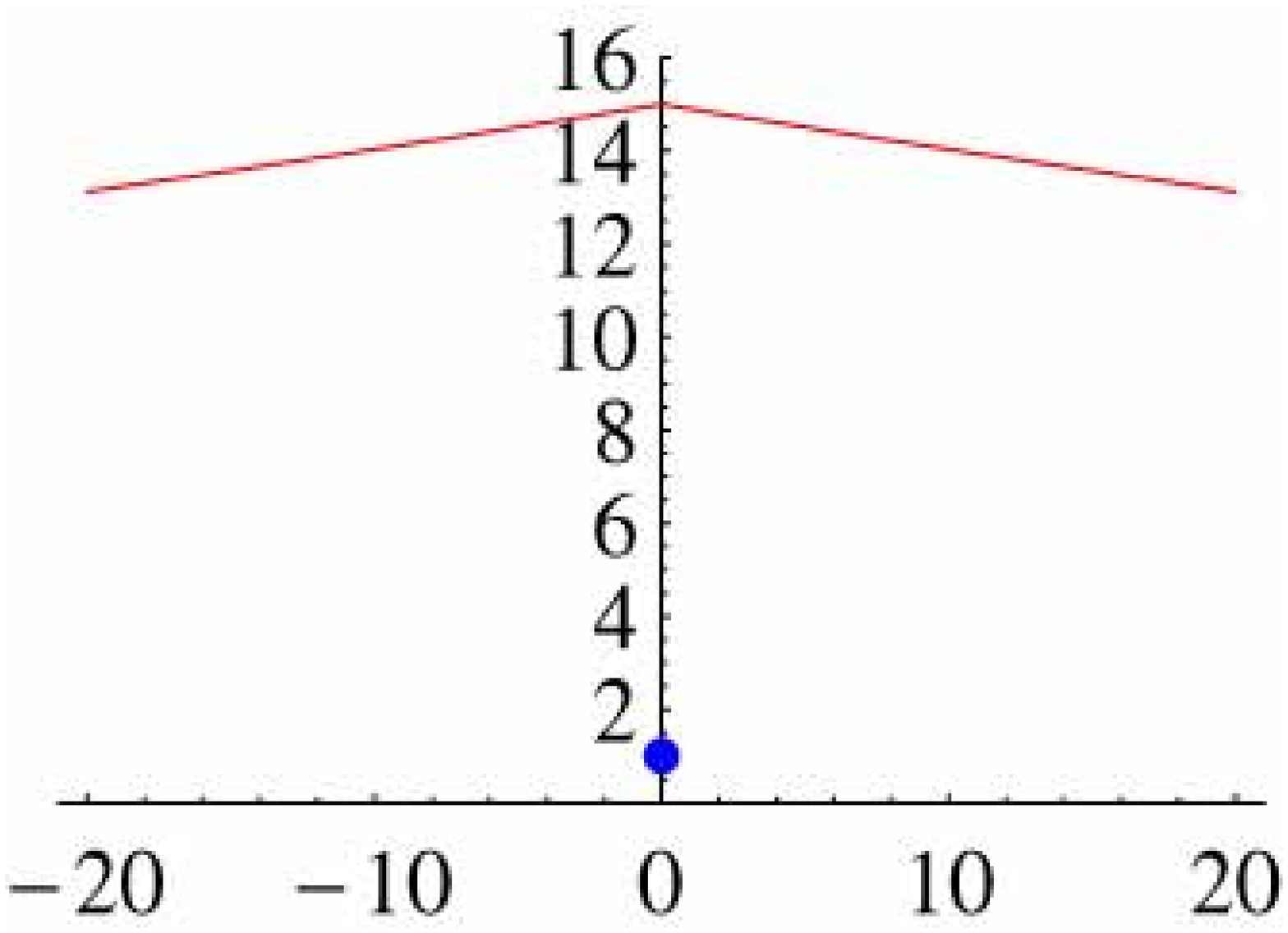}
\hfill
\includegraphics[width=0.24\textwidth]{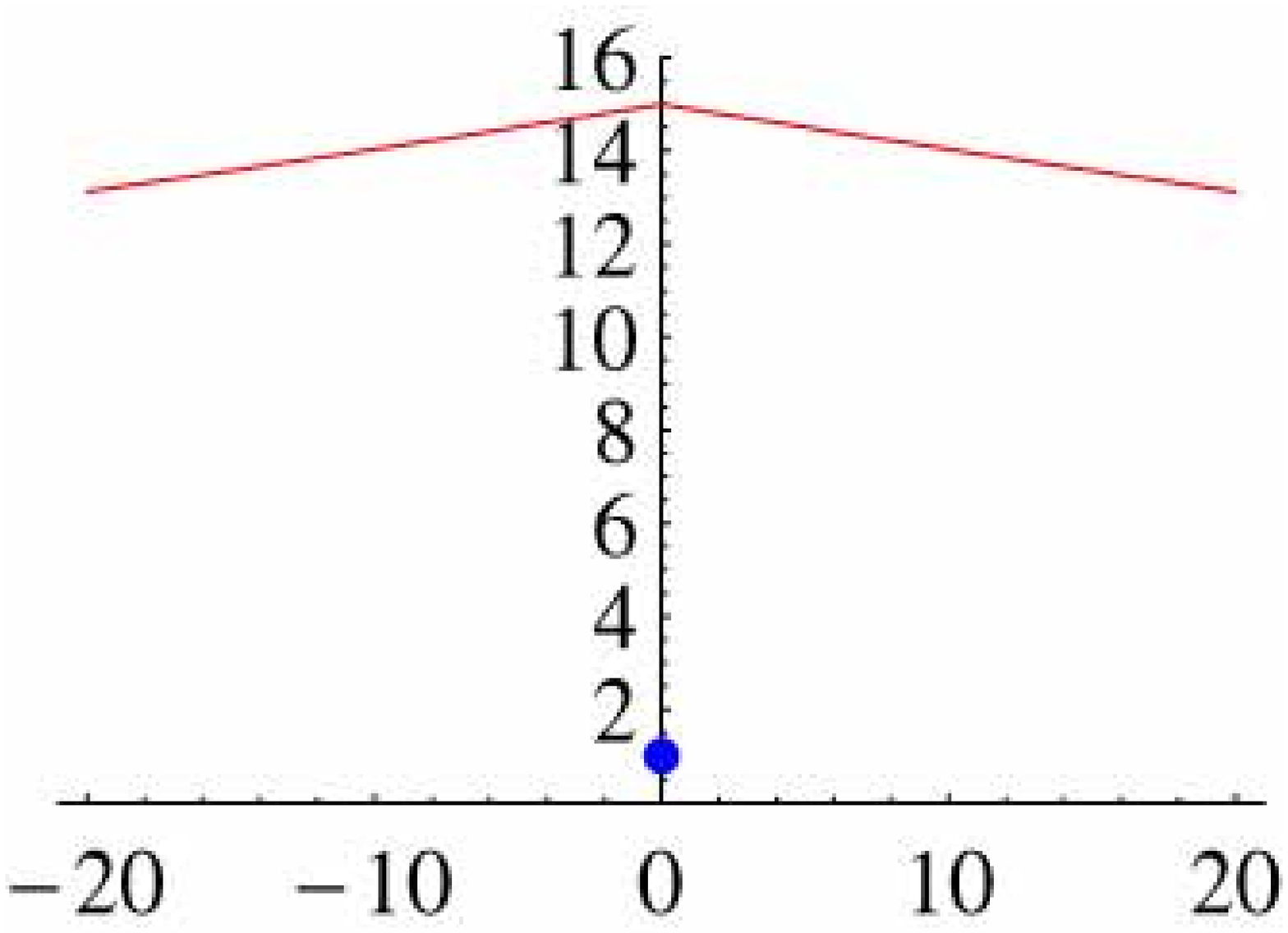}
\hfill
\includegraphics[width=0.24\textwidth]{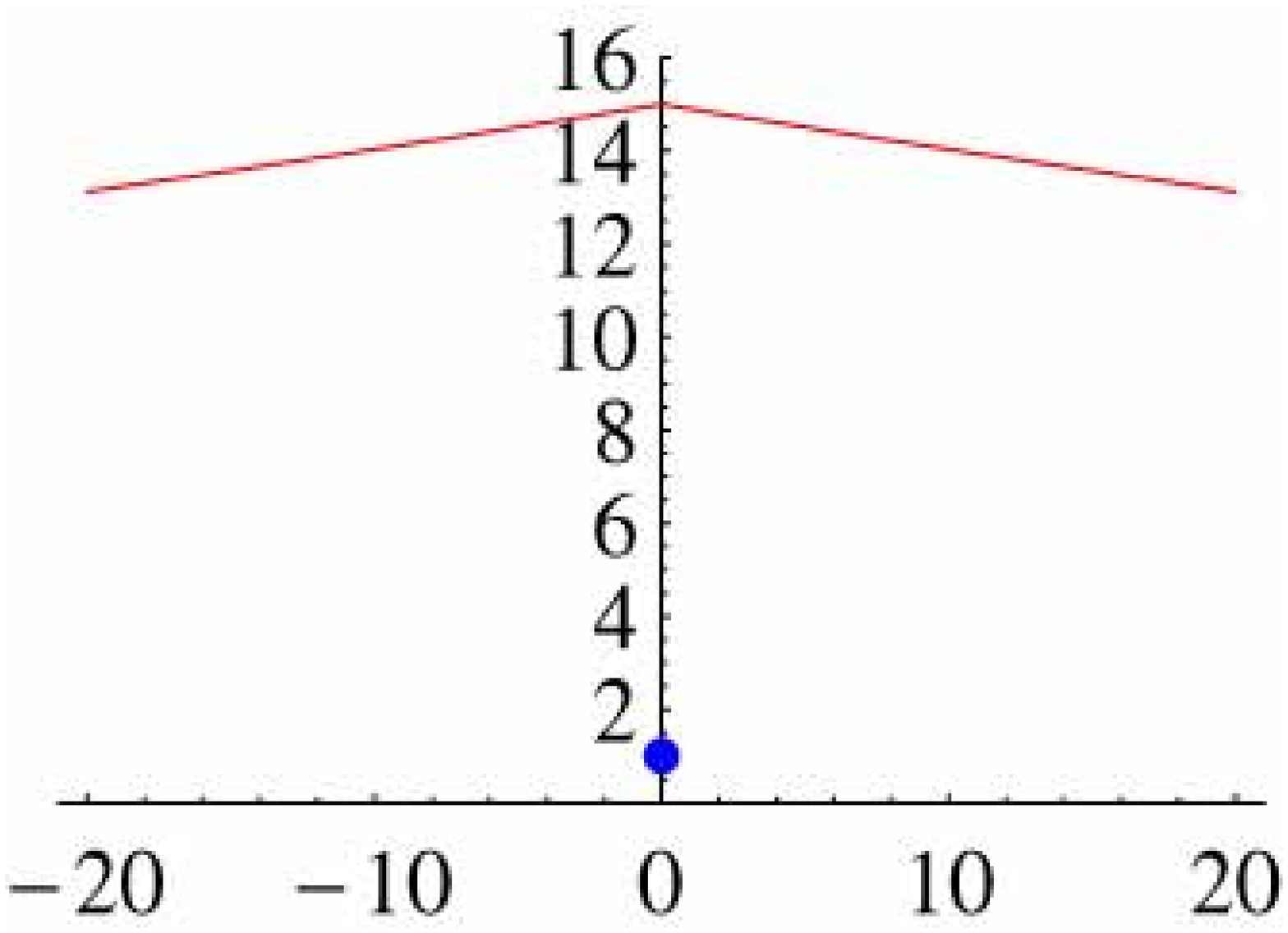}
\\
\vskip \baselineskip
\includegraphics[width=0.24\textwidth]{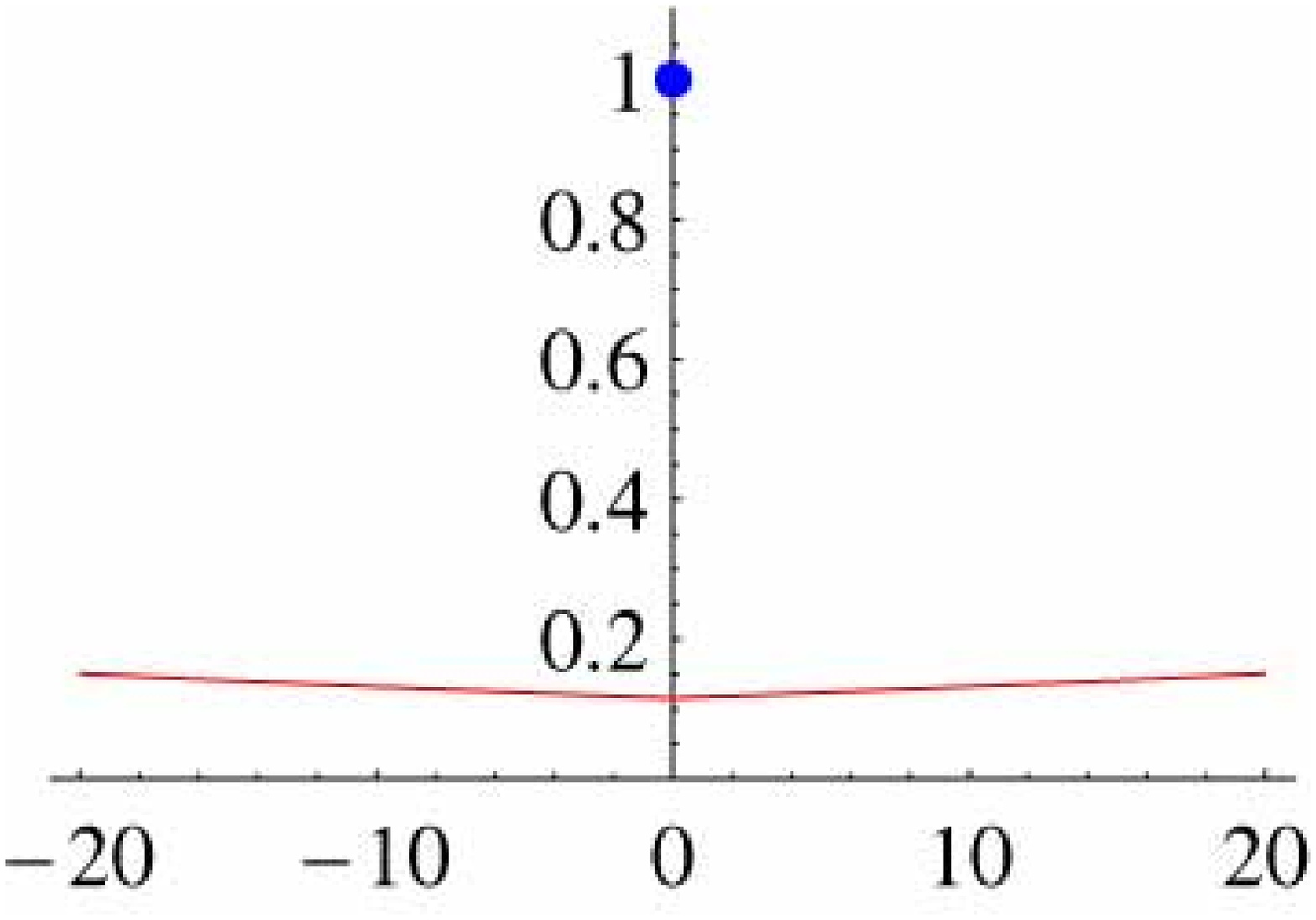}
\hfill
\includegraphics[width=0.24\textwidth]{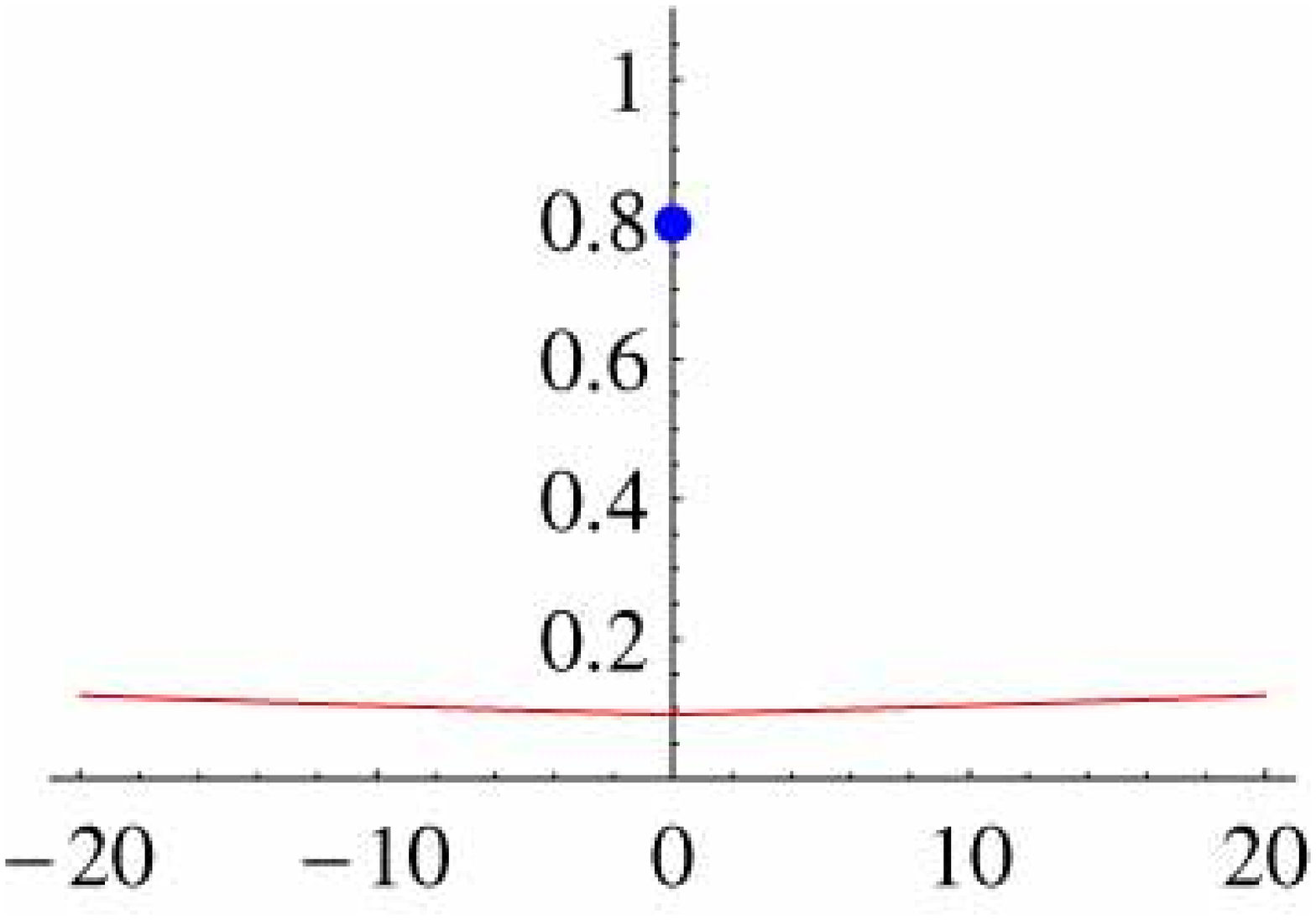}
\hfill
\includegraphics[width=0.24\textwidth]{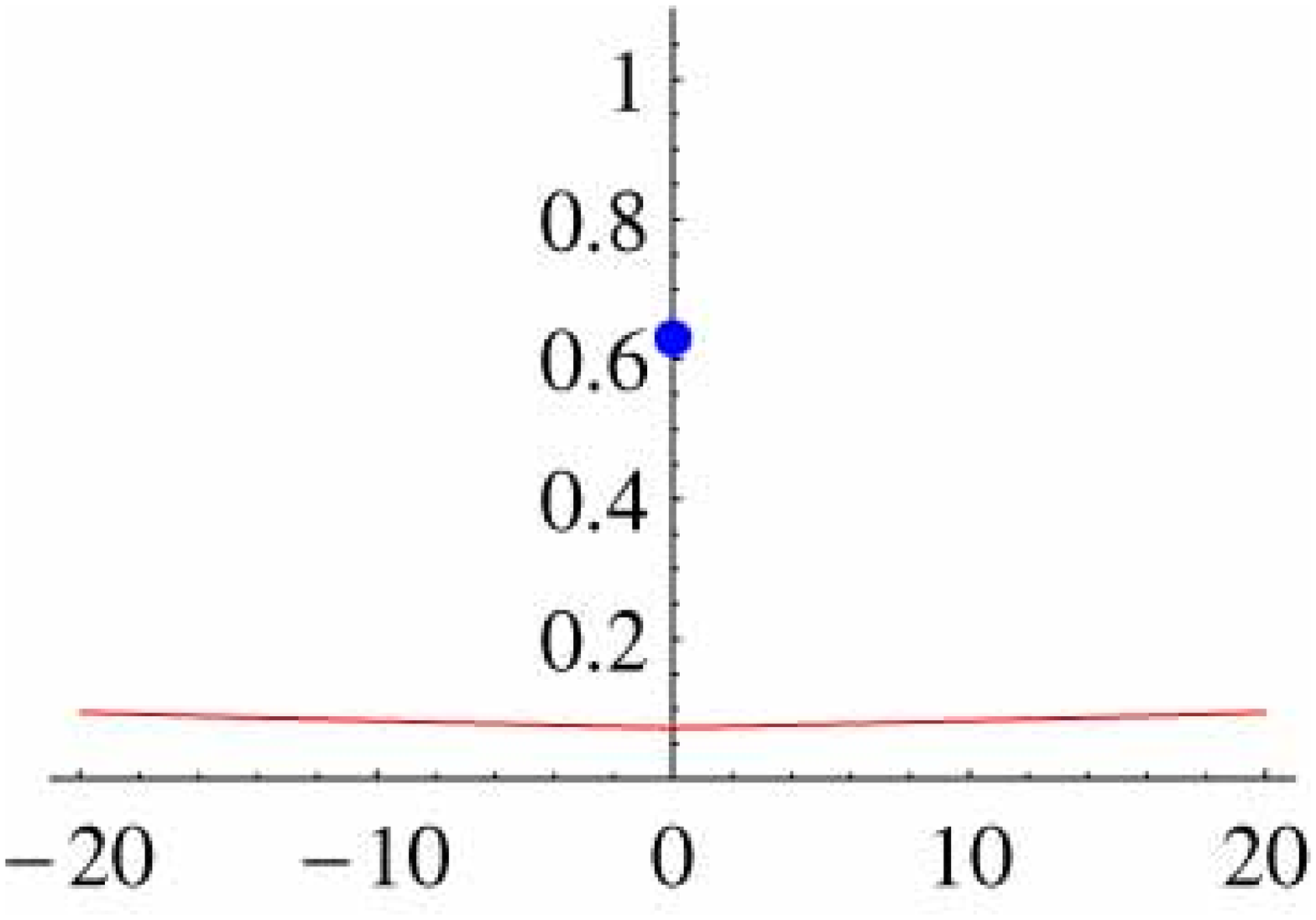}
\hfill
\includegraphics[width=0.24\textwidth]{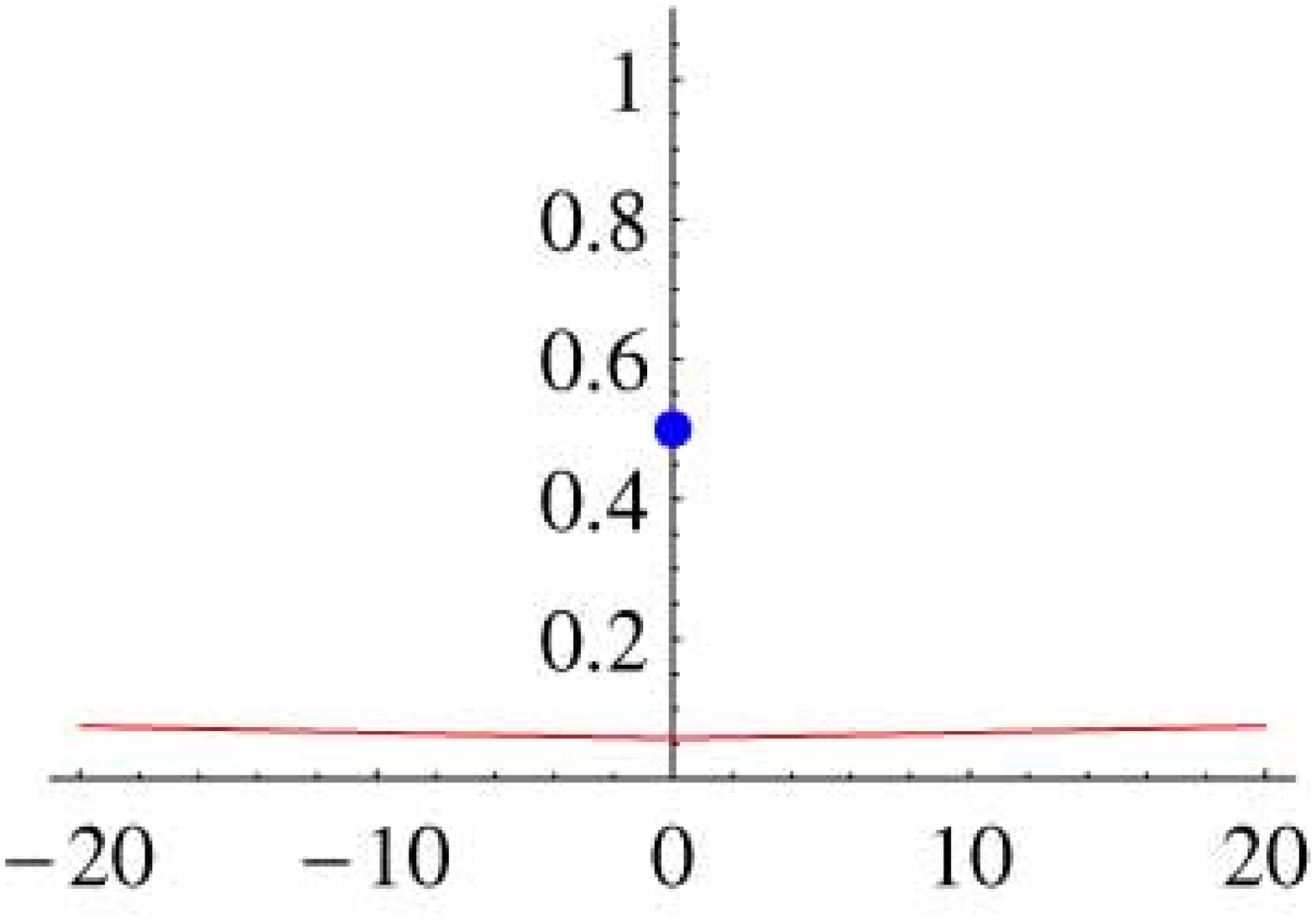}
\\
\vskip \baselineskip
\includegraphics[width=0.24\textwidth]{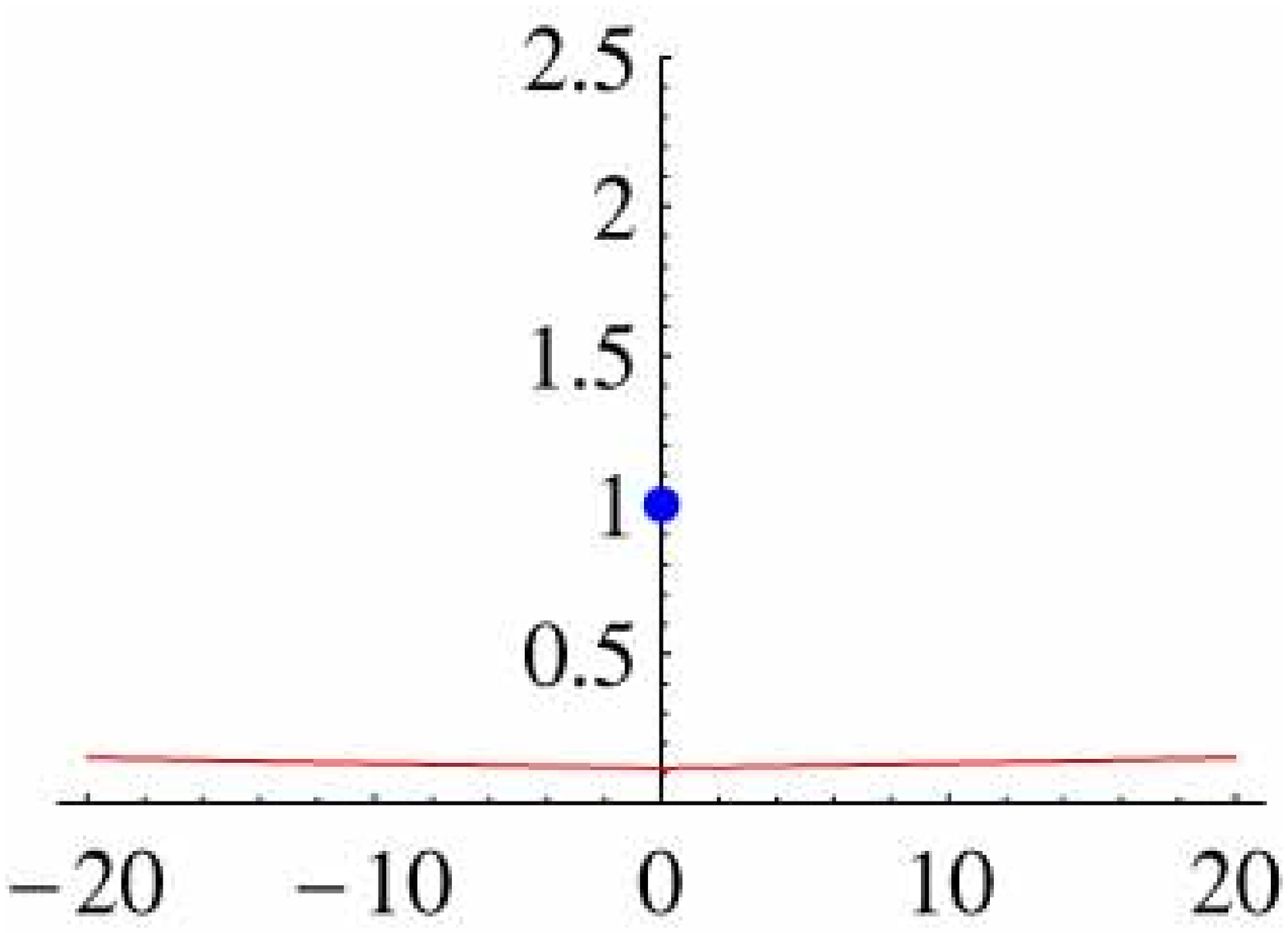}
\hfill
\includegraphics[width=0.24\textwidth]{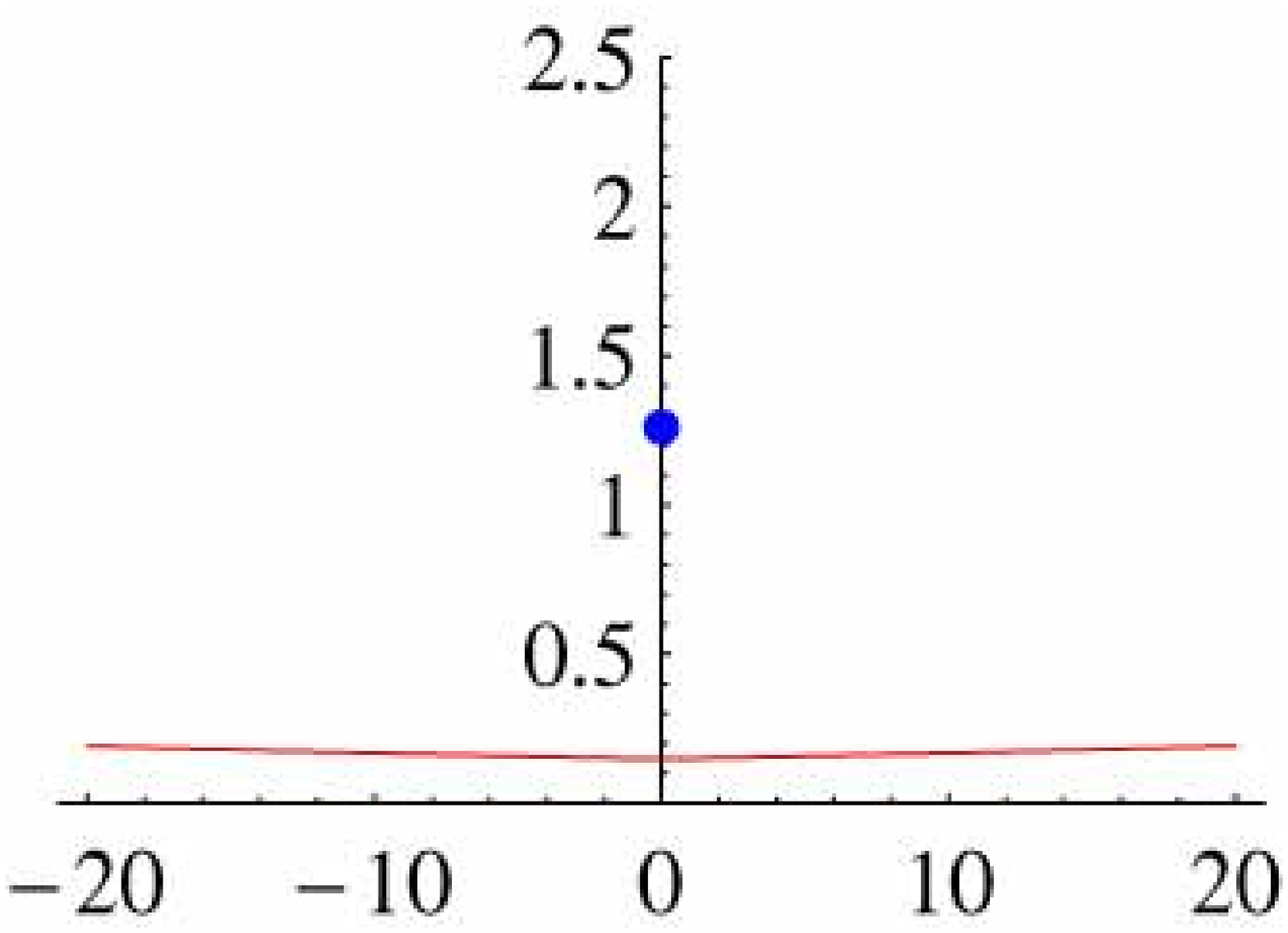}
\hfill
\includegraphics[width=0.24\textwidth]{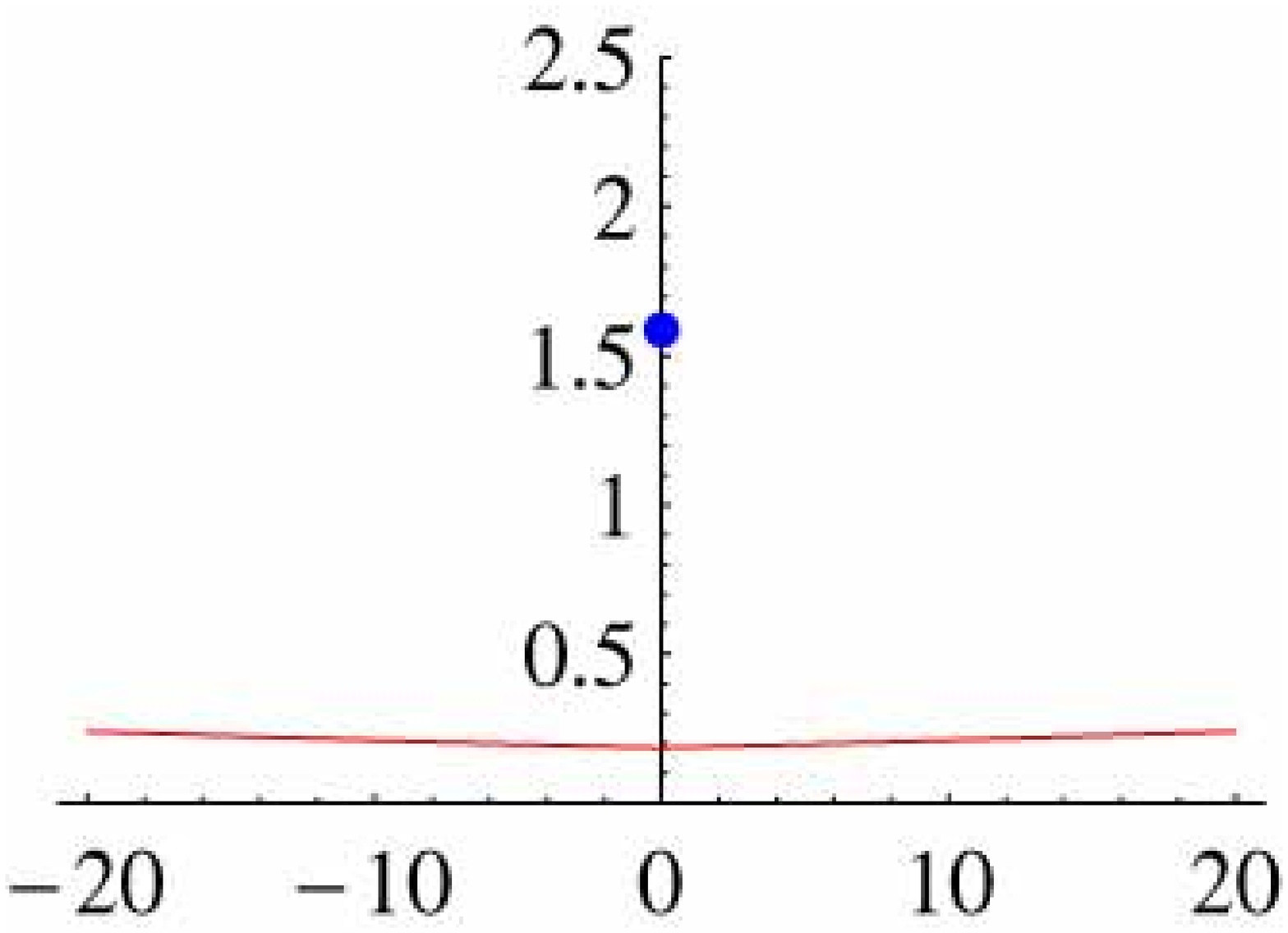}
\hfill
\includegraphics[width=0.24\textwidth]{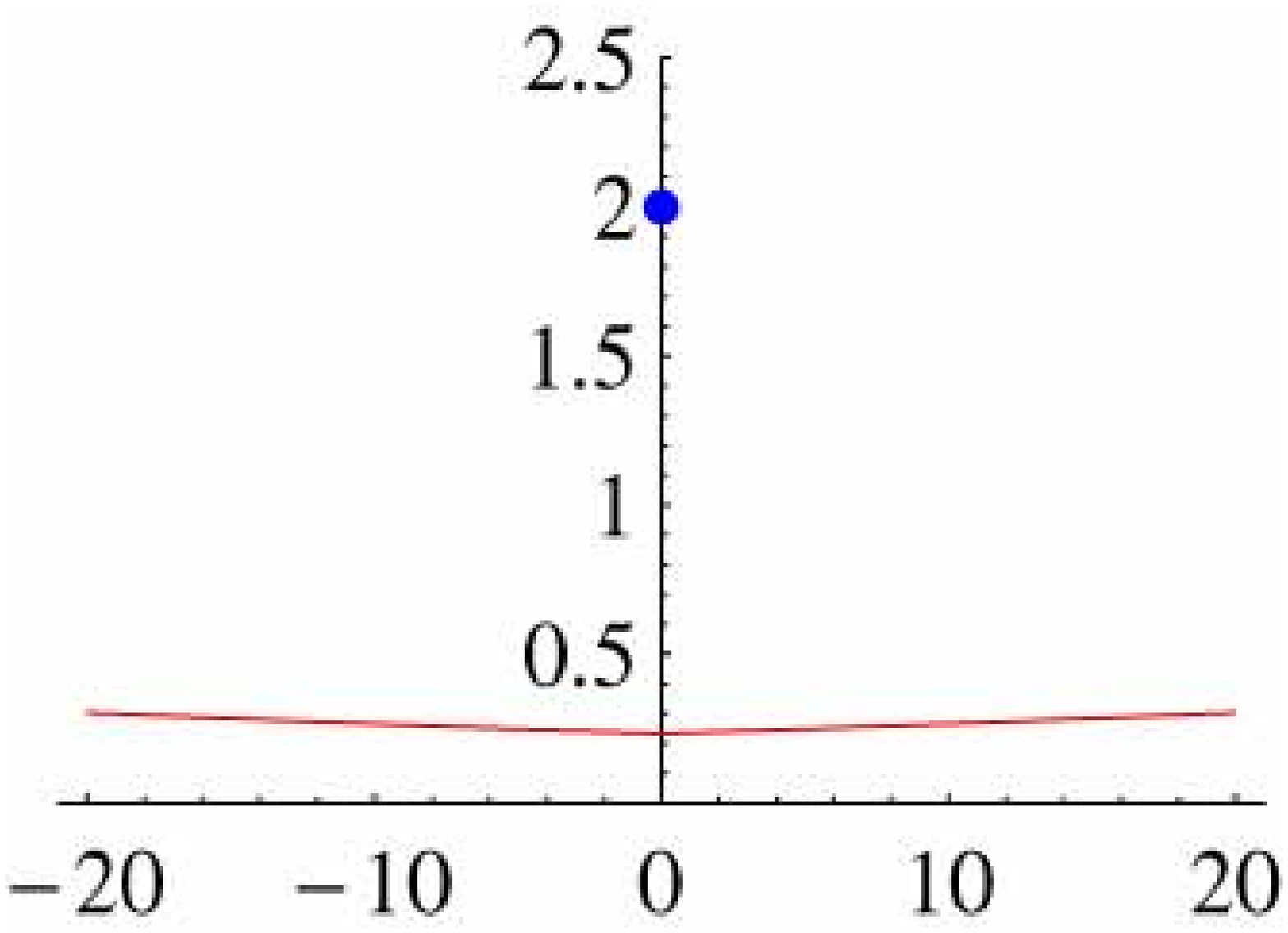}
\caption{The time evolution of the eigenfunctions at $t=0,20,40,60$ for $\tilde{g}=0.1$ and $\tilde{E}_d=-1/2$ is shown in the region $-20\leq x \leq 20$:
from top to bottom, the time evolution of the solid line (a), the broken line (b), the dotted line (c) and the chained line (d) in Fig.~\ref{fig-sols}, or the solutions a, b, c and d in Fig.~\ref{fig-sol4}, respectively; from left to right $t=0,20,40,60$.
The solutions a and b (the first and second rows) are bound states, and hence there is no time evolution in fact.
The solutions c and d (the third and fourth rows) are resonant states.
In each panel, the dot indicates $\Psi_{\mathrm{res},n}(d,t)$ and the curve indicates $|\Psi_{\mathrm{res},n}(x,t)|$.}
\label{fig-evol}
\end{figure*}

We then solved the time-dependent Schr\"{o}dinger equation
\begin{eqnarray}\label{eq6-180}
\lefteqn{
i\hbar\frac{\partial}{\partial t}\Psi_{\mathrm{res},n}(x,t)
=\hat{\mathcal{H}}_\mathrm{eff}\Psi_{\mathrm{res},n}(x,t)
}
\nonumber\\
&=&
\left\{\begin{array}{l}
\displaystyle
-\frac{t_\mathrm{h}}{2}\Psi_{\mathrm{res},n}(x+\Delta x,t)
\\
\qquad
+V_{\mathrm{eff},n}\Psi_{\mathrm{res},n}(x,t)
\quad\mbox{for $\displaystyle x=-L$}
\\
\\
\displaystyle
-\frac{t_\mathrm{h}}{2}\left(\Psi_{\mathrm{res},n}(x-\Delta x,t)+\Psi_{\mathrm{res},n}(x+\Delta x,t)\right)
\qquad\quad
\\
\qquad
+V(x)\Psi_{\mathrm{res},n}(x,t)
\quad\mbox{for $\displaystyle -L<x<L$}
\\
\\
\displaystyle
-\frac{t_\mathrm{h}}{2}\Psi_{\mathrm{res},n}(x-\Delta x,t)
\\
\qquad
+V_{\mathrm{eff},n}\Psi_{\mathrm{res},n}(x,t)
\quad\mbox{for $\displaystyle x=L$}
\end{array}\right.
\end{eqnarray}
in the range $-L\leq x \leq L$ by the Runge-Kutta method, taking the eigenfunctions $\{\psi_{\mathrm{res},n}(x)\}$ as the initial conditions.
The time evolution is partly shown in Fig.~\ref{fig-evol} for $\tilde{g}=0.1$ and $\tilde{E}_d=-1/2$.
The bound states a and b (the first and the second rows in Fig.~\ref{fig-evol}) are time-independent and stable, while the resonant state c in the third row decays in time and the state d in the fourth row grows exponentially in time.

It should be emphasized that if we would solve the original Schr\"{o}dinger equation simply by truncation of the domain of $x$ without introducing the effective potential, we could not obtain such accurate numerical results as in Figs.~\ref{fig-init} and~\ref{fig-evol}.
This is especially true for the resonant eigenstates with the complex eigenvalues, which diverge in $x$.

We have applied the above technique to a more complicated model on a ladder lattice.
As a result, we have found a quasi-stable resonant state with a very long lifetime for quite a wide range or parameters.
The state appears to be a bound state around the impurity, diverging only far away from the scattering center, and barely decays for quite a long time.
This finding will be reported elsewhere~\cite{Nakamura07}.

\section{Summary}
\label{sec7}

In the present paper, we have presented some properties on resonant states of open quantum systems in two parts.
In the first part, we mostly gave a review on properties of quantum-mechanical resonant states.
We nevertheless included two new points: 
(i) the imaginary part of the energy expectation is proportional to a momentum flux going out of the system for \textit{arbitrary} wave functions;
(ii) the number of particles in a resonant state is \textit{conserved} when we count the number in an expanding volume.
We also give an example where a ``resonance peak'' of the conductance is not accompanied by a resonant state.

In the second part, we gave numerical methods of finding a resonant state and computing its dynamics, using the effective Hamiltonian.
The method of finding a resonance pole is independent of the complex scaling and is applicable to singular potentials and lattice models.
This method is very efficient, showing the rapid convergence to the exact solution.
The method of computing the time evolution of the divergent resonant eigenfunctions was demonstrated for the Friedrichs-Fano model.

We have intentionally avoided discussing the completeness relation and the resonant-state expansion~\cite{Hokkyo65,Romo68,Berggren70,Gyarmati71,Romo80,Berggren82,Berggren96,Madrid05}.
We are planning to report discussions on the topic elsewhere.

\section*{Acknowledgments}
The authors thank Prof.~Satoshi Tanaka for useful discussions.
One of the authors (N.H.) expresses his sincere gratitude to Prof.~Kiyoshi Kato and Dr.~Takayuki Myo for helpful discussions and comments at the early stage of the present study and particularly Dr.~Myo for helpful comments on the final draft.
N.H.~also thanks Prof.~Takeo Kato for useful discussions.
The work is supported partly by the Murata Science Foundation as well as by the National Institutes of Natural Sciences undertaking Forming Bases for Interdisciplinary and International Research through Cooperation Across Fields of Study and Collaborative Research Program (No. NIFS06KDBT005).
One of the authors (N.H.) acknowledges support by Grant-in-Aid for Scientific Research (No.~17340115) from the Ministry of Education, Culture, Sports, Science and Technology as well as support by Core Research for Evolutional Science and Technology (CREST) of Japan Science and Technology Agency, by the Sumitomo Foundation and by the Casio Science Promotion Foundation.
The computation in this work was carried out partly on the facilities of the Supercomputer Center, Institute for Solid State Physics, University of Tokyo.
One of the authors (T.P.) is grateful to the Foundation for the Promotion of Industrial Science for its support during his stay at the Institute of Industrial Science, University of Tokyo.

\appendix

\section{Expressions in the three-dimensional case}
\label{appA}

In the main text, we focused on the one-dimensional case~(\ref{eq2-10}).
In the present Appendix, we give the corresponding expressions in the three-dimensional case,
$\hat{\mathcal{H}}=\hat{\mathcal{K}}+\hat{\mathcal{V}}$,
where
\begin{equation}\label{eqA-20}
\hat{\mathcal{K}}\equiv\frac{{\vec{p}}^2}{2m}=-\frac{\hbar^2}{2m}\vec{\nabla}^2,
\quad\mbox{and}\quad
\hat{\mathcal{V}}\equiv V(\vec{x}).
\end{equation}

\subsection{The imaginary part of the resonant eigenvalue}
\label{appA-1}

Instead of Eq.~(\ref{eq2-40}), we define the Hamiltonian expectation as
\begin{equation}\label{eqA-40}
\langle\psi|\hat{\mathcal{H}}|\psi\rangle_\Omega
\equiv
\int\!\!\!\!\int\!\!\!\!\int_\Omega \psi(\vec{x})^\ast \hat{\mathcal{H}} \psi(\vec{x}) dV,
\end{equation}
where we consider an integration volume $\Omega$ that includes the support of the potential support, $\Omega_\mathrm{pot}$.
The transformation in Eq.~(\ref{eq2-70}) then corresponds to Gauss's theorem:
\begin{eqnarray}\label{eqA-50}
\langle\psi|\hat{\mathcal{K}}|\psi\rangle_\Omega
&=&
-\frac{\hbar^2}{2m}\int\!\!\!\!\int\!\!\!\!\int_\Omega \psi(\vec{x})^\ast \vec{\nabla}^2 \psi(\vec{x}) dV
\nonumber \\
&=&
\frac{\hbar^2}{2m}\int\!\!\!\!\int\!\!\!\!\int_\Omega\vec{\nabla}\psi(\vec{x})^\ast\cdot
\vec{\nabla}\psi(\vec{x})
dV
\nonumber\\
&&
-\frac{\hbar^2}{2m}\int\!\!\!\!\int_{\partial\Omega}\psi(\vec{x})^\ast \vec{\nabla}\psi(\vec{x}) \cdot d\vec{S}.
\end{eqnarray}
In the second term of the last line, the integration is carried out over the surface $\partial\Omega$ of the volume $\Omega$ and $d\vec{S}$ is a vector normal to the surface and of the magnitude of the surface element;
hence only the normal component of the differentiation $\vec{\nabla}$ contributes to the integral.
Subtracting from Eq.~(\ref{eqA-50}) its complex conjugate, we have
\begin{equation}\label{eqA-70}
2i\mathop{\mathrm{Im}}
\langle\psi|\hat{\mathcal{H}}|\psi\rangle_\Omega
=
-\frac{i\hbar}{m}\mathop{\rm Re}\int\!\!\!\!\int_{\partial\Omega}
\psi(\vec{x})^\ast \vec{p} \psi(\vec{x}) \cdot d\vec{S},
\end{equation}
which is followed by the formula~(\ref{eq2-110}), where we define
\begin{equation}\label{eqA-75}
\langle\psi|\hat{p}_\mathrm{n}|\psi\rangle_{\partial\Omega}
\equiv
\int\!\!\!\!\int_{\partial\Omega}
\psi(\vec{x})^\ast \vec{p} \psi(\vec{x}) \cdot d\vec{S}.
\end{equation}
The expressions in Sec.~\ref{sec2-2} do not differ much in the three-dimensional case.

\subsection{The $S$ matrix in three dimensions}
\label{appA-2}

We here show in three dimensions the equivalence stated in Sec.~\ref{sec3-1}, that is, the equivalence of seeking the singularities of the $S$ matrix and solving the Schr\"{o}dinger equation under the boundary condition of the outgoing wave only.

Let us first review the theory of the $S$ matrix in three dimensions.
We show that singularities of the $S$ matrix can have large contributions to the scattering cross section.
Consider scattering on the basis of the Schr\"{o}dinger equation
\begin{equation}\label{eqA-750}
\left(-\frac{\hbar^2}{2m}\vec{\nabla}^2+V(\vec{x})\right)\psi(\vec{x})=\frac{\hbar^2k^2}{2m}\psi(\vec{x}).
\end{equation}
By assuming the solution of the form
\begin{equation}\label{eqA-755}
\psi(\vec{x})\simeq
e^{ikz}+\frac{e^{ikr}}{r}f(\theta;k)
\quad\mbox{as}\quad
\left|\vec{x}\right|\to\infty,
\end{equation}
we have the differential cross section as
\begin{equation}\label{eqA-758}
\frac{d\sigma}{d\Omega}=\left|f(\theta;k)\right|^2.
\end{equation}
Thus we need to obtain the function $f(\theta;k)$.
For the purpose, we expand the wave function~(\ref{eqA-755}) in the partial waves.
First, the incident wave is expanded as
\begin{eqnarray}\label{eqA-760}
e^{ikz}&=&\sum_{l=0}^\infty
(2l+1)i^lj_l(kr)P_l(\cos\theta)
\\ \label{eqA-765}
&\simeq&
\frac{1}{2ikr}\sum_{l=0}^\infty
(2l+1)\left[e^{ikr}-(-1)^le^{-ikr}\right]P_l(\cos\theta)
\nonumber\\
&&\qquad\qquad\qquad\qquad\qquad\mbox{as}\quad
r\to\infty,
\end{eqnarray}
where $j_l$ denotes the spherical Bessel function and $P_l$ is the Legendre polynomial.
We used an asymptotic form of the spherical Bessel function in moving from Eq.~(\ref{eqA-760}) to Eq.~(\ref{eqA-765}).
Next, the scattered wave is expanded as
\begin{equation}\label{eqA-770}
\frac{e^{ikr}}{r}f(\theta;k)
=-\frac{e^{ikr}}{2ikr}\sum_{l=0}^\infty
(2l+1)a_l(k)P_l(\cos\theta)
\end{equation}
with some coefficients $\{a_l(k)\}$.
By summing up Eqs.~(\ref{eqA-765}) and~(\ref{eqA-770}), we have
\begin{eqnarray}\label{eqA-775}
\lefteqn{
\psi(\vec{x})\simeq
\frac{1}{2ikr}\times
}
\nonumber\\
&&
\sum_{l=0}^\infty
(2l+1)\left[S_l(k)e^{ikr}-(-1)^le^{-ikr}\right]P_l(\cos\theta)
\nonumber\\
&&
\qquad\qquad\qquad\qquad\qquad\quad\mbox{as}\quad
r\to\infty,
\end{eqnarray}
where
\begin{equation}\label{eqA-780}
S_l(k)\equiv 1-a_l(k)
\end{equation}
is called the $S$ matrix.
The cross section~(\ref{eqA-758}) is calculated from the $S$ matrix as
\begin{equation}\label{eqA-790}
\frac{d\sigma}{d\Omega}=\frac{1}{4k^2}\sum_{l=0}^\infty(2l+1)\left|1-S_l(k)\right|^2.
\end{equation}
The singularities of the $S$ matrix thus can affect the cross section largely.
In the conventional theory, the resonance is defined as the singularity of the $S$ matrix.

We next review how we can compute the $S$ matrix.
For the purpose, we introduce the Jost solutions and the Jost functions.
We show that the singularities of the $S$ matrix is caused by the zeros of a Jost function.
Consider a solution of the Schr\"{o}dinger equation with the momentum $k$ and the angular momentum $l$:
\begin{equation}\label{eqA-800}
\psi(r,\theta,\phi)=\frac{\chi(r;k)}{r}Y_{lm}(\theta,\phi).
\end{equation}
For simplicity, we drop the subscript $l$ for various quantities hereafter.
The equation for $\chi(r;k)$ is given by
\begin{equation}\label{eqA-810}
\frac{d^2}{d r^2}\chi(r;k)
-\left[\frac{2m}{\hbar^2}V(r)+\frac{l(l+1)}{r^2}\right]\chi(r;k)=k^2\chi(r;k),
\end{equation}
where we consider only bounded potentials satisfying
\begin{equation}\label{eqA-815}
\lim_{r\to0}r^2V(r)=0,
\quad\mbox{and}\quad
\lim_{r\to\infty}rV(r)=0.
\end{equation}

Near the origin $r\simeq 0$, Eq.~(\ref{eqA-810}) is reduced to
\begin{equation}\label{eqA-820}
\frac{d^2}{d r^2}\chi(r;k)\simeq\frac{l(l+1)}{r^2}\chi(r;k),
\end{equation}
and hence there are two solutions of the forms $r^{l+1}$ and $r^{-l}$.
The physical solution must be regular at the origin.
We thus choose the solution which satisfies
\begin{equation}\label{eqA-445}
\chi(r;k)\simeq r^{l+1}
\quad\mbox{as}\quad
r\to0.
\end{equation}

Far away from the origin, on the other hand, there are two potential-free solutions of the forms $\exp(\pm ikr)$.
Let $f_\pm(r;k)$ denote the solutions that satisfy
\begin{equation}\label{eqA-840}
f_\pm(r;k)\simeq e^{\pm ikr}
\quad\mbox{as}\quad
r\to\infty.
\end{equation}
These are called the Jost solutions.
The physical solution $\chi(r;k)$ is a superposition of these two Jost solutions:
\begin{equation}\label{eqA-850}
\chi(r;k)=a_+(k)f_+(r;k)+a_-(k)f_-(r;k)
\end{equation}
with some coefficients $a_\pm(k)$.
For later convenience, we change the notation as follows:
\begin{equation}\label{eqA-855}
\chi(r;k)=\frac{1}{2ik}\left(
f_-(k)f_+(r;k)-f_+(k)f_-(r;k)
\right);
\end{equation}
that is,
\begin{equation}\label{eqA-856}
f_\mp(k)=\pm 2ika_\pm(k),
\end{equation}
which are called the Jost functions.
Note that the Jost solutions~(\ref{eqA-840}), near the origin, are in return superpositions of the solutions of the forms $r^{l+1}$ and $r^{-l}$:
\begin{equation}\label{eqA-860}
f_\pm(r;k)\simeq
b_\pm(k) r^{l+1}+c_\pm(k) r^{-l}
\quad\mbox{as}\quad
r\to0
\end{equation}
with some coefficients $b_\pm(k)$ and $c_\pm(k)$.
The superposition~(\ref{eqA-855}), in the limit $r\to\infty$, behaves as
\begin{equation}\label{eqA-865}
\chi(r;k)\simeq
\frac{f_+(k)}{2ik}\left(\frac{f_-(k)}{f_+(k)}e^{ikr}-e^{-ikr}\right)
\quad\mbox{as}\quad
r\to\infty.
\end{equation}
Comparing the asymptotic forms~(\ref{eqA-865}) and~(\ref{eqA-775}), we know that the $S$ matrix is calculated from the Jost functions as
\begin{equation}\label{eqA-868}
S_l(k)=(-1)^l\frac{f_-(k)}{f_+(k)}.
\end{equation}
Therefore large contributions to the cross section come from the zeros of the Jost function $f_+(k)$.

We can show that the Jost functions, or the superposition coefficients $f_\pm(k)$ in Eq.~(\ref{eqA-855}), is obtained from the intercept of the Jost solutions at $r=0$ as
\begin{equation}\label{eqA-930}
f_\pm(k)=(2l+1)c_\pm(k)=(2l+1)\lim_{r\to0}r^lf_\pm(r;k).
\end{equation}
For the present purpose, however, we simply put $f_+(k)=0$ in Eq.~(\ref{eqA-855}), which is reduced to
\begin{equation}\label{eqA-450}
\chi(r;k)\simeq \frac{1}{2ik}f_-(k)f_+(r;k)\propto e^{ikr}.
\end{equation}
This means that we should seek a particular form of the solution $\chi(r;k)$ under the two boundary conditions~(\ref{eqA-445}) and~(\ref{eqA-450}).
This is nothing but solving the Schr\"{o}dinger equation under the boundary condition of the outgoing wave only.
The only difference is that we normally solve the wave function~$\chi(r;k)$ \textit{outward} (starting from the inner boundary condition~(\ref{eqA-445}) and ending with the outer boundary condition~(\ref{eqA-855})), while we normally obtain the Jost function~$f_\pm(r;k)$ \textit{inward} (starting from the outer boundary condition~(\ref{eqA-840}) and ending with the inner boundary condition~(\ref{eqA-860})).

Note that in solving the above problem with the two conditions~(\ref{eqA-445}) and~(\ref{eqA-450}), we have two undetermined variables $k$ and $f_-(k)$.
Hence we obtain the solutions for discrete values of $k$.

\end{document}